\documentclass[twocolumn]{aastex631}

\usepackage{braket}
\usepackage{bm}
\usepackage{amsmath,amssymb}
\usepackage{subfigure}
\usepackage{xcolor}
\usepackage[normalem]{ulem}
\usepackage{graphicx}
\usepackage{float}
\usepackage{comment}
\usepackage{sidecap}
\usepackage{ulem}
\usepackage{multirow}
\usepackage{natbib}

\begin{document}

\newcommand{\gshvec}[3]{\left(\begin{matrix} #1 \\ #2 \\ #3 \end{matrix}\right)\,}

\newcommand{\sbh}[1]{{#1}}

\title{Detectability of axisymmetric magnetic fields from the core to the surface of oscillating post-main sequence stars}

\correspondingauthor{Shatanik Bhattacharya}\leavevmode\\
\email{shatanik.bhattacharya@tifr.res.in}

\author[0009-0002-8330-3391]{Shatanik Bhattacharya}
\affil{Department of Astronomy and Astrophysics \\
Tata Institute of Fundamental Research \\
Mumbai, India}

\author[0000-0003-0896-7972]{Srijan Bharati Das}
\affiliation{Center for Astrophysics | Harvard \& Smithsonian \\
60 Garden Street, Cambridge, MA 02138, USA\\}
\affiliation{Institute of Science and Technology Austria (IST Austria) \\
Am Campus 1, Klosterneuburg, Austria}

\author[0000-0003-0142-4000]{Lisa Bugnet}
\affiliation{Institute of Science and Technology Austria (IST Austria) \\
Am Campus 1, Klosterneuburg, Austria}

\author{Subrata Panda}
\affil{Department of Astronomy and Astrophysics \\
Tata Institute of Fundamental Research \\
Mumbai, India}

\author[0000-0003-2896-1471]{Shravan M. Hanasoge}
\affil{Department of Astronomy and Astrophysics \\
Tata Institute of Fundamental Research \\
Mumbai, India}

\begin{abstract}

Magnetic fields in the stellar interiors are key candidates to explain observed core rotation rates inside solar-like stars along their evolution.
Recently, asteroseismic estimates of radial magnetic field amplitudes near the hydrogen-burning shell (H-shell) inside about 24 red-giants (RGs) have been obtained by measuring frequency splittings from their power spectra. 
Using general Lorentz-stress (magnetic) kernels, we investigated the potential for detectability of near-surface magnetism in a $1.3 \, M_{\odot}$  star of super-solar metallicity as it evolves from a mid sub-giant to a late sub-giant into an RG. Based on these sensitivity kernels, we decompose an RG into three zones --- deep core, H-shell, and near-surface. The sub-giants instead required decomposition into an inner core, an outer core, and a near-surface layer. 
Additionally, we find that for a low-frequency $g$-dominated dipolar mode in the presence of a typical stable magnetic field, $\sim$25\% of the frequency shift comes from the H-shell and the remaining from deeper layers. 
The ratio of the subsurface tangential field to the radial field in H-burning shell decides if subsurface fields may be potentially detectable. For $p$-dominated dipole modes close to $\nu_{\rm{max}}$, this ratio is around two orders of magnitude smaller in subgiant phases than the corresponding RG. Further, with the availability of magnetic kernels, we propose lower limits of field strengths in crucial layers in our stellar model during its evolutionary phases. The theoretical prescription outlined here provides the first formal way to devise inverse problems for stellar magnetism and can be seamlessly employed for slow rotators.

\end{abstract}


\section{Introduction} \label{sec:intro}

Magnetic fields in stars are one of the largest sources of uncertainty in stellar modeling, as both the amplitude and the topology matter when accounting for their effect on the evolution of stars. As a result of a lack of magnetic field observation below the stellar surface, magnetic fields and their impact on dynamical processes have been largely excluded from stellar evolution models in the past decades. 
\sbh{However, there have been extensive studies} discussing how such magnetic fields can be formed and how they evolve through the evolutionary phases and processes \citep[eg. ][and reviews by \citeauthor{Donati_Review_2009} \citeyear{Donati_Review_2009} and \citeauthor{Braithwaite_Review2017} \citeyear{Braithwaite_Review2017}]{Braithwaite2004_Nature, BraithwaiteNorlund2006,Ferrario_2009,Mestel2010,Takahashi2021}, starting with models of star formation from the interstellar medium \citep[eg.][]{Mestel_1966,Mestel1967} or from later convective dynamo episodes \citep[eg.][]{tobias_2021,ConvDynamoYanetal2022}. 
A fraction of the magnetic flux of primordial origin or dynamo origin (from convective-core dynamo in intermediate-mass main-sequence stars, hereafter MS) might be conserved during stellar core contraction \citep{Toutetal2004} and stabilized in the radiative cores of evolved stars such as red giant (RG) stars before their envelopes are eventually shed. About 12\% of white dwarfs, descendants of RGs, present strong large-scale magnetic fields at the surface \citep[with amplitudes ranging from $10^6 - 10^9$ G, see][]{Landstreetetal2012,Bagnulo2021, WD}. They might represent proof of the survival of stable fields inside stellar radiative interiors.  
While spectropolarimetry allows us to characterize surface magnetism \citep[e.g.][]{Donati,Auriere2015,Marsden2014}, we need asteroseismology to unveil fields below the surface.

Study of mixed acoustic $p$ and gravity $g$ modes \citep[hereafter called mixed modes,][]{Becketal2011}, which simultaneously probe processes in the core and the envelope, reveal unexpectedly slow internal rotation profiles \cite[][]{Deheuvels2012, Deheuvels2014, Mosseretal2012, Mosser2017_vs_Fuller, Gehanetal2018} in post-MS stars like subgiants and red-giant branch stars (hereon SG and RGB respectively). 
Among other candidates \citep[e.g. transport by waves and modes as in][]{Talon2005, Rogers2015, Belkacem2015a, Belkacem2015b, Pincon2017}, \sbh{\cite{garaud2008dynamics}}
magnetic fields might lead to such a slow rotation rate in stellar interiors \citep{Mestel1987, Spruit2002, MathisZahn2004,MathisZahn2005, Fuller2019, Eggenberger2022, Moyano2022}.
This scenario of magnetized radiative interiors is supported by a significant fraction of RG showing anomalously low amplitudes of dipolar and quadrupolar mixed modes \citep{Garciaetal2014, Stelloetal2016a, Stelloetal2016b}. 
\sbh{Indeed, magnetic fields in RG cores could trap energies of the $g$-dominant mixed modes via processes like the {magnetic greenhouse effect} or conversion to slow magnetoacoustic waves when their strengths are beyond a threshold \citep[][]{Fuller2015,Lecoanet2017} thus removing \citep[or decreasing;][]{Mosser2017_vs_Fuller} their contribution in the oscillation amplitude observed at the surface. We also refer the readers to \cite{LoiPapa2017,LoiPapa2018,Weinberg2021} for other mechanisms that could also explain the observed low amplitude dipolar mixed modes.}

Studies investigating the effects of magnetic fields on mixed-mode oscillation frequencies have therefore become fundamental for the understanding of stellar evolution. \cite{Unno1989} laid a detailed groundwork for using perturbation theory to compute alterations to oscillation frequencies, and calculated the splittings as would be seen for a pure dipolar magnetic field. \cite{GoughThompson} proceeded to deduce the same for purely toroidal and purely poloidal magnetic fields with rotation, and also computed sensitivity kernels for acoustic modes to pure toroidal fields in the Sun. \cite{Hasanetal05} neglected structural alterations in the star due to Lorentz forces and derived an expression for splittings in $g$ modes. This was later implemented by
\cite{GomesLopes2020} using poloidal topologies. 
\cite{Loi2020}, \cite{Loi2021}, \cite{Bugnetetal2021}, \cite{MathisBugnetetal2021}, \cite{Bugnet2022} have incorporated some of these aforementioned concepts to investigate the impact of a buried stable mixed poloidal and toroidal configuration \citep[][]{Braithwaite2008, Duez2010A, Duez2010B} on mixed-mode frequencies. \cite{GangLietal2022} extended the theoretical landscape with the prescription for a general magnetic field topology trapped inside the radiative interior. \cite{Bugnetetal2021, MathisBugnetetal2021} demonstrated that the dominant contribution of internal magnetic fields in the frequency shifts is attributed to the radial component of the field. \cite{GangLietal2022} then showed that the sensitivity of the modes peaks at the Hydrogen-burning shell: observed magnetic field signatures in \cite{GangLietal2022, Deheuvels2023_mag, GangLietal2023_mag_13} are attributed to radial magnetic field amplitudes in the vicinity of the H-burning shell.

However, magnetic fields in the convective envelope might also affect mixed-mode oscillation frequencies, due to the acoustic component being sensitive to mostly the azimuthal and latitudinal components of the field \citep{Bugnetetal2021, MathisBugnetetal2021}. Depending on the magnetic field topologies in the radiative and convective zones, different locations in the stars could be probed. We investigate the \sbh{potential for detection} of magnetic fields along the SG and RGB phases at three key radial locations in each of them - one region among them contains the H-burning shell and another contains the \sbh{star's subsurface layers}.
 Using traditional tools of {normal-mode coupling} prevalent in geophysical literature \citep[such as the application of generalized spherical harmonics for tensorial perturbations and splitting functions approach for which we refer the reader to][]{DahlenTromp1999}, \cite{Dasetal20} laid down the framework to compute frequency splittings due to different components of the Lorentz stress for any chosen magnetic field topology. In Section~\ref{Self_coupling_of_normal_mode_multiplets}, we \sbh{discuss} magnetic field sensitivity kernels of mixed modes as observed in post-MS stars by invoking the {isolated-multiplet approximation} and using their {self-coupling} in the presence of an axisymmetric magnetic perturbation, a specific case of the more generalized work by \cite{Dasetal20}.
 \sbh{For proof of concept, we design models for post-MS stars with mass 1.3 $M_{\odot}$ of super-solar metallicity $Z=0.025$ in section~\ref{sec:results} and propose how to parameterize an inverse problem at different evolutionary stages, and investigate the detectability of magnetic fields from the magnetic inversion kernels.} We identify the evolutionary stage and typical modes that enable us to distinctly measure the magnetic fields in the different regions within the star.
We conclude on the prominence of the near-surface field as compared to the effect of core fields on the observed signal along the evolution of the modeled star from the sub-giant phase to the RGB.

\section{Self-coupling of normal mode multiplets}

\sbh{
A majority of this section is a recapitulation of the formalism set in \cite{Dasetal20} and their direct implications. It has been provided to the reader for completeness and ease of reference for this paper.
}
\label{Self_coupling_of_normal_mode_multiplets}
  \subsection{General formalism for an axisymmetric perturbation to wave frequencies}
For a spherically symmetric, non-rotating, non-magnetic, adiabatic, isotropic 
reference star, the wave equation over a hydrostatic equilibrium is given by the following eigenvalue problem for a specific mode of stellar oscillation denoted by the subscript $k$ \citep{jcd_notes, lavely92, Astero:book}:
        \begin{equation}
    \label{eq1a}
        \mathcal{L}_{0}(\boldsymbol{\xi}_{k}) = \rho_{0} \, \omega_{k}^{2} \, \boldsymbol{\xi}_{k} \, , \\
        \end{equation}
        where
         \begin{equation}
     \mathcal{L}_{0}(\boldsymbol{\xi}_{k}) = -\boldsymbol{\nabla} \left( \rho_{0} \, c_{s}^2 \, \boldsymbol{\nabla}\cdot\boldsymbol{\xi}_{k} - \rho_{0} \, g \, \boldsymbol{\xi}_{k}\cdot\hat{r} \right) - \boldsymbol{\nabla}\cdot(\rho_0 \, \boldsymbol{\xi}_{k}) \, g \, \hat{r} \, .
    \end{equation}
In these expressions, $\rho_{0}$ is the equilibrium density of the reference star, $c_s$ is the sound speed, $\omega_k$ and $\boldsymbol{\xi}_k$ are the eigenfrequency and eigenfunction of the mode $k$,  the unit vector pointed radially outward is denoted by $\hat{r}$, and $g$ is gravity directed radially inward. Upon solving Eqn.~\eqref{eq1a}, the quantization of the three dimensional oscillations are represented by the label $k \equiv (n, \ell, m)$ where the eigenfunctions may be represented in the form \citep[][]{kendall}:
\begin{equation}
    \label{xi_nl}
    \boldsymbol{\xi}_{n\ell m}(r, \theta, \varphi) =
     U_{n\ell}(r) \, Y_{\ell m}(\theta,\varphi) \, \hat{r} + \: V_{n\ell}(r) \, \boldsymbol{\nabla}_{h}Y_{\ell m}(\theta,\varphi) \, .
\end{equation}
$(r,\theta,\varphi)$ are the common spherical coordinates, $n$ is radial order, $Y_{\ell m}(\theta,\varphi)$ are spherical harmonics of degree $\ell$ and azimuthal order $m \: \in[-\ell, \, -\ell+1, ..., \, 0 , ... , \ell-1, \ell]$, $\boldsymbol{\nabla}_{h}$ is the horizontal gradient, and $U_{n\ell}(r),\: V_{n\ell}(r)$ are profiles of the radial and horizontal components of the eigenfunctions as a function of stellar radius $r$. By virtue of the operator $\mathcal{L}_0$ being hermitian, the eigenfunctions are orthogonal by 
construction. The modes of oscillations in solar-like stars are stochastically excited and intrinsically damped which requires one to solve for the structure and energy equations using the quasi-adiabatic approximation.
The total displacement vector is a superposition of all possible $\boldsymbol{\xi}_{nlm}(r,\theta,\varphi,t)\equiv\boldsymbol{\xi}_{n \ell m}(r,\theta,\varphi)\exp\left[ -i\omega_k t \right]$, with $\omega_k=\omega_r+i\eta$ (where $\omega_r,\eta\in\mathbb{R}$) and amplitudes related to the excitation and damping.
 The contribution due to this intrinsic damping is accounted for in $\eta$.
For a reference star, the oscillation power spectrum is $2\ell+1$ fold degenerate for a given multiplet ($n,\ell$). All these $2\ell+1$ values of $m$ for the multiplet have the same angular frequency $\omega_{r}=\omega_{n\ell}=2\pi\nu_{n\ell}$, which for solar-like oscillators are given by the asymptotic theories \citep[eg.][Section 3.4.3 of \citeauthor{Astero:book} \citeyear{Astero:book}]{JCD-Barthomieu1991}.

Now, we introduce a perturbation, $\epsilon\:\delta\mathcal{L}(\boldsymbol{\xi})$ to the system which modifies the resonant angular frequency of a mode to $\omega_{n\ell}+\epsilon\:\delta\omega$ and the displacement vector to $\boldsymbol{\xi}_{k}+\epsilon\:\delta\boldsymbol{\xi}_{k}$  (with $\epsilon\ll1$, indicating that the order of magnitude of the perturbation $\delta\mathcal{L}$ and angular frequency splitting $\delta\omega$ are $\mathcal{O}(\epsilon)$ smaller than the wave generator $\mathcal{L}_0$ and the background angular frequency $\omega_{n\ell}$ respectively), yielding
\begin{align}
    \label{QDPT1}
    \rho_0 (\omega_{n\ell}+\epsilon\:\delta\omega)^2(\boldsymbol{\xi}_{k}+\epsilon\:\delta\boldsymbol{\xi}_{k}) &= (\mathcal{L}_0 + \epsilon\:\delta\mathcal{L})(\boldsymbol{\xi}_{k}+\epsilon\:\delta\boldsymbol{\xi}_{k}).
\end{align}

Using the orthogonality of $\boldsymbol{\xi}_k$ and retaining terms in Eqn.~\eqref{QDPT1} up to first order in $\epsilon$, an axisymmetric perturbation $\delta \mathcal{L}$ induces the angular frequency splittings given by
\begin{align}
     \label{spl1}
    \delta\omega_{n \ell m} 
     &= \dfrac{1}{I_{n \ell}} \int \mathrm{d}^{3}r \; \boldsymbol{\xi}_{n \ell m}^* \cdot \delta\mathcal{L}(\boldsymbol{\xi}_{n \ell m}) \, ,
\end{align}
where $I_{n\ell}=2\omega_{n\ell}\;\int\limits_0^{R_{\rm{star}}} \rm{d}r \; 4\pi \: r^{2}\rho_0 \:[U_{n\ell}^2+\ell(\ell+1)V_{n\ell}^2]$.

\subsection{Formulation of the general magnetic splittings}

We consider a general magnetic field $\boldsymbol{B}$ which is represented in terms of the generalized spherical harmonics $Y_{pq}^{\mu}(\theta,\varphi)$ \citep{DahlenTromp1999,Phinneyetal1973} as 
       \begin{align}
           \boldsymbol{B}(r,\theta,\varphi)=\sum\limits_{p=0}^{\infty}\sum\limits_{t=-s}^s\sum\limits_{\mu} B_{pq}^{\mu}(r)Y_{pq}^{\mu}(\theta,\varphi)\hat{e}_{\mu}.
       \end{align}
       Here $\mu \in \lbrace -1, 0, 1 \rbrace$ and $\hat{e}_-=(\hat{\theta}-i\hat{\varphi})/\sqrt{2}$, $\hat{e}_0=\hat{r}$, $\hat{e}_+=-(\hat{\theta}+i\hat{\varphi})/\sqrt{2}$. The
        configuration of the perturbing field is specified by angular degree $p$ and azimuthal order $q\in\lbrace -p,-p+1,...,p-1,p \rbrace$.
       When the magnetic field $\boldsymbol{B}$ is the perturbation to the stellar model \citep[\sbh{$\epsilon \approx |\boldsymbol{B}|^{2}\left( 4\pi GM^2/R^4 \right)^{-1}$, a ratio of magnetic pressure to the gas pressure}][]{GoughThompson}{}{}, then $\delta\mathcal{L}(\boldsymbol{\xi})\equiv\delta\mathcal{L}_{B}(\boldsymbol{\xi})$, as derived in \cite{goedbloed2004} and subsequently used in \cite{Dasetal20}, is
    \begin{align}
        \label{dL}
        4\pi \: \delta\mathcal{L}_{B}(\boldsymbol{\xi}) &= 
            \boldsymbol{B} \times \lbrace \boldsymbol{\nabla} \times [ \boldsymbol{\nabla} \times (\boldsymbol{\xi} \times \boldsymbol{B} ) ] \rbrace \nonumber\\  
            &- (\boldsymbol{\nabla} \times \boldsymbol{B}) \times [ \boldsymbol{\nabla} \times (\xi \times \boldsymbol{B}) ] \\ 
            &- \boldsymbol{\nabla} \lbrace \boldsymbol{\xi} \cdot [ (\boldsymbol{\nabla} \times \boldsymbol{B}) \times \boldsymbol{B} ] \rbrace \nonumber
            .
    \end{align}

    For solar-like oscillators, the eigenfunctions become evanescent beyond the stellar surface. This renders the oscillation frequencies of such modes insensitive to the perturbations above the surface. Using Eqns.~\eqref{xi_nl}~and~\eqref{dL}, neglecting contributions due to surface boundary terms,
    and considering self-coupling in an isolated multiplet $(n,\ell)$, Eqn.~\eqref{spl1}  \citep[refer Appendix C of ][]{Dasetal20} takes the following form
    \begin{align}
        \label{eq10}
        \delta\omega_{n\ell m} 
         &=\frac{1}{I_{n\ell}}
         \int_{0}^{R_{\rm{star}}} dr \sum\limits_{s,t}\sum\limits_{\mu,\sigma} \: r^{2}\mathcal{B}_{st}^{\mu\sigma}(r;n,\ell,m) \; h_{st}^{\mu\sigma}(r) \, ,
    \end{align}
    where $\mu,\sigma\in\lbrace -1,0,1 \rbrace$, $s$ and $t$ are the angular degree and azimuthal order of the Lorentz stress perturbation, 
    $R_{\rm{star}}$ is the radius of the star, and $\mathcal{B}_{st}^{\mu\sigma}(r;n,\ell,m)$ are the Lorentz-stress sensitivity kernels which are dependent on $U_{n\ell}(r)$ and $V_{n\ell}(r)$. $h_{st}^{\mu\sigma}(r)$  
    denote components of the Lorentz-stress tensor \citep{Dasetal20} in the generalized spherical harmonics basis, which captures the amplitude and topology of the magnetic field
    \begin{align} \label{eq: hmunustr}
        h_{st}^{\mu\sigma} &= \sum_{p_1,q_1,p_2,q_2} B_{p_1 q_1}^{\mu}B_{p_2 q_2}^{\sigma}
        \int d\Omega \; Y_{st}^{*\:\mu+\sigma}Y_{p_1q_1}^{\mu}Y_{p_2 q_2}^{\sigma} \, .
    \end{align}  

Theoretical studies on the calculation of splittings, as mentioned in the introduction, have been carried out for red-giant stars. They have been employed often to compute the impact due to some dominant terms in the magnetic perturbation operator \citep[eg.][]{Bugnetetal2021, Loi2021, MathisBugnetetal2021, Bugnet2022, GangLietal2022}{}. However, most studies have considered the effect of only the radial field strength $B_r$ in the H-shell burning region to find frequency splittings.
The formalism in \cite{Dasetal20} can be put to use for any magnetic field topology, provided the oscillation eigenfunctions retain their form as in Eq.~\eqref{xi_nl} in the presence of other perturbations (which breaks down when traditional approximation of rotation is considered for fast rotators) and the magnetic field amplitude remains small enough for first order perturbations to remain valid. We use this formalism to calculate the effects of axisymmetric magnetic fields on frequency splittings in a non-rotating red-giant and its corresponding sub-giants. For both these cases, using a thorough calculation of kernels, we estimate threshold strengths for the \sbh{potential for detection} of different field components at different layers which could let us constrain their configurations.

  \subsection{Constructing the Lorentz-stress tensor from a model axisymmetric magnetic field}
\label{Lorentz_stress_tensor}
  
\cite{Duez2010A} and \cite{Duez2010B} discuss in detail the stable magnetic field configurations that may be present inside \sbh{a radiative interior}. Earlier works \citep[eg.][]{Tayler_Tor_unstable, Braithwaite_tor_unstable, Markey_Pol_unstable,Braithwaite_pol_unstable} show that pure toroidal or poloidal magnetic field models are physically unstable on dynamical timescales inside a star, and instead, a mix of the two is stable \citep[eg.][]{Tayler_stable,Braithwaite_mixed}. The first solution to the stable field configuration obtained semi-analytically in Appendix B of \cite{Bugnetetal2021} is seen to have a strong poloidal magnetic field component and a weak toroidal component, with the maximum amplitude of the toroidal field being much smaller than that of the maximum radial field amplitude. 
Hence, we consider an axisymmetric solenoidal magnetic field that may be decomposed into toroidal and poloidal components on the spherical coordinates
\begin{align}
    \label{eq:field_config_general}
    \boldsymbol{B}(r,\theta) &= B_{0}
    \Big[
        2\beta(r)\cos\theta \; \hat{r}  \nonumber \\ 
        &-  \frac{1}{r} \frac{\partial}{\partial r}\left\lbrace r^{2}\beta(r) \right\rbrace \sin\theta \; \hat{\theta} - \alpha(r)\sin\theta \hat{\varphi}
    \Big],
\end{align}
where $B_0$ is the magnetic-field amplitude scaling factor, and $\beta(r)$ and $\alpha(r)$ are functions parameterizing the radial variations of the poloidal and toroidal counterparts, respectively. Following the same decomposition method as employed in Appendix D.1 of \cite{Dasetal20}, this magnetic field $B_{p_{0}q_{0}}(r)$ is represented in the general spherical harmonics basis as 
\begin{eqnarray}
    \label{eq14}
    \begin{cases}
        -i\frac{B_0}{\gamma_1}
        \begin{pmatrix}
            \alpha(r)
            \\
            0
            \\
            -\alpha(r)
        \end{pmatrix}
        +\frac{B_{0}}{\gamma_{1}} 
        \begin{pmatrix}
            2\beta(r)+r\frac{d\beta(r)}{dr}
            \\
            2\beta(r)
            \\
            2\beta(r)+r\frac{d\beta(r)}{dr}
        \end{pmatrix} & p_{0},q_{0}=1,0
        \\
        0 & \rm{otherwise}.
    \end{cases}
\end{eqnarray}
The topmost component belongs to $Y_{10}^{-1}\hat{e}_-$, the middle one to $Y_{10}^{0}\hat{e}_0$, the bottom-most component belongs to $Y_{10}^{+1}\hat{e}_+$. Using Eqn.~\eqref{eq: hmunustr}, the Lorentz-stress tensor for this model magnetic field can be expressed as
\begin{align}
    \label{eq15}
    h_{s0}^{\mu\sigma}(r) = &3\gamma_{s} B_{10}^{\mu}(r) B_{10}^{\sigma}(r) (-1)^{\mu+\sigma} \: \nonumber\\
    &\begin{pmatrix}
        1 & s & 1
        \\
        \mu & -(\mu+\sigma) & \sigma
    \end{pmatrix}\:
    \begin{pmatrix}
        1 & s & 1
        \\
        0 & 0 & 0
    \end{pmatrix},
\end{align}
where  $\gamma_{j}=\sqrt{\frac{2j+1}{4\pi}}$, $\begin{pmatrix}
  \ell_1 & \ell_2 & \ell_3
  \\
  m_1 & m_2 & m_3
\end{pmatrix}=\mathcal{W}_{m_1\,m_2\,m_3}^{\ell_1\,\ell_2\,\ell_3}$ is a Wigner 3j symbol \citep{Wigner1993} which takes a value of 0 \citep[Appendix C of][]{DahlenTromp1999} when it fails to satisfy one or more of the following selection rules:
\begin{enumerate}
    \item $m_i\in\lbrace -\ell_i,-\ell_i+1,...,\ell_i-1,\ell_i \rbrace$,
    \item $|\ell_1 - \ell_2|\leq \ell_3 \leq \ell_1+\ell_2$,
    \item $m_1+m_2+m_3=0$,
    \item $\ell_1+\ell_2+\ell_3$ is an even integer if $m_1=m_2=m_3=0$.
\end{enumerate}
According to these selection rules, $h_{st}^{\mu\sigma}=0$ when $s \notin \lbrace 0,1,2 \rbrace$ and $t\neq0$. 
Also, we have $h_{10}^{\mu\sigma}~=~0\;\forall \mu,\sigma$ since it is proportional to
$\mathcal{W}_{0\,0\,0}^{1\,1\,1}$ which yields zero. For a dipolar magnetic field, due to the selection rule imposed by Wigner 3-$j$ in Eqn.~(\ref{eq15}), we are only required to compute the sensitivity kernels for $s\in\lbrace 0,2 \rbrace$ and $t=0$ for this particular axisymmetric model of stellar magnetic field (since $h_{st}^{\mu\sigma}=0$ when $t\neq0$, or $s\neq 0\:\rm{or}\:2$).

  \subsection{General expression of the sensitivity Kernels for axisymmetric magnetic fields (t=0)}

The kernels $\mathcal{B}_{st}^{\mu\sigma}(r;n,\ell,m)\equiv\mathcal{B}_{st}^{\mu\sigma}(k,m)$ depend on $U_{n\ell}(r)\equiv U_k$ and $V_{n\ell}(r)\equiv V_k$ and their first and second order radial derivatives. They describe how different oscillation modes respond to the Lorentz stresses due to stellar internal magnetic fields of different strength and geometric configurations. 
For axisymmetric magnetic fields as defined in Section \ref{Lorentz_stress_tensor}, we are only required to calculate $\mathcal{B}_{st}^{\mu\sigma}$ for $t=0$ (since $h_{st}^{\mu\sigma}=0$ when $t\neq0$).
The sensitivity kernels for self-coupled modes in the presence of an axisymmetric magnetic field in the star, as derived in Appendix C of \cite{Dasetal20}, can be expressed as
\begin{align}
    \label{eq:r2B}
    r^2 \mathcal{B}_{s0}^{\mu\sigma}(k,m)
    =
    4\pi (-1)^{m}\gamma_{\ell}^{2}\gamma_{s}
    \begin{pmatrix}
        \ell & s & \ell
        \\
        -m & 0 & m
    \end{pmatrix}
    \tilde{\mathcal{G}}_{s}^{\mu\sigma}(k),
\end{align}
where $\Tilde{\mathcal{G}}_{s}^{\mu\sigma}(k)=r^{2}\mathcal{G}_{s}^{\mu\sigma}(k)$ \citep[$\mathcal{G}_{s}^{\mu\sigma}(k)$ has been derived in Appendix C.1 of][]{Dasetal20} 
and the kernel components were found to satisfy the following identities:
\begin{align}
    r^2 \mathcal{B}_{s0}^{\mu\sigma}(k,m) &= r^2 \mathcal{B}_{s0}^{\sigma\mu}(k,m), \label{eq: kern_identity_flip}
    \\
    r^2 \mathcal{B}_{s0}^{--}(k,m) &= (-1)^{s} \; r^2 \mathcal{B}_{s0}^{++}(k,m), \label{eq: kern_identity_ppmm}
    \\
    r^2 \mathcal{B}_{s0}^{0-}(k,m) &= (-1)^{s} \; r^2 \mathcal{B}_{s0}^{+0}(k,m), \label{eq: kern_identity_0mp0}
\end{align}
As a consequence of these symmetries and relations, we have only four independent components $\tilde{\mathcal{G}}_{s}^{\mu\sigma}$ to compute. 
Using the expressions for $\mathcal{G}_{s}^{\mu\sigma}(k)$ derived in Appendix C.1 of \cite{Dasetal20}, the expressions for these independent $\tilde{\mathcal{G}}_{s}^{\mu\sigma}$ are:
\begin{align}
    \label{eq:Gs--}
    \tilde{\mathcal{G}}_{s}^{--}(k) = &-\dfrac{1}{2}
    \Bigg[
        \left\lbrace
        \begin{pmatrix}
            \ell & s & \ell 
            \\
            2 & -2 & 0
        \end{pmatrix}
        +
        \begin{pmatrix}
            \ell & s & \ell 
            \\
            0 & -2 & 2
        \end{pmatrix}
        \right\rbrace
        \chi_{1}^{--}(k)\nonumber \\
        &+
        2
        \begin{pmatrix}
            \ell & s & \ell 
            \\
            1 & -2 & 1
        \end{pmatrix}
        \chi_{2}^{--}(k)
        \Bigg] \nonumber
        \\
        &-
        \dfrac{1}{2}
        \left\lbrace
        \begin{pmatrix}
            \ell & s & \ell 
            \\
            3 & -2 & -1
        \end{pmatrix}
        +
        \begin{pmatrix}
            \ell & s & \ell 
            \\
            -1 & -2 & 3
        \end{pmatrix}
        \right\rbrace
        \chi_{3}^{--}(k),
\end{align}

\begin{align}
    \tilde{\mathcal{G}}_{s}^{0-}(k) &= \dfrac{1}{4}
    \Bigg[
        \Bigg\lbrace
        \begin{pmatrix}
            \ell & s & \ell 
            \\
            1 & -1 & 0
        \end{pmatrix}
        +
        \begin{pmatrix}
            \ell & s & \ell 
            \\
            0 & -1 & 1
        \end{pmatrix}
        \Bigg\rbrace
        \chi_{1}^{0-}(k)\nonumber \\
        &+
        \Bigg\lbrace
        \begin{pmatrix}
            \ell & s & \ell 
            \\
            -1 & -1 & 2
        \end{pmatrix}
        +
        \begin{pmatrix}
            \ell & s & \ell 
            \\
            2 & -1 & -1
        \end{pmatrix}
        \Bigg\rbrace
        \chi_{2}^{0-}(k)
        \Bigg],\\
    \tilde{\mathcal{G}}_{s}^{00}(k) &=
    \dfrac{1+(-1)^s}{2}
    \Bigg[
        \begin{pmatrix}
            \ell & s & \ell 
            \\
            0 & 0 & 0
        \end{pmatrix}\chi_{1}^{00}(k)\nonumber \\
        &+
        2
        \begin{pmatrix}
            \ell & s & \ell 
            \\
            -1 & 0 & 1
        \end{pmatrix}\chi_{2}^{00}(k)
    \Bigg],
    \\\label{eq:Gs+-}
    \tilde{\mathcal{G}}_{s}^{+-}(k) &=
    \dfrac{1+(-1)^s}{4}
    \Bigg[
        \begin{pmatrix}
            \ell & s & \ell 
            \\
            0 & 0 & 0
        \end{pmatrix}\chi_{1}^{+-}(k)\nonumber \\
        & +2
        \begin{pmatrix}
            \ell & s & \ell 
            \\
            -2 & 0 & 2
        \end{pmatrix}\chi_{2}^{+-}(k)\nonumber \\
        &+2
        \begin{pmatrix}
            \ell & s & \ell 
            \\
            -1 & 0 & 1
        \end{pmatrix}\chi_{3}^{+-}(k)
    \Bigg],
\end{align}
where expressions for the ten $\chi_{i}^{\mu\sigma}(k)$ are shown in the Appendix \ref{appendix:a}. 

\subsubsection{Case of an odd angular degree s}
We now consider the case of odd $s$ for self-coupled multiplets in a magnetic field that has 
azimuthal symmetry, i.e., $t=0$.
One can directly notice that $1+(-1)^s=0$ for odd $s$, which makes $\Tilde{\mathcal{G}}_{s}^{00}(k)$ and $\Tilde{\mathcal{G}}_{s}^{+-}(k)$ equal to 0.
\cite{Wigner1993} shows that a Wigner 3-j symbol is multiplied by $(-1)^{\ell_1+\ell_2+\ell_3}$ for odd permutations and by 1 for even permutations. Hence, we obtain the relation
\begin{align}
\label{odd_permutation}
\mathcal{W}_{m_1\;-(m_1+m_3)\;m_3}^{\ell\,s\,\ell} = (-1)^{s}\:\mathcal{W}_{m_3\;-(m_1+m_3)\;m_1}^{\ell\,s\,\ell}.
\end{align} 
Owing to this identity, all the kernels become zero for all odd $s$. This means that the odd $s$ components of $h_{s0}^{\mu\sigma}$ leave behind no signature onto the frequency splittings, and only even values of $s$ impact the frequency shifts and splittings.

\subsubsection{Case of an even angular degree s}
One can clearly note that the second column of all Wigner 3-j symbols in $\Tilde{\mathcal{G}}_{s}^{--}(k)$ have $\begin{pmatrix}
            s 
            \\
            -2
        \end{pmatrix} $
which is 0 for $s<2$, which results in $\mathcal{B}_{s0}^{--}(k)=0$ for $s=0$. In a similar fashion $\mathcal{B}_{s0}^{0-}(k)=0$ for $s=0$.
Setting $m_3~=~-m_1~=~m$ in Eqn.~\eqref{odd_permutation} and considering only even values of $s$, we note that 
$\mathcal{W}_{-m\,0\,m}^{\ell\,s\,\ell}=\mathcal{W}_{m\,0\,-m}^{\ell\,s\,\ell}$. Also, since $(-1)^{m}=(-1)^{-m}$, the $m$-dependent part of $\mathcal{B}_{s0}^{\mu\sigma}(k,m) \propto
(-1)^{m}\:\mathcal{W}_{-m\,0\,m}^{\ell\,s\,\ell}$
is same for both $m$ and $-m$. Therefore, while rotation introduces symmetric splitting in $m$, an axisymmetric magnetic field (aligned with the rotation axis) introduces an asymmetric splitting \citep[see Figure~7 in][]{Bugnetetal2021}.

\subsubsection{Contribution of the magnetic kernels to frequency shifts}
The $m$-dependent part of sensitivity kernels for $s=0$ is 
$(-1)^{m}\:\mathcal{W}_{-m\,0\,m}^{\ell\,0\,\ell}$. 
This factor according to Eqn.~32(a) in page 627 of \cite{Wigner1993} is equal to $(-1)^{\ell}/\sqrt{2\ell+1}$, which is independent of the value of $m$.    
Hence, the $\mathcal{B}_{00}^{\mu\sigma}$ add a net shift in the unperturbed frequency $\nu_{n\ell}$ in the presence of the perturbation. The $\mathcal{B}_{20}^{\mu\sigma}$ contributes to the actual splitting among different $m$ in each multiplet.
The construction of the magnetic field kernels also conveys to us that for the kind of magnetic field topologies discussed above, for any $\ell(\geq 2)$, one can easily obtain the splittings for $|m|\in\lbrace 2,..,\ell \rbrace$ if one can calculate the splittings for $m=0$ and $|m|=1$ following steps similar to ones shown in Appendix~\ref{l2_factor_of_4}.
Also, thanks to the fact that 
$\sum_{m=-\ell}^{\ell}\:(-1)^m\:\mathcal{W}_{-m\,0\,m}^{\ell\,s\,\ell}= (-1)^\ell\sqrt{2\ell+1}\:\delta_{s0}$ (where $\delta_{jk}\equiv$ \sbh{Kronecker} delta), 
the mean value of the total splittings within a multiplet is equal to 
the net shift solely due to $s=0$ terms, and independent of the splitting among the different $m$-components.

A simple yet special set of kernels worth noting are those for the $\ell=0$ pure radial modes, which contain only $m=0$. 
Then the Wigner 3-j symbol in Eqn.~\eqref{eq:r2B} becomes 
$\mathcal{W}_{0\,0\,0}^{0\,s\,0}$
which is equal to 0 for $s\neq 0$ (criterion 2 mentioned above for the Wigner 3-j to be non-zero is violated in other cases), thus leaving us with kernels for $s=0$ to be calculated. Also from Appendix \ref{appendix:a}, we note that for $\ell=0$, $\Omega_{0\ell}=0$, which takes all value of $\chi_{i}^{\mu\sigma}(k)$ to 0 except $\chi_1^{00}(k)=2\left[ -2rU_k\dot{U}_k + 3U_k^{2} \right]$ and $\chi_1^{+-}(k)=2\left[ -2rU_k\dot{U}_k + U_k^{2} - r^2\dot{U}_k^2 \right]$. Thus, for $\ell=0$ modes, the only two non-zero kernels are $r^{2}\mathcal{B}_{00}^{00}(k,0)=\dfrac{1}{\sqrt{\pi}}\:\left[ -2rU_k\dot{U}_k + 3U_k^{2} \right]$  and $r^{2}\mathcal{B}_{00}^{+-}(k,0)=\dfrac{1}{\sqrt{\pi}}\:\left[ -2rU_k\dot{U}_k + U_k^{2} -r^{2}\dot{U}_k^2 \right]$. 
$U_k$ and $\dot{U}_k$ for being extremely small in the core \sbh{(containing $g$-component of mixed modes)} and large in the envelope \sbh{(containing $p$-component)}, the kernels and their resultant shift for the $\ell=0$ modes are mostly sensitive to the envelope magnetic field instead of that inside the core.

Appendix~\ref{l2_factor_of_4} \sbh{outlines a method} which enables us to calculate splittings for all other $m$ belonging to an $\ell$ once those for $m=0$ and $1$ are individually computed, thus saving significant computation time.

\subsection{Brief discussion on non-axisymmetric fields}

The formalism of \cite{Dasetal20} is by construction independent of the inherent symmetries of the internal magnetic field topology. As discussed in Section 2.4.1 of \cite{Dasetal20}, the axisymmetric nature of the magnetic field topologies simplifies our calculations by setting $t=0$ for all cases. For cases where non-zero values of $t$ are possible, the selection rules for the Wigner 3-j symbol render it mandatory for us to take into consideration coupling among the different degenerate $m$ (say, $m,m'\in\lbrace 
-\ell,-\ell+1,..,0,..,\ell-1,\ell \rbrace$) belonging to the same multiplet. This will result in a more complicated forward model involving eigenvalue problems to calculate the splittings. For slowly rotating stars, such non-axisymmetric magnetic fields could arise in cases where there is a finite obliquity between rotational and magnetic axes.

\section{Results}
\label{sec:results}

\subsection{Modeled stars}
\label{sec:modelling_star_and_fields}

\begin{figure}
    \centering
    \includegraphics[width=1\linewidth]{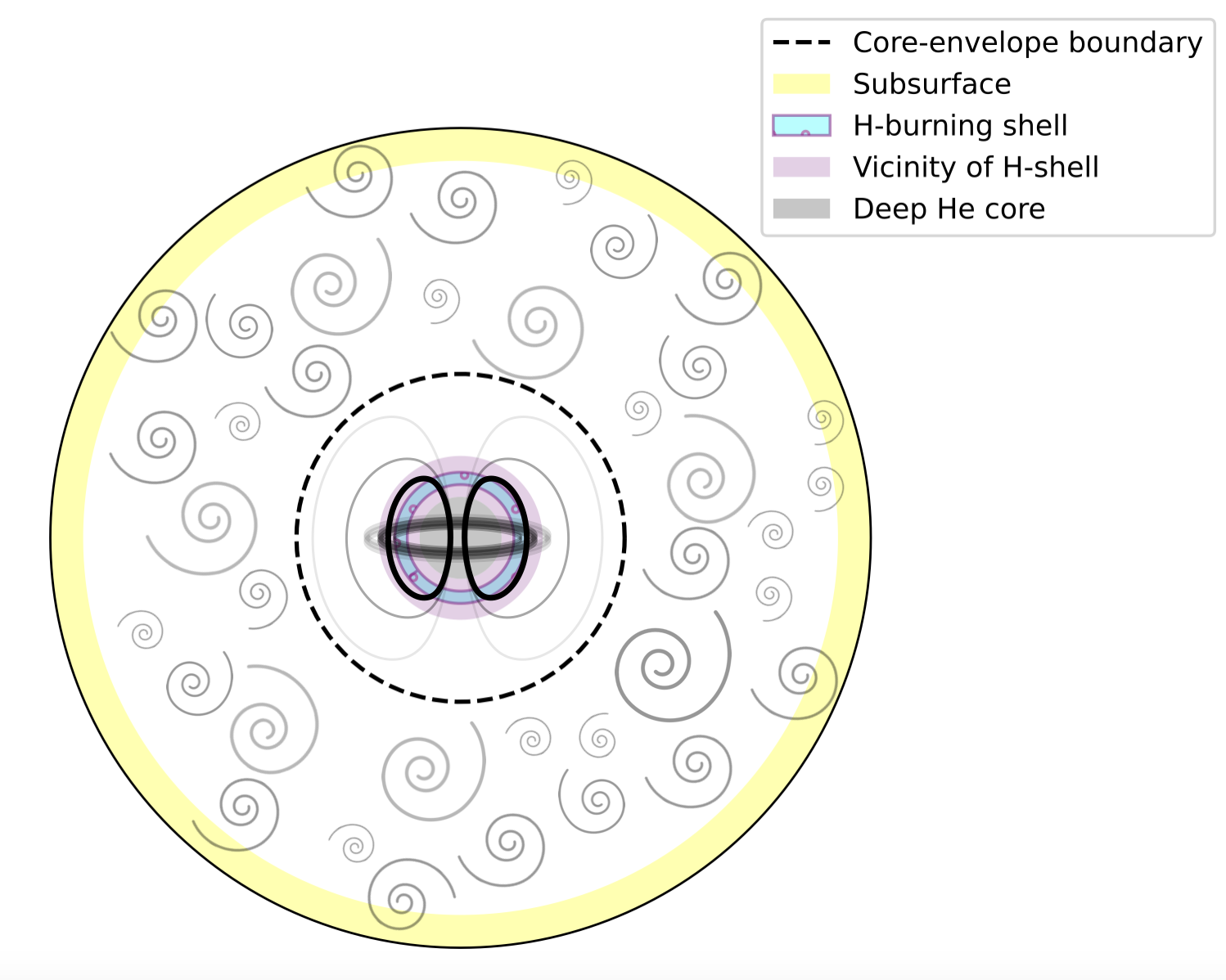}
    \includegraphics[width=1\linewidth]{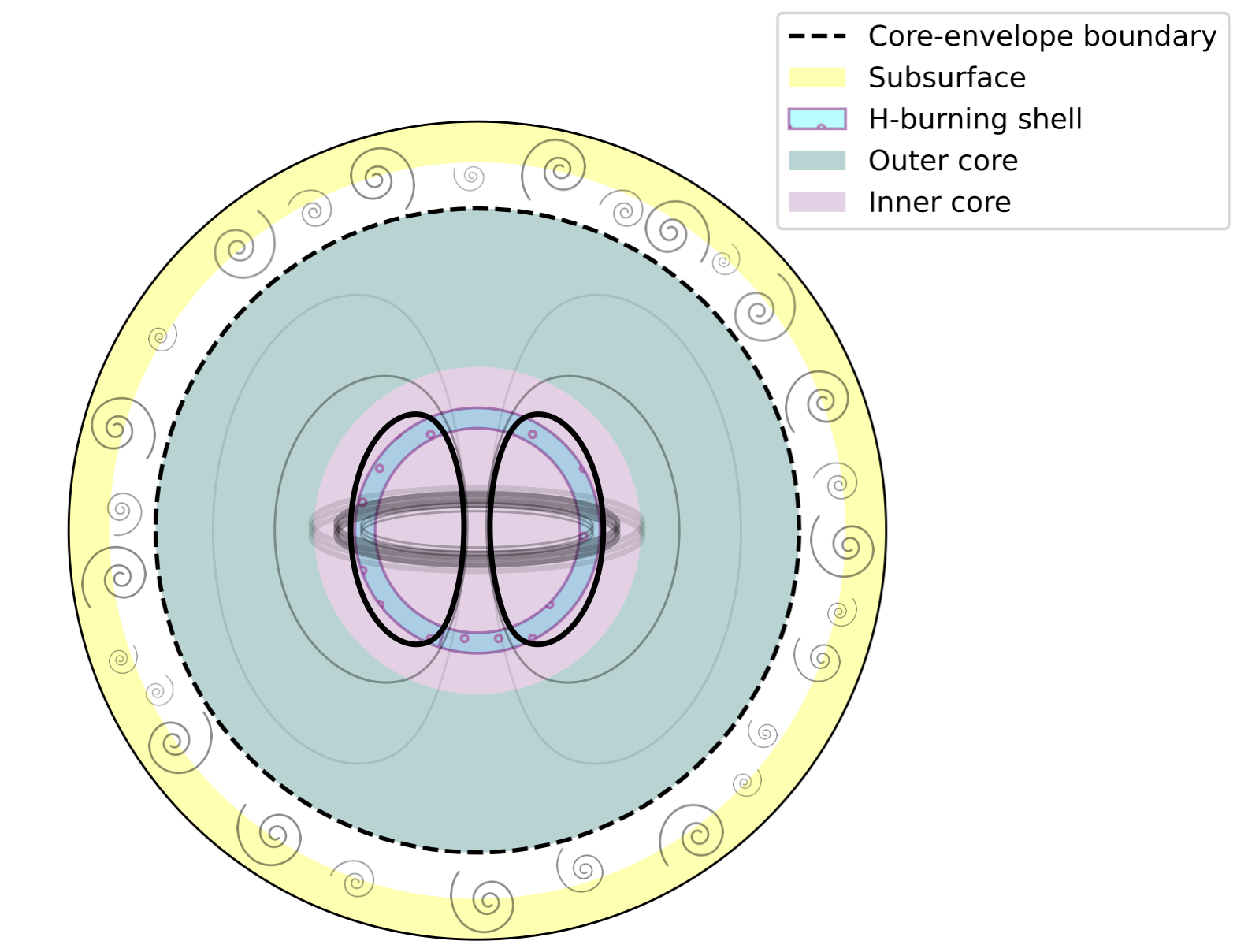}
    \caption{
    \sbh{
    Scheme (not to scale) of red giant branch star (\textsl{top}) and a subgiant star (\textsl{bottom}). A large-scale stable magnetic field given by the Eqn.~\eqref{B_Bugnet_eqn} is represented in the cores, and dynamo action is present in the convective envelopes. Colored areas indicate the different regions probed by mixed modes considered in this study.
    }
    }
    \label{fig:schemes}
\end{figure}

   \begin{figure*}
    \centering
    \begin{subfigure}[]{
        \includegraphics[width=0.3575\linewidth]{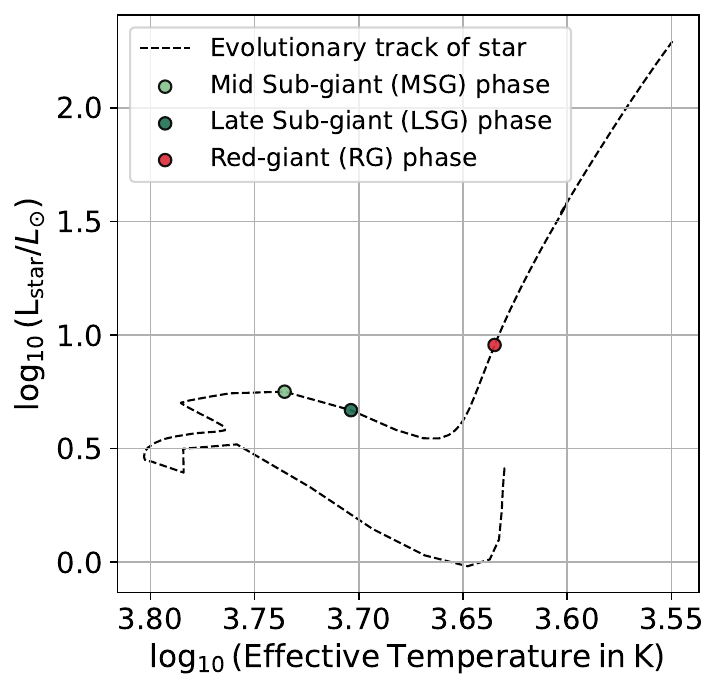}
        \label{fig:HRD_all_phases}}
    \end{subfigure}
    \begin{subfigure}[]{
        \includegraphics[width=0.445\linewidth]{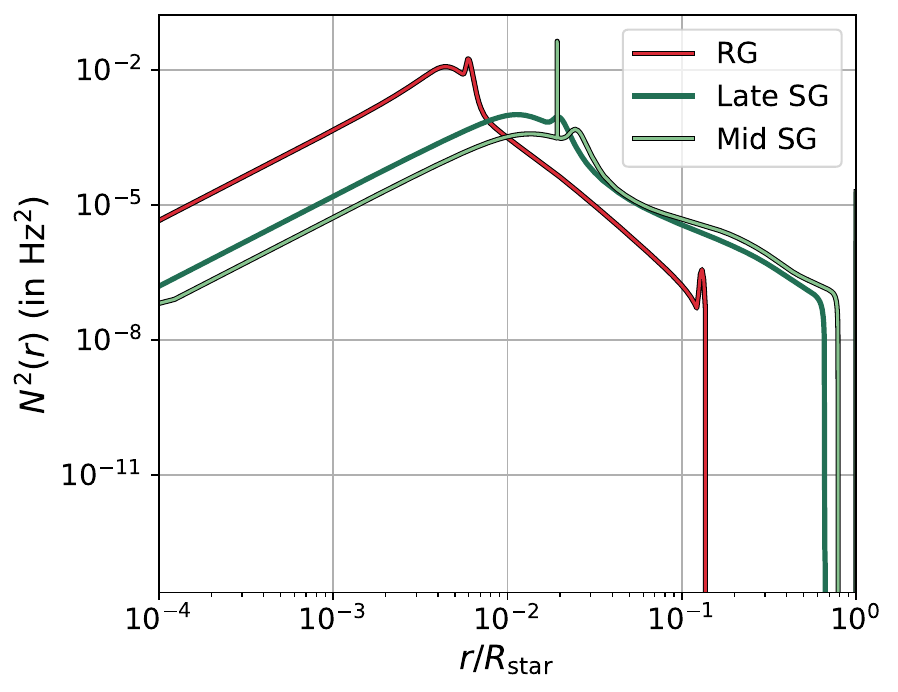}
        \label{fig:N2_all_phases}}
    \end{subfigure}
    \caption{(a) Hertzprung-Russell diagram indicating the three evolution stages of the 1.3 $M_\odot$ star used in this study. (b) Squared Brunt-V\"ais\"al\"a frequency profiles for the 3 stages of the star. We notice that the H-burning shell and the outer part of the radiative zone move closer to the star's centre, indicating the deepening of the outer convective envelope. Note: $R_{\rm{star}}$ is different at each stage of evolution and is scaled accordingly.}
   \end{figure*}

   For the major part of this work, we focus on a model of a typical 4.056 Gyr red-giant branch star with mass $M=1.3 \, M_{\odot}$, metallicity $Z=0.025$, radius $R_{\rm{star}}= 5.383 \, R_{\odot}$. This RGB structure model is generated using 
   {\texttt{MESA}} \citep[][]{Paxton2011, Paxton2013, Paxton2015, Paxton2018, Paxton2019}{}{} 
   We also generate two models of the same star in its sub-giant phase using \texttt{MESA} at ages of 3.624 Gyr and 3.702 Gyr\footnote{\sbh{The data is available on Zenodo under an open-source 
Creative Commons Attribution license: 
\dataset[doi:10.5281/zenodo.10913776]{https://doi.org/10.5281/zenodo.10913776}} }. 
   \sbh{Figure~\ref{fig:schemes} shows schematic diagrams of the stellar structure in the RG and SG phases.} 
   The star in the former SG stage has reached a peak luminosity in the SG evolution, beyond which luminosity starts dropping due to the narrowing H-burning shell around the core, which includes the latter \citep[][]{Iben1967_Off_MS}. These phases will henceforth be referred to as the mid sub-giant phase (MSG) and the late sub-giant phase (LSG) respectively (see Figure~\ref{fig:HRD_all_phases}). Their frequencies of maximum power ($\nu_{\rm{max}}$) are $\nu_{\rm{max,MSG}}=581.73\:\mu\rm{Hz}$ and $\nu_{\rm{max,LSG}}=542.99\:\mu\rm{Hz}$ respectively, which are much larger than that of the red giant ($\nu_{\rm{max,RG}}=160.928\:\mu\rm{Hz}$). The eigenfunctions and resonant frequencies of $\ell=0,1,2$ are computed with 
   {\texttt{GYRE}} \citep[][]{Townsend2013}, generated in the range of [$120\mu\rm{Hz},200\mu\rm{Hz}$] for the RG and $[\nu_{\rm{max}}-3 \Delta \nu, \nu_{\rm{max}}+3 \Delta \nu ]$ for the SGs, $\Delta \nu$ being their large frequency separations. The three stages used in this study have been indicated in Figure~\ref{fig:HRD_all_phases}.

   We then use unperturbed frequencies, displacement eigenvectors, and relevant structure parameters of the model stars as input to compute the field kernels and the splittings induced by a given magnetic field having a topology mentioned in Section \ref{Lorentz_stress_tensor}. 
 
   The Brunt-V\"ais\"al\"a frequency $N$, commonly known as buoyancy frequency, is defined as
   \begin{align}
       N^{2}(r) = g \left( \dfrac{1}{\Gamma_1} \dfrac{d \ln P_{0}}{dr} - \dfrac{d \ln \rho_{0}}{dr} \right),
   \end{align}
   with $P_{0}$ being the equilibrium pressure profile, and $\Gamma_1$ being the adiabatic exponent of the material in the stellar interior. $N^{2}$ is the square of the frequency of oscillation of a blob of plasma in the statically stable stellar interior when it is displaced radially. Hence, it takes negative values in convective regions and positive values in radiative zones. From Figure~\ref{fig:N2_all_phases}, we note that $N^{2}$ goes to 0 at $R_{\rm{rad}}=0.1364\:R_{\rm{star}}$ for the RG, indicating the boundary between the radiative core and convective envelope.

   As we have discussed in the introduction of this work, red-giant stars show the presence of mixed modes, which are formed due to considerable coupling between the $p$-modes in the convective envelope and the $g$-modes within the radiative interior. These mixed modes thus have part of their inertia in the core and part of it in the envelope. For such modes, \citep[][]{Deheuvels2012, Goupil2013} defined :
   \begin{align}
    \label{zeta}
    \zeta_{n\ell} = \dfrac{\int_{g\rm{-mode\:cavity}}^{} dr \: \rho(r) \: r^2 \: \left[ U_{n\ell}^2+\ell(\ell+1)V_{n\ell}^2 \right]}{ \int_{0}^{R_{\rm{star}}} dr \: \rho(r) \: r^2 \: \left[ U_{n\ell}^2+\ell(\ell+1)V_{n\ell}^2 \right] }.
\end{align}
The $g$-mode cavity of the RG extends from the centre to its $R_{\rm{rad}}$.

\begin{figure}
    \centering
    \includegraphics[width=0.97\linewidth]{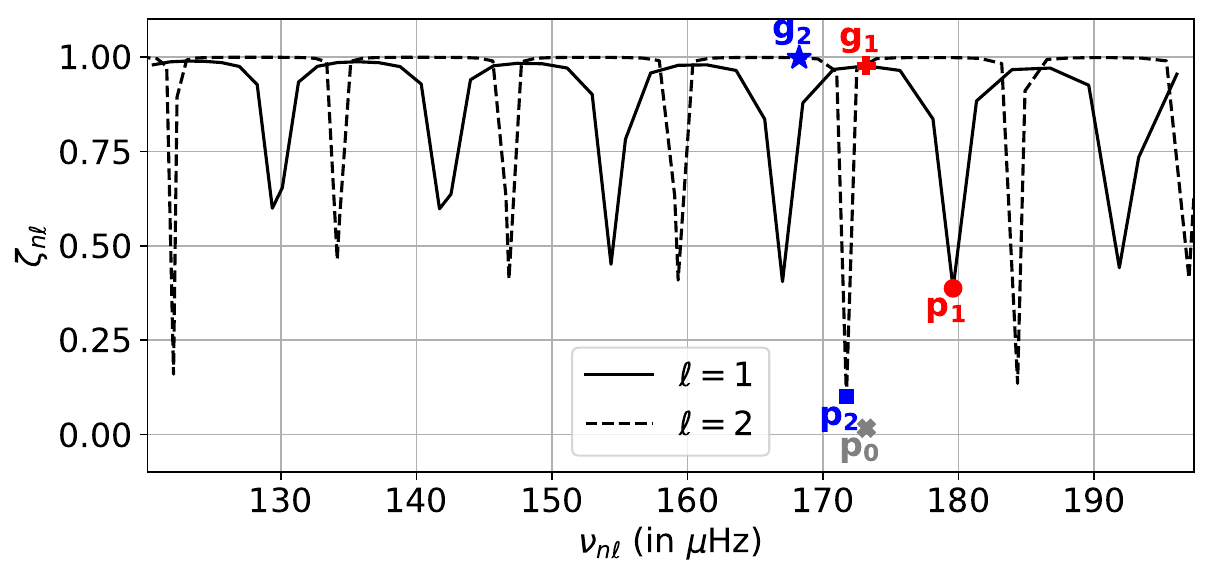} 
    \caption{ 
    Fraction of inertia of the oscillation eigenmodes in the radiative interior of the RG model, calculated using the eigensolutions to the structure equations of the \texttt{MESA} stellar model. Five oscillation modes have been chosen and marked $p_0$, $p_1$, $p_2$, $g_1$ and $g_2$. The red and blue symbol indicates that the selected mode $\ell=1$ and $\ell=2$ respectively. The kernels for $g$-dominant modes $g_1$ and $g_2$, $p$-dominant modes $p_1$ and $p_2$, and an $\ell=0$ pure $p$-mode marked $p_0$ in grey, are plotted in Figures~\ref{fig:f17}, \ref{fig:f61}, \ref{fig:f14}, \ref{fig:f58}, \ref{fig:f3}, respectively.
    }\label{Zeta_plot}
\end{figure}

The $\zeta_{n\ell}$ (hereafter simplified as $\zeta$ for notational convenience) is the fraction of the total mode inertia existing inside the $g$-mode cavity, which usually lies within the star's radiative interior. For $g$-dominant modes, $\zeta_{n\ell}\approx 1$ and for a $p$-dominated mode, $\zeta_{n\ell} \ll 1$.
Hence, $\zeta$ quantifies the $p$/$g$-mode nature of a mixed mode \citep[e.g.][]{Mosser2015, Vrard2016}. $\zeta_{n\ell}$ for the RG is plotted in Figure~\ref{Zeta_plot}.

 By definition, the kernel $r^{2}\mathcal{B}_{s0}^{\mu\sigma}(k,m)$ corresponding to different modes of oscillations $(k,m)$ teach us how frequency splittings depend on an axisymmetric Lorentz-stress component $h_{s0}^{\mu\sigma}(r)$. This identifies which $B_{i}B_{j}$ ($i,j\in\lbrace r,\theta,\varphi\rbrace$) components in different regions of the star dominantly contribute to the magnetic splittings.
 
\begin{table} 
\begin{tabular}{|c|c|}
\hline
\textbf{Kernel component}                       & \textbf{Field component}                      \\ \hline
$r^2 \, \mathcal{B}_{s0}^{00}$ (even $s$)                       & $B_r^2$                      \\ \hline
$r^2 \, \mathcal{B}_{s0}^{+-}$ (even $s$)                      & $B_{\theta}^2 + B_{\varphi}^2$                      \\ \hline
$r^2 \, \mathcal{B}_{s0}^{0-}$ (even $s$)                       & $B_r \, B_{\theta}$                      \\ \cline{2-2} 
$r^2 \, \mathcal{B}_{s0}^{0-}$ (odd $s$)                       & $B_r \, B_{\varphi}$                      \\ \hline
\multicolumn{1}{|c|}{$r^2 \, \mathcal{B}_{s0}^{--}$ (even $s$)} & \multicolumn{1}{c|}{$B_{\theta}^2 - B_{\varphi}^2$} \\ \cline{2-2} 
\multicolumn{1}{|c|}{$r^2 \, \mathcal{B}_{s0}^{--}$ (odd $s$)} & \multicolumn{1}{c|}{$B_{\theta} \, B_{\varphi}$} \\ \hline
\end{tabular}
\caption{Lorentz-stress kernel components and the respective components of $\textbf{B}\textbf{B}$ which they are sensitive to \citep[see Eq.~32 and Eq.~39-44 in][]{Dasetal20}.}
\label{table:kernel_vs_field}
\end{table}

Table~\ref{table:kernel_vs_field} summarizes which component of Lorentz-stress $\mathbf{B} \mathbf{B}$ is sensed by which kernel component. By virtue of Eqns.~(\ref{eq: kern_identity_flip})-(\ref{eq: kern_identity_0mp0}), we have only used the four independent kernel components. We have also demarcated the parity in $s$ (even or odd) for these kernel sensitivities based on self-coupling assumption. The interested reader is referred to the discussion in Section.~3.2 of \cite{Dasetal20} where these parity-based selection rules are extensively laid out. Restricting the analysis to $\ell=1$ dipolar modes, the following section investigates the sensitivity of the 6 magnetic kernels (two for $s=0$ and four for $s=2$) to various locations in the modeled stars.
    
\begin{figure*}
    \centering
    \subfigure{
    \includegraphics[width=1.001\textwidth]{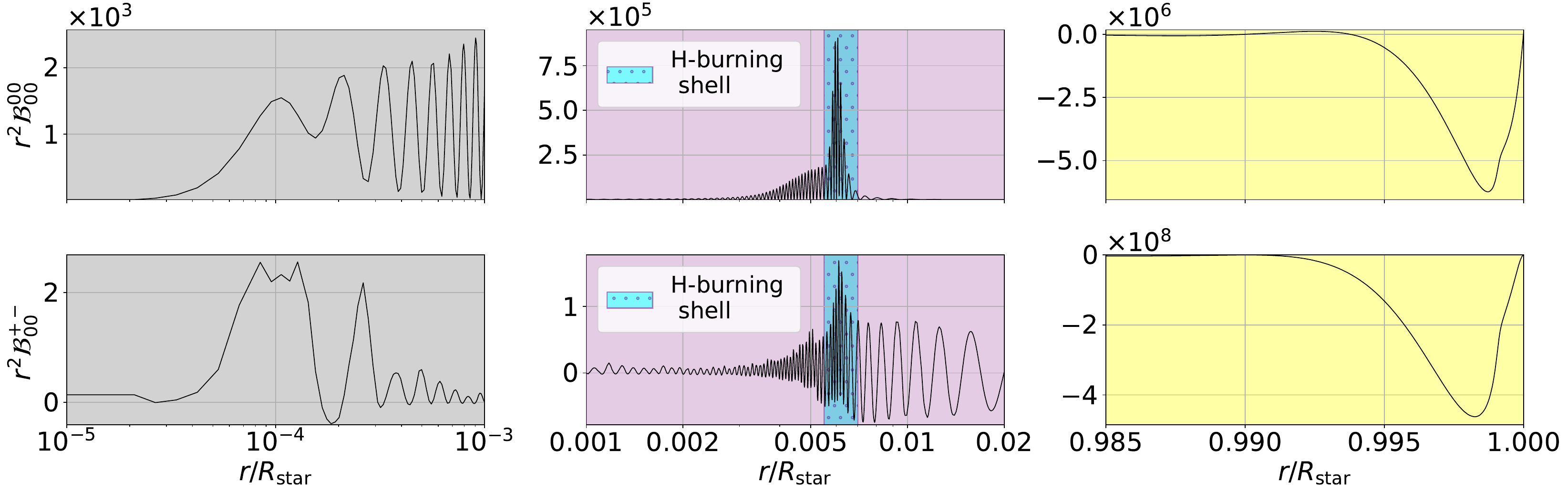}
    \label{fig:f17a}
    }
    \subfigure{
    \includegraphics[width=1.0\textwidth]{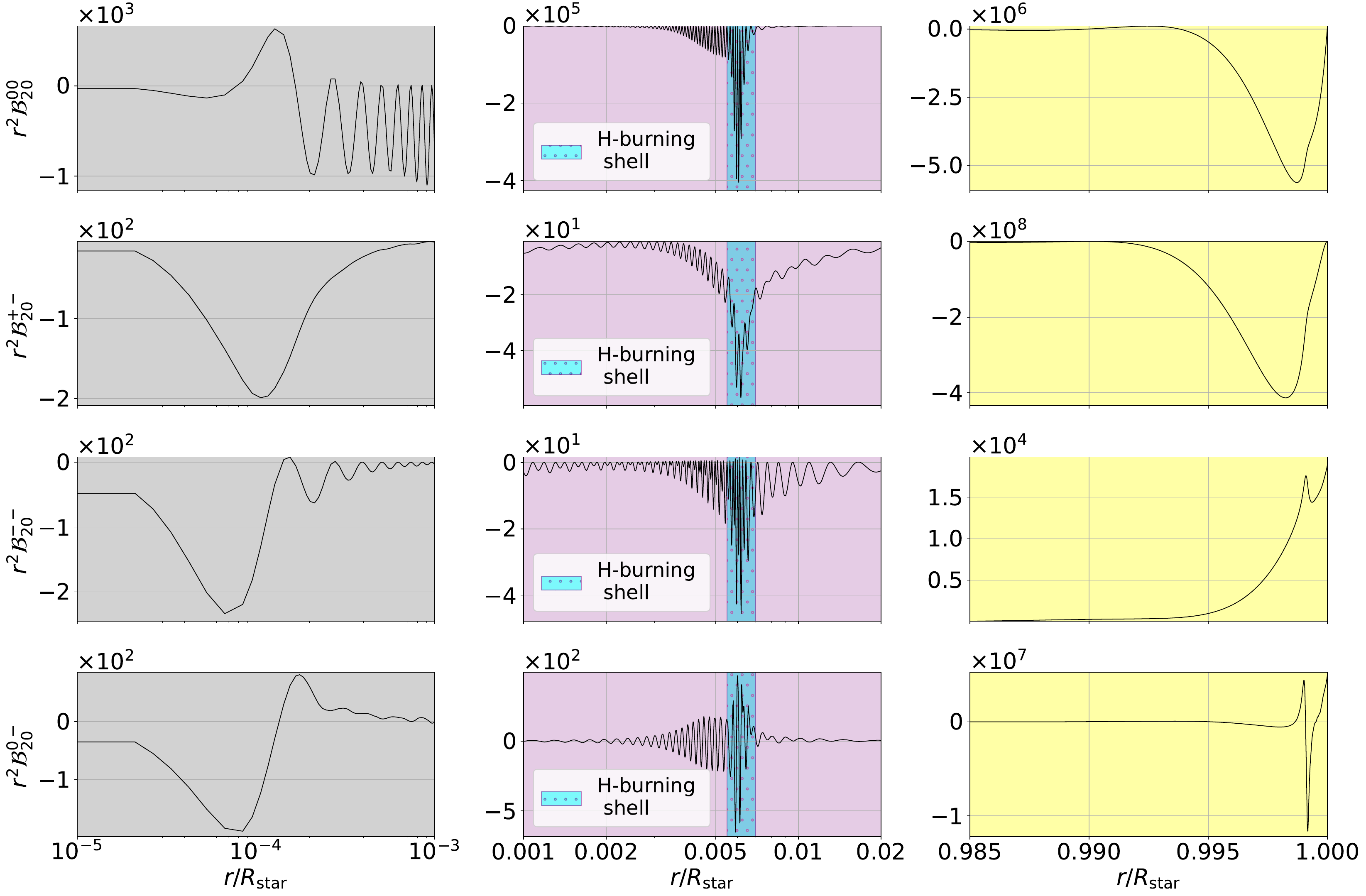}
    \label{fig:f17b}
    }
\caption{Prominent trends in the magnetic field sensitivity kernels $r^{2}\mathcal{B}_{s0}^{\mu\sigma}$ in $\rm{Hz}^{2}cm^{-1}G^{-2}$
of the RG which contribute to the splittings in the $m=0$ component for different models of axisymmetric poloidal + toroidal magnetic field configurations. In this figure, we choose a $g$-dominated $\ell=1$ mode with an unperturbed frequency of $173.191\:\mu\rm{Hz}$ (point $\color{red}\mathbf{g_1}$ in \ref{Zeta_plot}). The top two rows and bottom four rows show the $s=0$ and $s=2$ components of the kernels respectively in the deep regions of the He-core (\textit{left-most column}), around the H-burning shell (\textit{middle column}) and near the surface (\textit{right-most panel}). \sbh{The kernels $r^{2}\mathcal{B}_{s0}^{00}$, $r^{2}\mathcal{B}_{s0}^{+-}$, $r^{2}\mathcal{B}_{s0}^{--}$, and $r^{2}\mathcal{B}_{s0}^{0-}$ correspond to the sensitivities to $B_{r}^{2}$, $B_{\theta}^{2}+B_{\varphi}^{2}$, $B_{r}B_{\theta}$, and $B_{\theta}^{2}-B_{\varphi}^{2}$ respectively (refer Table~\ref{table:kernel_vs_field}).}
}
    \label{fig:f17}
\end{figure*}

\subsection{Study on the red-giant phase} \label{sec:red_giant}
\subsubsection{Peak sensitivities of different Lorentz-stress components as a function of depth}
\label{subsec:varkern}

We first study the 6 independent sensitivity kernels for the modes generated for the red giant stellar model. The kernels corresponding to modes labelled $g_1,\,p_1,\,g_2,\,p_2,\,p_0$ in Figure~\ref{Zeta_plot}, are plotted in Figures~\ref{fig:f17}, \ref{fig:f14}, \ref{fig:f61}, \ref{fig:f58}, \ref{fig:f3}, respectively. It is found that the kernels take significant values in three different parts of the star: the deep He core ($r/R_{\rm{star}}\in\left[ 10^{-5},10^{-3} \right]$), the vicinity of the H-burning shell ($r/R_{\rm{star}}\in\left[ 10^{-3},2\times10^{-2} \right]$), and the subsurface layers ($r/R_{\rm{star}}\in [0.985,1]$). In the H-shell's vicinity, the $r^{2}\mathcal{B}_{s0}^{00}$ components take the largest values for all dipolar and quadrupolar modes. Hence, the shift in $\nu_{n\ell}$ (due to the $s=0$ Lorentz-stress component) and the splitting among different values of $m$ (due to the $s=2$ Lorentz-stress component) result primarily from the radial pressure of the magnetic field (dominant in the radiative interior) on the plasma. 
    This is consistent with previous inversions which attributed splitting asymmetries on $g$-dominated modes to $B_r^2$ \citep{Bugnetetal2021, MathisBugnetetal2021, GangLietal2022, Deheuvels2023_mag, GangLietal2023_mag_13}.
    For all modes, we observe that the $r^2 \mathcal{B}_{s0}^{+-}$ kernels take the largest values in the near-surface layers. This implies that the tangential magnetic field pressure on the plasma in these layers should have the largest impact on the splittings of $p$-dominated modes, as demonstrated in \citep{Bugnetetal2021, MathisBugnetetal2021}. 
    This is especially the case during the subgiant phase (see Figures~\ref{fig:kern_mid}~and~\ref{fig:kern_late}). We will explore this phase in more detail in section~\ref{subg}.

\subsubsection{Contribution to splittings from fields in various regions of the star}
\label{h_mins_Section}

In Appendix \ref{subsec:Values_Splittings} we verify that the contribution to the splittings due to poloidal parts of $\boldsymbol{B}$-fields in the core, especially their radial component, dominate significantly over those due to their toroidal counterparts, and show an approximate $\nu_{n\ell}^{-3}$ variation, as also found in \cite{Bugnetetal2021}.
As an aside, we use a modification of the formula for magnetic splittings reported by \cite{Hasanetal05} to cater to mixed modes and report that it can be safely employed for faster computation of splittings for dipolar modes, but not suggested for calculating those for the $p$-dominant quadrupolar modes, which are often prominent in the power spectrum of stars as obtained by telescopes like \textit{Kepler}.
The computation of splittings (ref. Appendix~\ref{App:Stable_field_splitting}) for a theoretically deduced stable mixed magnetic field profile in the star shows that the contribution from its toroidal component (in any part of the star) to the splittings is negligible compared to the poloidal field in the vicinity of the H-shell.

Having established consistencies with previous studies, we then focus on quantifying the minimum values of different Lorentz-stress components, and therefore of minimum magnetic field amplitudes in each of the directions, required to be detectable. Such an exercise is valuable from the point of understanding (a) if any field components/layers other than $B_r$ at the H-burning shell could realistically be detectable from selected modes, and conversely (b) if one should be careful about ignoring all other layers and field components except $B_r^2$ at the H-shell. We carry out this analysis for the model stars prescribed in Section~\ref{sec:modelling_star_and_fields}.

In Figure~\ref{fig:f17}, the plots of the kernels for the different modes of the RGB model used here tell us that the radial (resp. tangential) magnetic pressure in the radiative interior (resp. outer envelope) dominantly impacts the splittings. Having access to these sensitivity kernels allows us to ask (a) whether the observed mode frequencies contain information from the core or the envelope or both, and, (b) what is the dominant component of the magnetic field in these regions that contributes to the perturbed eigenfrequencies.
As seen in earlier studies of RGB magnetoasteroseismology \citep{Bugnetetal2021,MathisBugnetetal2021,GangLietal2022,Deheuvels2023_mag,GangLietal2023_mag_13}, $B_r^2$ around the H-shell region is believed to be the component that contributes dominantly to observed frequency splittings. 
In this study, we want to investigate this further by finding the minimum strengths of the other Lorentz stress components in \sbh{different regions of the star} so as to contribute significantly to the observed $\delta \omega_{n \ell m}$.

\subsubsection{Sensitivity ratios}
\label{sec:Sensitivity_ratios}

We are looking for the minimum strengths of the various Lorentz stress components in the He core, H-burning shell, and envelope so as to contribute significantly to the observed $\delta \omega_{n \ell m}$. The ratio of the respective Lorentz stress components in the deep core and near the surface to those in the radiative interior near the H-burning shell is instrumental in answering these questions and may be estimated using the kernel peak heights in these regions.

\begin{figure*}
\centering
\includegraphics[width=0.98\linewidth]{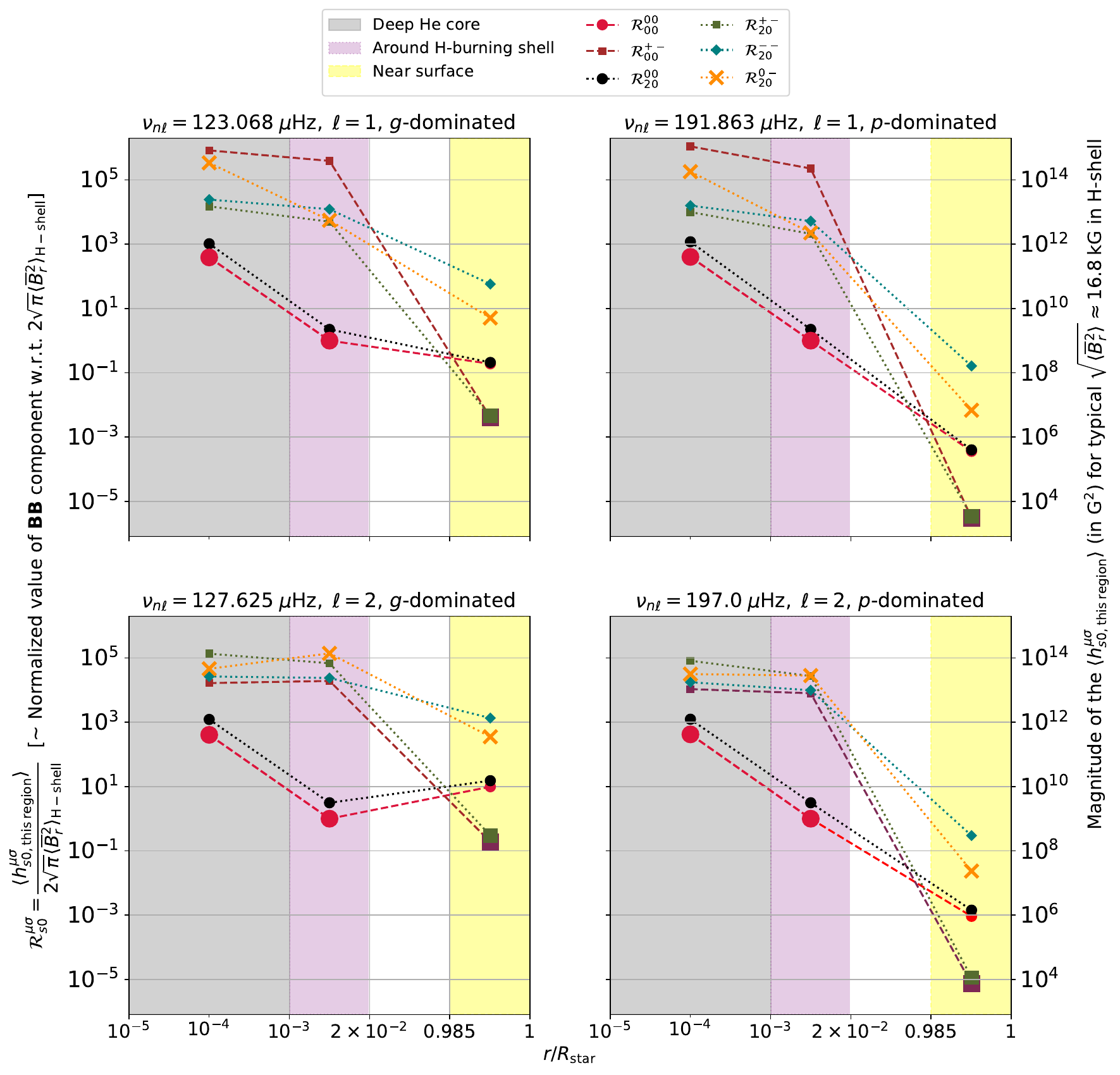}
\caption{Minimum ratios $\mathcal{R}_{s0}^{\mu\sigma}$ plotted for 4 different modes (\textit{top left}: a low-frequency $g$-dominated dipole mode, \textit{top right}: a high frequency $p$-dominated dipole mode, \textit{bottom left}: a low-frequency $g$-dominated $\ell=2$ mode, \textit{bottom right}: a high frequency $p$-dominated $\ell=2$ mode) showing the minimum local value of $\langle h_{s0}^{\mu\sigma} \rangle$ required to obtain the same splitting as that due to $\langle h_{00}^{00} \rangle$ component of the field around the H-burning shell. 
 \sbh{$h_{s0}^{00}$, $h_{s0}^{+-}$, $h_{s0}^{--}$, and $h_{s0}^{0-}$ correspond to the fields $B_{r}^{2}$, $B_{\theta}^{2}+B_{\varphi}^{2}$, $B_{r}B_{\theta}$, and $B_{\theta}^{2}-B_{\varphi}^{2}$ respectively (refer Table~\ref{table:kernel_vs_field}).}
$\mathcal{R}_{s0}^{\mu\sigma}$ for the dominantly sensitive components in each zone (discussed further in the text below) are marked with bigger symbols. The dashed/dotted connecting lines are used to guide the eye to the different $\mathcal{R}_{s0}^{\mu\sigma}$ obtained in each region. The ``$\;\langle h_{s0,\rm{this\:region}}^{\mu\sigma} \rangle\;$" used in the plot is denoted as $\langle h_{s0,\square}^{\mu\sigma} \rangle_{\rm{crit}}$ in the text for conciseness.}
\label{h_mins_graph}
\end{figure*} 
Let us define the absolute value of the local kernel-weighted average (for a mode with frequency $\nu_{n\ell}$) of a Lorentz stress component $h_{s0}^{\mu\sigma}$ in a region (denoted as $\square$, chosen among the 3 regions of interest: deep core, H-burning shell's vicinity, and sub-surface layers) as
\begin{align}
    \langle h_{s0,\rm{\square}}^{\mu\sigma} \rangle &= 
    \Bigg|
     \dfrac{\int_{\rm{\square}} dr \; r^{2} \: \mathcal{B}_{s0}^{\mu\sigma}(r;n,\ell,m) \: h_{s0}^{\mu\sigma}}{\int_{\rm{\square}} dr \; r^{2} \: \mathcal{B}_{s0}^{\mu\sigma}(r;n,\ell,m) }
     \Bigg|\nonumber \\
    &= 
    \left|
    \dfrac{ \left(\delta\omega_{n\ell m}\right)_{s0,\rm{\square}}^{\mu\sigma} }{ \frac{1}{I_{n\ell}} \: \int_{\rm{\square}} dr \; r^{2} \: \mathcal{B}_{s0}^{\mu\sigma}(r;n,\ell,m) }
    \right|,
\end{align}
where the frequency splittings due to Lorentz stress components $\langle h_{s0,\rm{\square}}^{\mu\sigma} \rangle$ in a region $\square$ is  $\left(\delta\omega_{n\ell m}\right)_{s0,\rm{\square}}^{\mu\sigma}$. 

Following calculations as shown in Appendix \ref{app:Critical_ratio_approach}, \sbh{we are inspired to introduce the parameter} $\mathcal{R}_{s0}^{\mu\sigma}$ as:
\begin{align} \label{eq: R_definition}
    \mathcal{R}_{s0}^{\mu\sigma} &= \left|\dfrac{ \int_{\rm{H}} dr \; r^{2} \: \mathcal{B}_{00}^{00}(r;n,\ell,0) }{ \int_{\rm{\square}} dr \; r^{2} \: \mathcal{B}_{s0}^{\mu\sigma}(r;n,\ell,0) } \right| = \dfrac{ \langle h_{s0,\rm{\square}}^{\mu\sigma} \rangle_{\rm{crit}} }{\langle h_{00,\rm{H-shell}}^{00} \rangle} \, ,\nonumber \\
    &= \dfrac{ \langle h_{s0,\rm{\square}}^{\mu\sigma} \rangle_{\rm{crit}} }{2 \sqrt{\pi}\langle \bar{B}_{r}^{2}\rangle_{\rm{H-shell}}} \, .
\end{align}
Here, we have used the identity from Eqn.~(\ref{eq: h00_to_Br}) in Appendix~\ref{app:Critical_ratio_approach}, $\langle h_{00,\rm{H-shell}}^{00} \rangle = 2 \sqrt{\pi}\langle \bar{B}_{r}^{2}\rangle_{\rm{H-shell}}$, where $\langle\bar{B}_{r}^{2}\rangle_{\rm{H-shell}}$ is the kernel-weighted average of the horizontally averaged squared radial magnetic field around the H-burning shell region. The factor $\mathcal{R}_{s0}^{\mu\sigma}$ in Eqn.~(\ref{eq: R_definition}) enables us to calculate \sbh{the minimum absolute value of the local kernel-weighted Lorentz stress} component required for the splitting (of $m=0$) corresponding to that component in a particular region of the star $\square$ to be equal in magnitude to the net shift in the multiplet due to radial magnetic field pressure in the vicinity of the H-burning shell region (denoted as H in the equations).

This inspires us to construct Lorentz stress component contribution diagrams for typical modes. In Figure~\ref{h_mins_graph} we show the 
$\mathcal{R}_{s0}^{\mu\sigma}$ ratio for each of the 6 kernels, in the three regions, representing the minimum value of $\langle h_{s0,\rm{\square}}^{\mu\sigma} \rangle$ in the area to have signature equivalent to the one due to the radial field at the H-burning shell. To facilitate the analysis, we also represent in the right axis the corresponding minimum field amplitude associated with $\langle h_{s0,\rm{\square}}^{\mu\sigma} \rangle$ in the given region to have signature equivalent to the one due to a 16.8 kG radial field at the H-burning shell. These ratios for the 6 sensitivity kernels are represented in Figure~\ref{h_mins_graph}, in the deep He core, at the H-burning shell, and near the surface.

For Figure~\ref{h_mins_graph}, we have chosen modes with the maximum contrast in their $g$ and $p$ characters. This was done in order to demonstrate the cases of the strongest H-shell $B_r^2$ sensitivity as compared to the strongest near-surface $(B_{\theta}^2 + B_{\varphi}^2)$ sensitivity. The choice of these modes is motivated by the analysis in Appendix~\ref{sec:simplified_ratio_appendix} where we compute a measure of the critical ratio for detectability of near-surface tangential fields for dipolar and quadrupolar modes. It is to be noted that Figure~\ref{fig:kern_vs_zeta} is a simplified version of Figure~\ref{h_mins_graph} since it compares the detectability potential of only the surface tangential component as compared to the H-shell radial field component. However, Figure~\ref{h_mins_graph} conveys a more complete picture which should inspire how magnetic field should be parameterized when carrying out an inverse problem. For instance, Figure~\ref{h_mins_graph} suggests that a maximum of three zones could be considered when discretizing the radial dimension. Although, in practice, the number depends on the relative critical ratios of field strengths at different depths. This is elucidated in the following subsections~\ref{sec:deep_core} \& \ref{sec:surface}.

\subsubsection{Critical fields in the deep core} \label{sec:deep_core}

\begin{figure}
    \centering
    \includegraphics[width=0.5\textwidth]{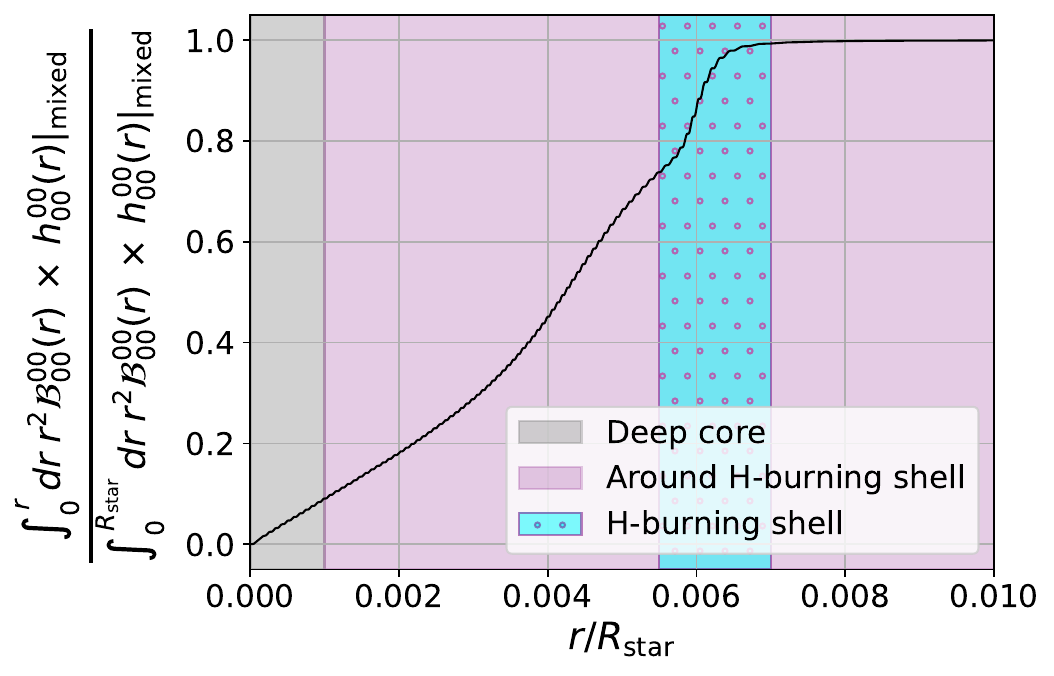}
    \caption{Plot of the relative cumulative contribution of $h_{00}^{00}(r)\bigr|_{\rm{mixed}}$ to the shift obtained in the dipolar mode with frequency $\nu_{n\ell}=123.068\:\mu\rm{Hz}$ (top left of Figure~\ref{h_mins_graph}) due to radial magnetic pressure, where $h_{00}^{00}(r)\bigr|_{\rm{mixed}}$ is calculated using the mixed field as mentioned in Eqn.~\ref{B_Bugnet_eqn}.}
    \label{fig:h0000_times_r2B0000_49.pdf}
\end{figure}

In the deep He core and around the H-burning shell, the $\langle h_{s0,\rm{\square}}^{00} \rangle$ are seen to have the least cutoff compared to other Lorentz stress components, indicating that contributions to splittings from the other Lorentz stress components (considering that they all have the same order of magnitude) in that region are much smaller, in agreement with \cite{Bugnetetal2021, MathisBugnetetal2021, GangLietal2022}. For the radial field in the He core to have an effect similar to the field in the H-burning shell, it would have to be at least about 20 times larger. 
For the mixed-field configuration (refer Eqn.~\ref{B_Bugnet_eqn}), we calculated the $\langle h_{s0}^{00}\rangle$ in the deep core and H-burning shell's vicinity and found that the deep core contributed to nearly 10\% of the total splitting (for both $s=0$ and 2) due to radial magnetic fields in the star's interior which are verified by Figure~\ref{fig:h0000_times_r2B0000_49.pdf}. The plot also demonstrates that the H-burning shell accounts for only $\sim 25\%$ of the total shift in the multiplet due to the radial magnetic pressure, with the majority remaining contribution coming from layers beneath it.

\subsubsection{Critical field near the surface}\label{sec:surface}

The outer layers, on the other hand, usually have smaller cutoffs, especially for the $h_{s0}^{+-}$ Lorentz stress component, making the total splitting more susceptible to tangential magnetic field pressure than radial magnetic field pressure or anisotropic Lorentz stress components, which is expected as modes are mostly acoustic near the surface. The $\mathcal{R}_{s0}^{\mu\sigma}$ in the subsurface layers are observed to have much lower values for $p$-dominated modes than $g$-dominated modes and also have smaller values at higher frequencies than at lower frequencies, which is very characteristic of asymptotic acoustic modes \citep{Bugnetetal2021, MathisBugnetetal2021}. 
These panels convey to us that for dipolar $g$-dominated modes at lower frequencies to be equally sensitive to $B_r$ in the H-burning shell than the tangential magnetic field in the envelope, the kernel-weighted field (field calculated from the kernel-weighted Lorentz stress component) amplitude in the envelope should be at least about 0.1 times that at the H-burning shell.
Under the same assumption, high-frequency dipolar $p$-dominated modes only require the surface tangential field to be above 0.001 times the internal radial field. 

This makes high-frequency $p$-modes the best candidates for the inversion of magnetic fields in the subsurface layers of the RG star. For the effect of tangential surface magnetism to have a significant effect, the surface field amplitude should be higher than about $10^{-3} \times$ amplitude of the radial field in the H-burning shell.
In Appendix~\ref{appC}, we have computed splittings due to various typical magnetic field topologies. Among them, two distinct poloidal magnetic fields dominate in the radiative zone.
In both cases, the strongest contributions to the splittings in low-frequency modes, especially those which are $g$-dominated, come from the $h_{00}^{00}$ and $h_{20}^{00}$ components in the vicinity of the H-burning shell as the remaining Lorentz stress components do not qualify their corresponding critical criteria as given by $\mathcal{R}_{s0}^{\mu\sigma}$.

\subsection{Study on subgiant phases}
\label{subg}
\begin{figure*}
    \centering
    \begin{subfigure}[]{ \includegraphics[width=0.45\linewidth]{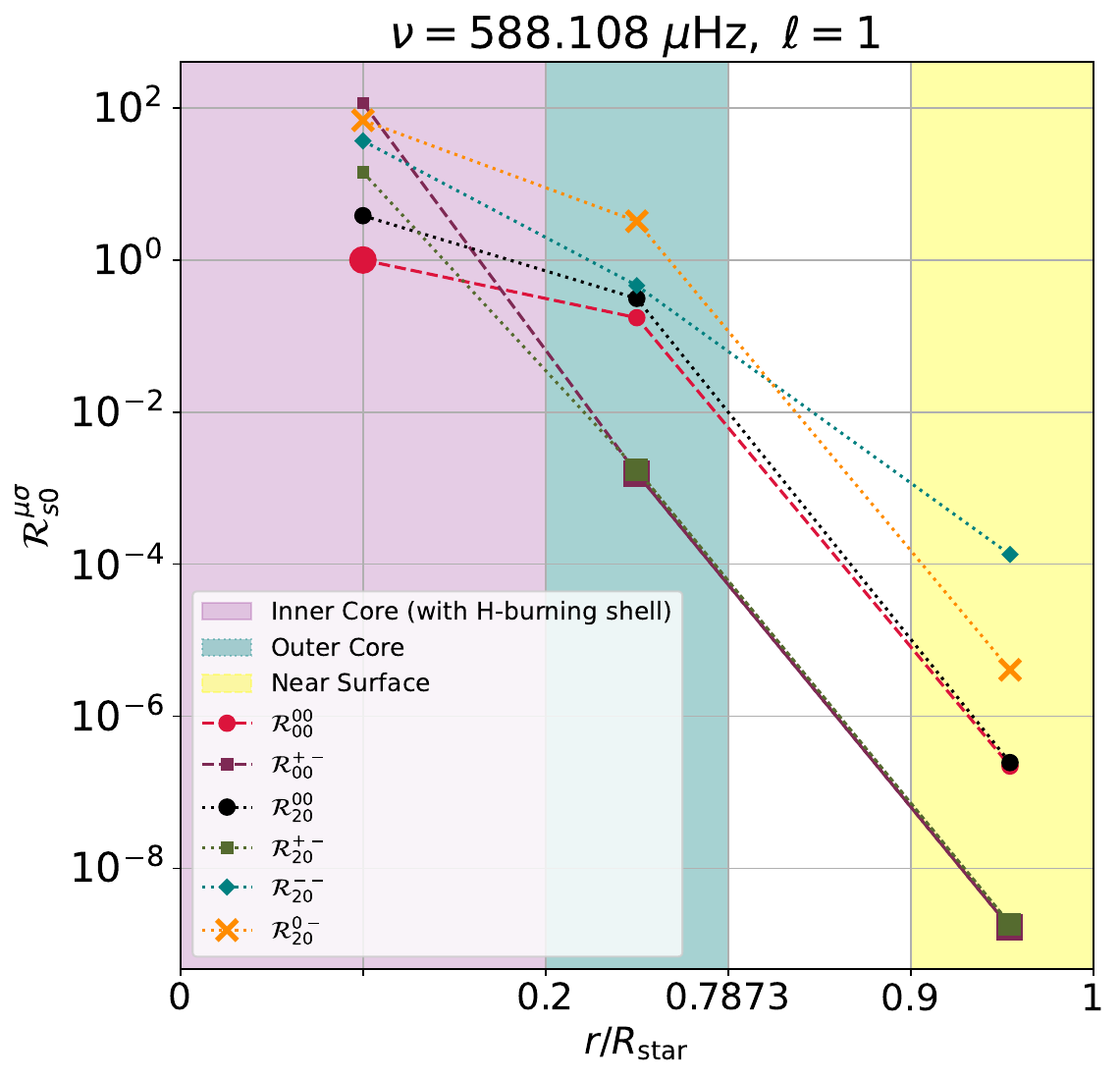} }
    \end{subfigure}
    \begin{subfigure}[]{ \includegraphics[width=0.45\linewidth]{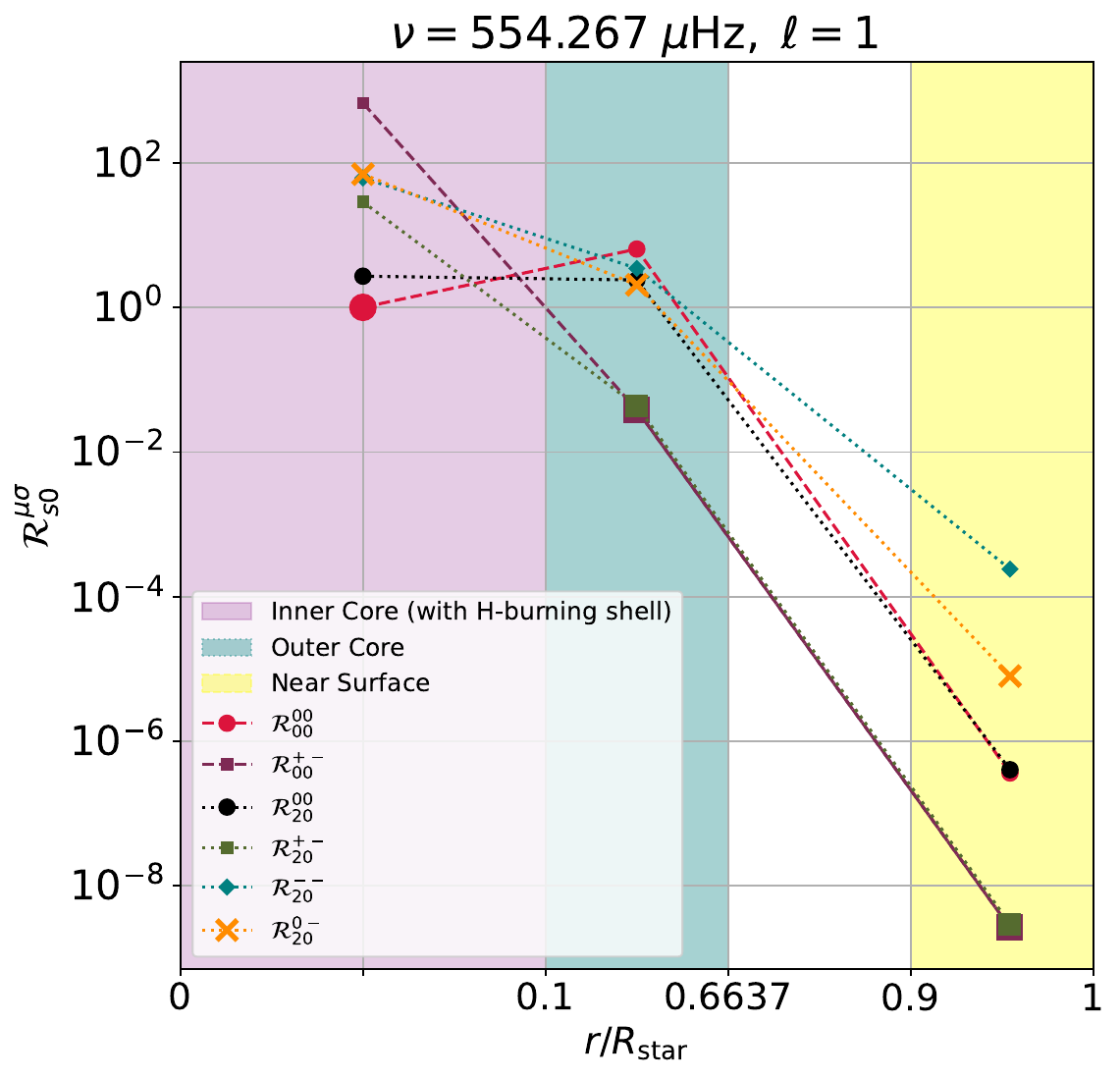} }
    \end{subfigure}
    
    \caption{Plot of $\mathcal{R}_{s0}^{\mu\sigma}$ measured for the two $p$-dominant dipole mixed modes (whose kernels are plotted in Figures~\ref{fig:kern_mid} and~\ref{fig:kern_late}) closest to their corresponding $\nu_{\rm{max}}$ in the (a) MSG, and (b) LSG stages.  \sbh{The kernels $r^{2}\mathcal{B}_{s0}^{00}$, $r^{2}\mathcal{B}_{s0}^{+-}$, $r^{2}\mathcal{B}_{s0}^{--}$, and $r^{2}\mathcal{B}_{s0}^{0-}$ correspond to the sensitivities to $B_{r}^{2}$, $B_{\theta}^{2}+B_{\varphi}^{2}$, $B_{r}B_{\theta}$, and $B_{\theta}^{2}-B_{\varphi}^{2}$ respectively (refer Table~\ref{table:kernel_vs_field}).}}
    \label{fig:h_plot_MSG_LSG}
\end{figure*} 
Having dealt with a red-giant star in the last subsection, we proceed to investigate the \sbh{potential for detection} of magnetic fields in the earlier phases of the star - the sub-giant phase.
This phase is identified as the transition period between the main sequence and red-giant characterized by the deepening of the convective envelope in low and intermediate-mass stars. In this stage, the $g$-mode trapped in the core starts coupling to the $p$-mode in the envelope and presents us with mixed modes for the first time in the evolution. We now explore which zones of these stars affect splittings the most and whether near-surface fields and core fields could be distinctly measured using different modes in different parts of the power spectrum.

Axisymmetric magnetic field sensitivity kernels were constructed (eg., as in Figures~\ref{fig:kern_mid} and~\ref{fig:kern_late}) using the aforementioned theory, and three significant zones were identified for each of them. The significant zones for the MSG stage are the inner core with $r/R_{\rm{star}}\in(0,0.2]$ containing the H-burning shell, the outer core with $r/R_{\rm{star}}\in(0.2,0.7873)$, and the near-surface layers with $r/R_{\rm{star}}\in(0.9,1)$, whereas those for the LSG stage are $r/R_{\rm{star}}\in(0,0.1]$ containing the H-burning shell, the outer core $r/R_{\rm{star}}\in(0.1,0.6637)$, and the near-surface layers with $r/R_{\rm{star}}\in(0.9,1)$.
As done in Section~\ref{h_mins_Section}, we investigate Lorentz stress component contribution diagrams similar to Figure~\ref{h_mins_graph}  (normalization was done with respect to the inner radiative zones of both the sub-giants which contain the H-burning shell) for all used modes. 
Lorentz stress component contribution diagrams for a dipole mode closest to $\nu_{\rm{max}}$ at each sub-giant stage are shown in Figure~\ref{fig:h_plot_MSG_LSG}.

\subsubsection{Critical field near the surface}
The plots for $\mathcal{R}_{s0}^{\mu\sigma}$ help us identify how much local kernel-weighted tangential magnetic field is required close to the star's surface to match the contribution to splittings due to the kernel-weighted radial magnetic field in the deepest parts of the radiative core, where the magnetic field is expected to be much higher than the rest of the star as suggested in \cite{Bugnetetal2021}.
As a result, the tangential subsurface magnetic amplitude for the discussed SGs only has to be above $10^{-4.5}\times$ the amplitude of the radial field in the vicinity of the H-burning shell to have a comparable effect on $p$-dominated dipolar modes close to their $\nu_{\rm{max}}$. This is a more favorable scenario than on the RGB for the detection of shift on $p$-dominated modes resulting from envelope magnetic fields.

\subsubsection{Critical field in the outer core}

The outer core, lying beneath the base of the convective zone, is most sensitive to the tangential magnetic pressure. The kernel-weighted tangential field in this region has to be nearly 3 and 3.5 orders of magnitude larger for the MSG and LSG phases respectively than that in the sub-surface layers to dominate over splittings due to the latter in $p$-dominated dipolar modes close to $\nu_{\rm{max}}$. Conservation of field flux in the core during post-MS evolution dictates the core field magnitudes to be much smaller than the red-giant. The evolution of sub-surface fields would then help quantify the splittings better. Hence, depending on the exact model of the internal magnetic field topology, any of the three regions in the SG stages could contribute to the majority of the splittings in their spectrum.

\sbh{
\subsection{Detectability of magnetic splittings in RG and SG phases from observations}
\label{section:Detectability}
}
Perturbations to the stellar reference model induce frequency splittings.
These splittings are detectable in the data obtained from a telescope only when certain criteria are satisfied. For the different cases discussed in this work (also see Appendix~\ref{appC}), we calculate $y_{n\ell m}=\delta\nu_{n\ell m}\times \left( B_{0,\rm{ref}}/B_{0} \right)^{2}$, $B_{0,\rm{ref}}$ being a reference value for field strength $B_0$ that induces a splitting of $y_{n\ell m}$. A necessary criterion to be satisfied is $\delta\nu_{n\ell m}>\nu_{\rm{min}}$, where $\nu_{\rm{min}}$ is some measure of minimum separation in frequency. This implies that for detectability, the inequality $B_{0}>B_{0,\rm{min}}$, needs to be satisfied. Here $B_{0,\rm{min}}=B_{0,\rm{ref}} \times \sqrt{\nu_{\rm{min}}/y_{n\ell m}}$.
The bare minimum and simplest of all such criteria is that the distance between the peaks on the power spectrum should be separated by at least 1 unit of the frequency resolution $\nu_{\rm{res}}$, i.e. $\nu_{\rm{min}}=\nu_{\rm{res}}$, which is typically the reciprocal of the total observation time for the star. While calculating the value of $B_{0,\rm{min}}$, one has to verify that $|\boldsymbol{B}|^2(4\pi GM^2/R^4)^{-1}<<1$ and that it is below its corresponding critical field as mentioned in \cite{Fuller2015}.
Since these splitting measurements are key to inferring the perturbations, in this subsection, we investigate the detectability of splittings in dipolar mixed modes closest to $\nu_{\rm{max}}$ during the 3 evolutionary stages of the star in consideration.

The configuration of the fields used for the calculation of the magnetic splittings is of the form given in Eqn.~\eqref{eq:field_config_general}, where for a poloidal field in the core (H-burning shell region in RG and inner core in SG) we have $\alpha(r)=0$ and
\begin{align} \label{eq: beta_expression}
    \beta(r) = \dfrac{1}{4} \left[ 1 - \tanh \left\lbrace \beta_{1} \left(\frac{r}{R_{\rm{star}}} - \beta_{2} \right) \right\rbrace \right],
\end{align}
and for a toroidal field in the envelope we have $\beta(r)=0$ and
\begin{align}\label{eq: alpha_expression}
    \alpha(r) = \dfrac{1}{4} \left[ 1 + \tanh \left\lbrace \alpha_{1} \left(\frac{r}{R_{\rm{star}}} - \alpha_{2} \right) \right\rbrace \right],
\end{align}
for a toroidal field in the envelope (refer Appendix~\ref{B_topologies} for details on the choice of such toy topologies and $B_0$), where the coefficients $\beta_1,\beta_2,\alpha_1,\alpha_2$ and $B_0$ (check Table~\ref{table:Detectability_configs}) were chosen keeping the structure of the star in mind.
The poloidal magnetic fields in the core have magnitudes $\sim B_0$ close to the centre and $\sim 0$ in the envelope, whereas the toroidal fields in the envelope have magnitudes $\sim B_0/2$ close to the surface and $\sim 0$ in the core.

\begin{figure*}[t]
    \centering
    \includegraphics[width=0.8\textwidth]{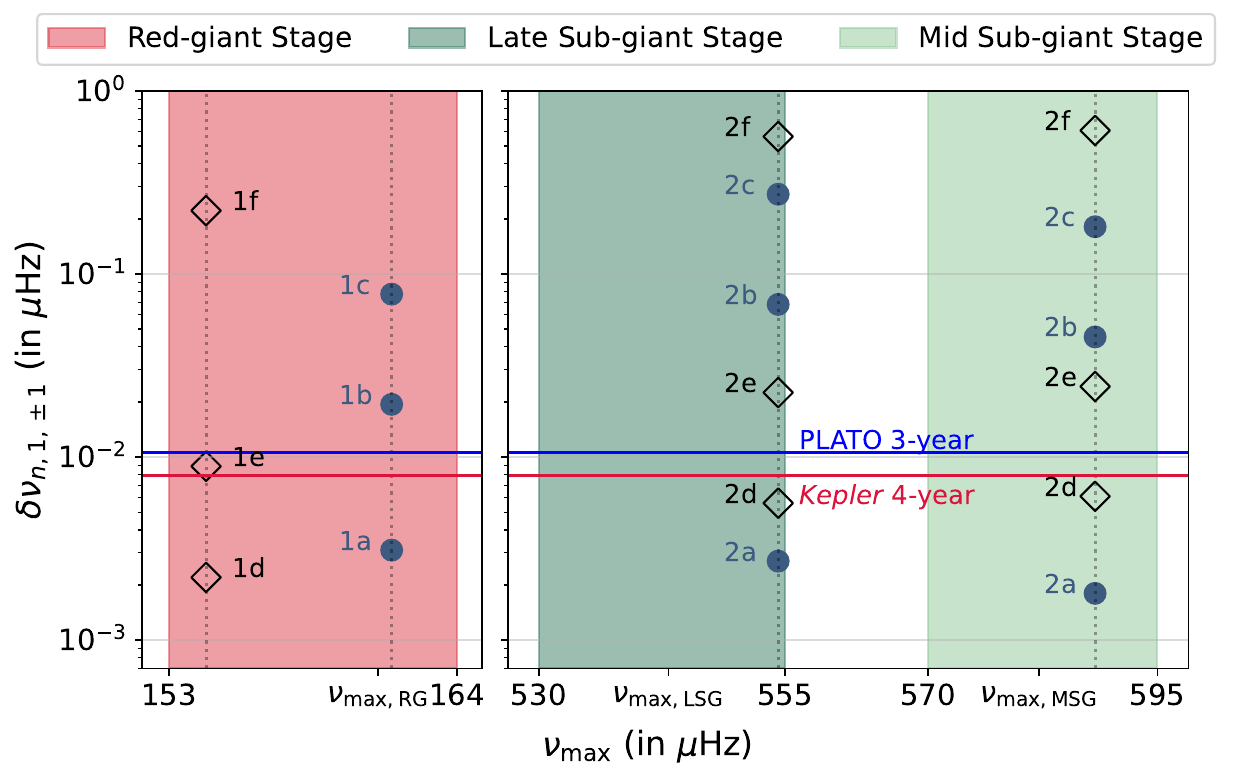}
    \caption{\sbh{Plot of the splittings in the $m=\pm 1$ of dipolar mixed modes closest to the $\nu_{\rm{max}}$ for all three stages of the star. The black-unfilled diamonds indicate toroidal fields in the envelope and the blue-filled circles indicate the effect of core poloidal fields. The configuration of the field used is indicated beside the points using \textbf{C.\#} from Table~\ref{table:Detectability_configs}. The \textit{Kepler} 4-year and upcoming PLATO's 3-year frequency resolutions have been plotted as a reference for detectability.}}
    \label{fig:detectbility_all_phases}
\end{figure*}

Figure~\ref{fig:detectbility_all_phases} demonstrates the detectability of magnetic splittings for the most visible dipolar mixed modes (closest to $\nu_{\rm{max}}$) in our 1.3 $M_{\odot}$ star during its MSG, LSG and RGB phases. In Figure~\ref{h_mins_graph} we already noted that specific field components at the H-burning shell and the near-surface envelope dominantly affect the splittings. Similarly, Figure~\ref{fig:h_plot_MSG_LSG} shows that in the MSG and LSG phases, magnetic splittings are primarily influenced by field components in the inner core H-burning shell, the outer core and near-surface envelope. To have a consistent model of magnetic field across the different evolutionary phases of the star, we use inner poloidal and outer toroidal fields as per Eqns.~(\ref{eq: beta_expression}) \& (\ref{eq: alpha_expression}) with values of $\alpha_1, \alpha_2, \beta_1, \beta_2$ specified in Table~\ref{table:Detectability_configs}. We already showed that for MSG and LSG phases, there are three zones of potential magnetic detectability. Therefore, the two-zone model of the magnetic field does not provide explicit control over the outer-core field. Nevertheless, it is easy to see that we could estimate the limit of detectability of field strength of the outer core by using a three-zone model instead of a two-zone model. 

For the RG, where we have clearly distinguishable $g$ and $p$-dominated modes, we have two vertical lines showing the detectability for the inner poloidal field (to which the $g$-dominated modes are sensitive) and the outer toroidal field (to which the $p$-dominated modes are sensitive). This is not the case for MSG and LSG and hence all the cases of splitting are plotted along one vertical line.

A minimum splitting of 7.9 $\rm{n}Hz$ and 10.6 $\rm{n}Hz$ is needed to be detectable by 4-years of Kepler and 3-years of PLATO. Treating these as the $\nu_{\rm{res}}$ and using $B_0^2$ proportionality to splitting strength, we calculate the threshold values for detectability of field strengths of different topologies at different layers of the star. For Kepler observations for our choice of RG and the model magnetic field, we find a threshold of $B_0=$32 kG poloidal field in the H-burning shell and 190 G toroidal field in the near-surface envelope to be detectable. Similarly, H-burning shell threshold of $B_0=$2.1 MG and 1.7 MG for MSG and LSG, respectively, and near-surface threshold of toroidal $B_0$ around 114 G and 119 G for MSG and LSG, respectively are required for these splittings to be detectable.

We want to emphasize that these limits are specific to our choice of stellar model and magnetic field model. But the procedure of constructing Lorentz-stress diagrams as in Figures~\ref{h_mins_graph} \& \ref{fig:h_plot_MSG_LSG} is general and can be followed in a similar fashion for the choice of star at hand to obtain estimates of magnetic field detectability thresholds for a given satellite measurement.

\begin{table}[]
\centering
\begin{tabular}{|c|c|c|c|c|c|c|}
\hline
\textbf{Phase}           & 
\textbf{$\beta_1$}      & \textbf{$\beta_2$}        & \textbf{$\alpha_1$}     & \textbf{$\alpha_2$}       & \textbf{$B_0 [G]$}       & \textbf{C.\#}\\ \hline
\multirow{6}{*}{RGB}     & \multirow{3}{*}{50} & \multirow{3}{*}{0.05} & \multirow{3}{*}{0}  & \multirow{3}{*}{0}    & $2 \times 10^4$  &1a       \\ \cline{6-7} 
                         &                     &                       &                     &                       & $5 \times 10^4$  &1b       \\ \cline{6-7} 
                         &                     &                       &                     &                       & $10^5$   &1c      \\ \cline{2-7} 
                         & \multirow{3}{*}{0}  & \multirow{3}{*}{0}    & \multirow{3}{*}{50} & \multirow{3}{*}{0.15} & $10^2$   &1d      \\ \cline{6-7} 
                         &                     &                       &                     &                       & $2 \times 10^2$    &1e     \\ \cline{6-7} 
                         &                     &                       &                     &                       & $10^3$    &1f     \\ \hline
\multirow{6}{*}{SG} & \multirow{3}{*}{20} & \multirow{3}{*}{0.1}  & \multirow{3}{*}{0}  & \multirow{3}{*}{0}    & $10^6$     &2a    \\ \cline{6-7} 
                         &                     &                       &                     &                       & $5 \times 10^6$   &2b      \\ \cline{6-7} 
                         &                     &                       &                     &                       & $10^7$    &2c     \\ \cline{2-7} 
                         & \multirow{3}{*}{0}  & \multirow{3}{*}{0}    & \multirow{3}{*}{50} & \multirow{3}{*}{0.85} & $10^2$    &2d     \\ \cline{6-7} 
                         &                     &                       &                     &                       & $2 \times 10^2$    &2e     \\ \cline{6-7} 
                         &                     &                       &                     &                       & $10^3$   &2f      \\ \hline
\end{tabular}
\caption{\sbh{Table showing the different values of parameters used for constructing field configurations according to Eqns.~(\ref{eq: beta_expression}) \& (\ref{eq: alpha_expression}) to analyze detectability in RGB and LSG/MSG. Note that since the same parameters are used for both MSG and LSG, we have referred to it as SG. Column C.\# is the configuration ID that is used to calculate the splittings in Figure~\ref{fig:detectbility_all_phases}. }}
\label{table:Detectability_configs}
\end{table}

\section{Discussions and Conclusions}

Deciphering magnetic field topology inside stars constitutes an outstanding challenge which is crucial to address the angular momentum transport problem in stellar interiors. Recent breakthrough studies have measured $B_r^2$ magnitudes in the vicinity of H-shell region in RGB stars \citep{GangLietal2022, Deheuvels2023_mag, GangLietal2023_mag_13}. However, insights on field topology at different depths in the star would provide invaluable constraints for simulating stellar evolution. Borrowing Lorentz-stress sensitivity kernels in its full glory from helioseismology \citep{Dasetal20}, our study (a) evaluates component-wise detectability of Lorentz-stress in red giant stars, (b) proposes simplified minimum ratios $\mathcal{R}_{s0}^{\mu\sigma}$ to quantify potential contributions of different field configurations from sensitive layers, (c) demonstrates that near-surface detectability in the subgiant phase is more favorable than its red giant phase,  (d) highlights that caution needs to be adopted when attributing field contributions to be coming from purely the H-shell vicinity, and (e) most importantly, given a stellar model, lays out a formal method to assess how to parameterize an inverse problem by constructing minimum ratio $\mathcal{R}_{s0}^{\mu\sigma}$ plots.
\sbh{Figure~\ref{fig:schemes} shows a schematic diagram for the various zones in a red-giant and sub-giant star we identified as important for the magnetic field inversion.}

By using the formalism provided in \cite{Dasetal20} for calculating the splittings when magnetic fields are incorporated in a model star, we can obtain thorough information about how different components of Lorentz-stress $\boldsymbol{B}\boldsymbol{B}$ impact these observables. A prerequisite for this is the knowledge of sensitivity kernels. For ease of reference, we provide the reader with Table~\ref{table:kernel_vs_field} connecting kernels to the corresponding Lorentz-stress components. However, not all components contribute to the total observed frequency splittings to the same extent. We demonstrate this by laying out the specific kernels for each component and drawing attention to the most sensitive components for a $1.3\,M_{\odot}$ red giant (in Figure~\ref{fig:f17}) and its corresponding mid and late subgiant phases (in Figures~\ref{fig:kern_mid} \& \ref{fig:kern_late}).

Since different modes can potentially sense different layers in the star, having access to these kernels for stellar magnetism allowed us to investigate which are the radial layers where magnetic fields dominantly contribute to the observed splittings. This is equivalent to asking the question --- for a given star, which are the layers and field components that are detectable? Quantifying detectability is not straightforward since it depends on multiple factors such as (1) the kernel sensitivity, (2) the strength of the corresponding field component, and (3) the data resolution of the satellite. It is known from previous studies that for red giants the $B_r^2$ in the vicinity of the H-shell has a dominant contribution. Therefore, in order to assess the potential detectability of a field component at a given depth, we assess its contribution to total splitting with respect to $B_r^2$ in the vicinity of the H-shell. We quantify this as a critical ratio $R_{s0}^{\mu\sigma}$ for Lorentz-stress components denoted by $\mu$ and $\sigma$ at zones where kernels are significant. The lower the value of $R_{s0}^{\mu\sigma}$, the higher is the detectability potential of that component at that depth. 

We first provide a detailed analysis for a RGB star considering magnetic fields act as a small perturbation to the background stellar model. In order to investigate component-wise contribution from internal layers, we construct plots for $\mathcal{R}_{s0}^{\mu\sigma}$ as in Figure~\ref{h_mins_graph}. We report that there are broadly three important regions of consideration in our red giant --- deep core, H-shell vicinity and near-surface. $B_r^2$ dominates in the deep core and the vicinity of the H-shell while the tangential field $(B_{\theta}^2 + B_{\varphi}^2)$ dominates in the near-surface. However, the near-surface tangential field would have to be at least a thousandth of the H-shell radial field in order to have a comparable contribution to the high-frequency dipolar $p$-dominated modes' splitting. The detectability potential for the near-surface tangential field is slightly less for high-frequency quadrupolar $p$-dominated modes and nearly impossible using $g$-dominated modes (which require impractically strong fields in the near-surface).

Having independently established that $B_r^2$ is indeed the dominant contributor for our red giant, we assess the relative contributions from the deep core, the H-shell and the region in between. Using the same field as in \cite{Duez2010B} which was also used in \cite{Bugnetetal2021}, in Figure~\ref{fig:h0000_times_r2B0000_49.pdf} we demonstrate that around 25\% of the total splitting comes from the H-shell, 10\% from the deep core and the rest from the intermediate region between deep core and H-shell. As a result, the asteroseismic signature cannot directly be attributed to the field in the H-burning shell but rather to a weighted average inside the radiative interior.

Similar analyses are performed for two sub-giant stages of the same star. The kernel plots in Figures~\ref{fig:kern_mid} \& \ref{fig:kern_late} suggest three important zones --- the inner core (which contains the H-shell), the outer core, and the near-surface region. We construct the corresponding $\mathcal{R}_{s0}^{\mu\sigma}$ plots in Figure~\ref{fig:h_plot_MSG_LSG} and find that the near-surface tangential field has to be only about $10^{-4} \times$ the \sbh{inner} core $B_r$ to contribute significantly to splittings in $p$-dominated \sbh{dipolar} modes. This allows for the potential detectability of a weaker field near the surface as compared to the RGB phase. However, to definitively quantify detectability of near-surface fields, more comprehensive models of magnetic fields in the near-surface regions of the star are necessary. This is also why we have consistently adopted caution in quantifying potential detectability in the form of ratio $R_{s0}^{\mu\sigma}$ instead of the absolute value of field strengths.
\sbh{
To address concerns about the detection of these different components from telescope data,
we have also evaluated the splittings for poloidal fields in the (inner) core and toroidal fields in the subsurface layers. This led us to determine minimum values of $B_0$ for these topologies in the different stages of evolution of our $1.3 M_{\odot}$ stellar model.
}

\begin{figure}[t]
    \centering
    \includegraphics[width=1.02\linewidth]{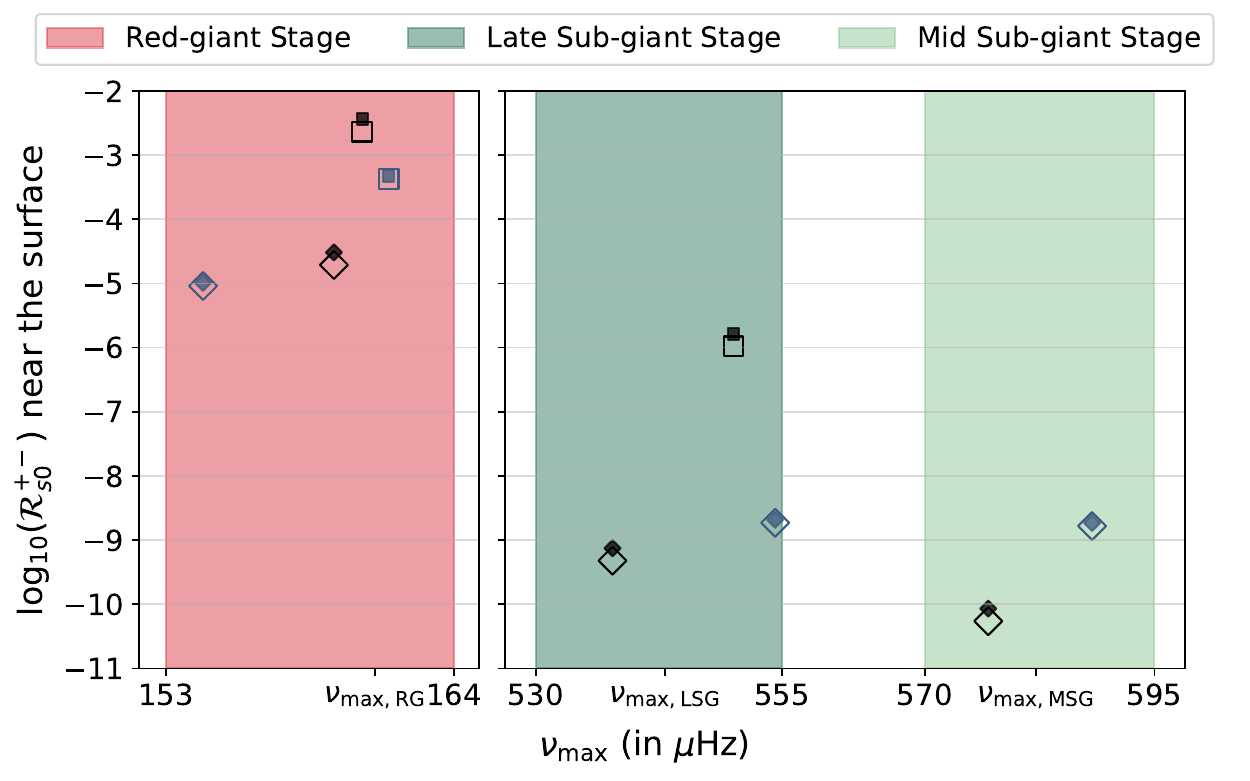}
    \caption{Plot of the log of $\mathcal{R}_{s0}^{+-}$ in the sub-surface layers for $p$ and $g$-dominated dipolar and quadrupolar mixed modes closest to the $\nu_{\rm{max}}$ for all three stages of the star. The square marker indicates a $g$-dominant mode whereas diamonds indicate $p$-dominant modes. The colors \sbh{blue} and black indicate dipolar and quadrupolar modes respectively. Big hollow symbols are used for $s=0$, and small filled symbols are used for $s=2$ terms \sbh{of the same modes}. }
    \label{fig:Rs0+-_all_phases}
\end{figure}

From Figure~\ref{fig:Rs0+-_all_phases}, we observe that for eigenmodes close to $\nu_{\rm{max}}$ it is easier to detect the effect of the near-surface field over that at the H-burning shell in the earlier stages. This gets increasingly harder as the star ages from the main sequence to the RG phase.
\sbh{
Figure~\ref{fig:Rs0+-_all_phases} also leads us to identify that the ratio of subsurface tangential magnetic field pressure and the radial field pressure in the H-burning shell in the RG phase has to be nearly $10^4$ times larger than in the SG phases for the splitting contribution from the respective subsurface and core fields to be equivalent.}
Upon considering that no convective core dynamo episodes occur beyond the terminal age main sequence phase, fossil field flux conservation in the core suggests that the core field amplitude should scale as 0.030 and 0.042 to that in the RG phase for the MSG and LSG stages respectively. If the order of magnitude of the near-surface field doesn't alter dramatically over the three discussed phases, then for the MSG and LSG, the contribution from the inner core will be negligible compared to contributions from tangential fields in the outer core and especially the subsurface layers.

Finally, we believe that the method of constructing the critical ratio $\mathcal{R}_{s0}^{\mu\sigma}$ plots should inspire future efforts in inferring stellar magnetism. In the same way, as has been done in the last few years to inverse rotation rates inside solar-type pulsators, magnetic fields might be probed through asteroseismic data inversion. Given a stellar model, an inversion effort is usually preceded by parameterization of the unknown model (in this case, the Lorentz-stress tensor). This requires knowing how many zones should we discretize the star into and which components we should include in the inverse problem at each sensitive zone. As a first step, the kernels may be plotted to obtain an initial guess about the number of important zones and the radial extent of each zone. In the second step, using the knowledge of the zones, $\mathcal{R}_{s0}^{\mu\sigma}$ plots may be constructed to obtain the most sensitive component for each zone (the one with the lowest value of $\mathcal{R}_{s0}^{\mu\sigma}$). Based on the number of available modes with reliable splitting measurements, relative contributions of different model parameters can be used as a guiding tool to decide which ones to reject from the magnetic inverse problem setup.

\section{Acknowledgements}

This project has received funding from the European Union’s Horizon 2020 research and innovation programme under the Marie Skłodowska-Curie grant agreement No 101034413. SBD acknowledges Prof.~Jeroen Tromp at Princeton University for supporting a part of this work. SMH, SB, and SP acknowledge support from the Department of Atomic Energy, Government of India, under Project Identification No. RTI 4002. 
\sbh{The authors would like to thank the reviewer(s) and data editor for their constructive comments and suggestions.}
The generation of the stellar models was done using the Modules for Experiments in Stellar Astrophysics \citep[\texttt{MESA}][]{Paxton2011, Paxton2013, Paxton2015, Paxton2018, Paxton2019}  (we have used \texttt{MESA} version r22.05.1 for RG and r23.05.1 for SG models, \texttt{MESA-SDK} version x86\_64-linux-22.6.1). The eigenfrequencies and eigenfunctions for this model were calculated using the \texttt{GYRE} \citep{Townsend2013}
code. The code to calculate the kernels and the splittings has been written completely in \texttt{Python 3.8.16}.

\bibliography{references}{}
\bibliographystyle{aasjournal}

\appendix
\label{section:Appendix}
    \section{Calculating the Lorentz stress sensitivity kernels}
    \label{appendix:a}

    The ten $\chi_{i}^{\mu\sigma}(k)$ are written as:
   \begin{align}
    \chi_{1}^{--}(k) &= \Omega_{0\ell}\Omega_{2\ell} \left[ V_k(3U_k-2\Omega_{2\ell}^{2} V_k +3r\dot{U_k})-rU_k\dot{V_k}\right],
    \\
    \chi_{2}^{--}(k) &= \Omega_{0\ell}^2 
    \left[
        3U_kV_k + (\Omega_{2\ell}^2- 2\Omega_{0\ell}^{2})V_k^{2} + rV_k\dot{U_k} - rU_k\dot{V_k} - U_k^{2}
    \right],
    \\
    \chi_{3}^{--}(k) &= \Omega_{0\ell}^2 \Omega_{2\ell} \Omega_{3\ell} V_k^2,
    \\
    \chi_{1}^{0-}(k)
    &=
    \Omega_{0\ell}
    \left[
        4\Omega_{0\ell}^2 V_k^2 - 4r\Omega_{0\ell}^2 V_k\dot{V_k} + 2r^2 \dot{U_k}\dot{V_k} + r^2 V_k \ddot{U_k} + U_k\lbrace 8U_k-6(\Omega_{0\ell}^2+1)V_k+r(4\dot{V_k}-r\ddot{V_k}) \rbrace
    \right],
    \\
    \chi_{2}^{0-}(k)
    &= \Omega_{0\ell}^2\Omega_{2\ell} \left[ U_kV_k + V_k(U_k-4V_k+3r\dot{V_k}) + rV_k\dot{V_k} \right],
    \\
    \chi_{1}^{00}(k) &= 2
    \left[
        -2rU_k\dot{U_k} + \Omega_{0\ell}^{2}r(V_k\dot{U_k} + U_k \dot{V_k}) - 5\Omega_{0\ell}^2 V_kU_k + 2\Omega_{0\ell}^4 V_k^{2} + 3U_k^{2}
    \right],
    \\
    \chi_{2}^{00}(k) &=
    -\Omega_{0\ell}^2 
    \left[
        - U_kV_k + V_k^2 + r(V_k\dot{U_k}+U_k\dot{V_k}) - 2rV_k\dot{V_k} + r^{2}\dot{V_k}^2
    \right],
    \\
    \chi_{1}^{+-}(k) &= 2 
    \left[
        -2r\dot{U_k}U_k + \Omega_{0\ell}^2 r( \dot{U_k}V_k + U_k\dot{V_k} ) - r^2 \dot{U_k}^2 
        - \Omega_{0\ell}^2U_kV_k + U_k^2
    \right],
    \\
    \chi_{2}^{+-}(k) &= -2\Omega_{0\ell}^2\Omega_{2\ell}^2 V_k^2,
    \\
    \chi_{3}^{+-}(k) &= \Omega_{0\ell}^2 
    \left[
        r(U_k\dot{V_k}-V_k\dot{U_k}) - U_kV_k + U_k^2
    \right].
    \end{align}
    where:
    \begin{eqnarray}
        \dot{U}_k = \dfrac{\partial U_k}{\partial r} , 
        \ddot{U}_k = \dfrac{\partial^2 U_k}{\partial r^2} , \dot{V}_k =\dfrac{\partial V_k}{\partial r} , 
        \ddot{V}_k = \dfrac{\partial^2 V_k}{\partial r^2} 
        \\
        \label{eq:omega}
        \Omega_{N\ell}=\begin{cases}0 & \text{if $|N|>\ell$}\\\sqrt{\frac{1}{2}(\ell+N)(\ell-N+1)} &\text{otherwise}\end{cases}
    \end{eqnarray}

\section{A first core-to-envelope sensitivity ratio}\label{sec:simplified_ratio_appendix}

Here, we present a simplified analysis of sensitivity comparison between H-shell $B_r^2$ and near-surface $(B_{\theta}^2 + B_{\varphi}^2)$ for a range of dipolar and quadrupolar modes for the RG discussed in Sec.~\ref{sec:red_giant}. For simplicity of notation, we henceforth define kernels as $K_{t,\rm{env}}=r^2\mathcal{B}_{s0}^{+-}$ in the envelope and $K_{r,\rm{core}}=r^2\mathcal{B}_{s0}^{00}$ in the radiative interior. Therefore, $K_{t,\rm{env}}$ is sensitive to the tangential magnetic field pressure in the envelope $B_{t,{\rm{env}}}^{2}$, and $K_{r,\rm{core}}$ is sensitive to the radial magnetic field pressure $B_{r,{\rm{core}}}^{2}$ around the H-shell burning region. 

The H-burning shell has a thickness of the same order the magnitude as the most sensitive zone close to the surface. Hence, for the splittings due to $B_{r,{\rm{core}}}$ to dominate over those due to $B_{t,{\rm{env}}}$, the following inequality needs to hold
\begin{equation}
    {\rm{max}}[|K_{r,\rm{core}}|] \, B_{r,{\rm{core}}}(s)^{2}\gtrapprox 
    {\rm{max}}[|K_{t,\rm{env}}|] \, B_{t,{\rm{env}}}(s)^{2},
\end{equation}
which leads to the definition of a critical field ratio
\begin{align}
    \left[ 
\frac{B_{r,{\rm{core}}}(s)}{B_{t,{\rm{env}}}(s)}  \right]_{\rm{crit}}\equiv \sqrt{\dfrac{{\rm{max}}[|K_{t,\rm{env}}|]}{{\rm{max}}[|K_{r,\rm{core}}|]}}.
\end{align}

\begin{figure*}
    \centering
        \includegraphics[width=0.98\linewidth]{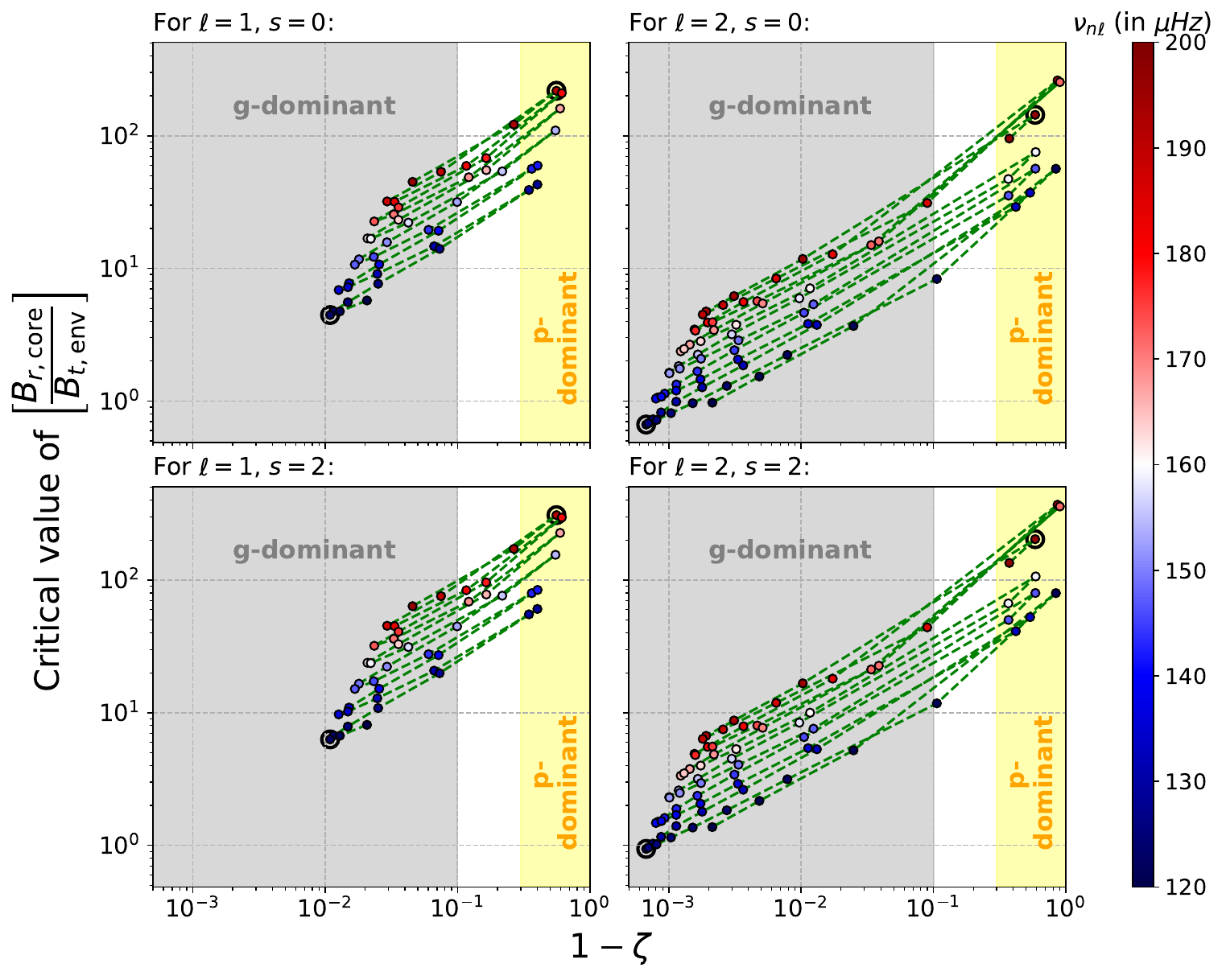}    
    \caption{Variation of the square root of ratios of the peak kernel values in the radiative interior and the envelope with $1-\zeta$ at different unperturbed normal mode frequencies, where $\zeta$, is defined in Eqn.~\eqref{zeta}. The modes marked with black circles in these four panels have been used for further analysis in Section~\ref{h_mins_Section} and the following results are presented in Figure~\ref{h_mins_graph}.
    }
    \label{fig:kern_vs_zeta}
\end{figure*}

A larger value of this critical ratio indicates that a weaker envelope $B_{t,\rm{env}}$ could potentially contribute to $\delta \omega_{n \ell m}$ at the same order as the $B_{r}^2$ in the radiative zone. Figure~\ref{fig:kern_vs_zeta} gives us an order of magnitude estimate for this critical ratio.
The higher the critical field ratios of the modes (which are the ordinates of the plotted points), the easier it is to detect the presence of near-surface tangential fields compared to internal radial fields. Modes with lower critical field ratios are dominated by the H-shell radial magnetic field. 

These left panels convey to us that for dipolar $g$-dominated modes to be equally sensitive to $B_r$ in the H-burning shell than $B_t=\sqrt{B_{\theta}^2+B_{\varphi}^2}$ in the envelope, the amplitude in the envelope should be at least about 0.1 times the amplitude at the H-burning shell. Dipolar (and quadrupolar, see right panels) $p$-dominated modes require the surface field to be about 0.01 times the internal field. For quadrupolar modes represented on the right panels, the surface field amplitude has to be equivalent to the internal field amplitude for its effect to be of the same order of magnitude on the $g$-dominated frequencies. From this first attempt to compare surface and core contributions, measuring surface magnetic fields from mixed-mode frequencies in red giant stars appears to be challenging.

From further inspection, we can also infer that low-frequency $g$-dominant modes in this model red-giant star are more sensitive to the $B_{r,{\rm{core}}}$, especially for $\ell=2$, whereas high-frequency $p$-dominant mode splittings are impacted more by the $B_{t,{\rm{env}}}$.
Quadrupolar $p$-modes have turning points closer to the surface of the star, they are less coupled to their $g$-mode counterparts \citep{JoelOng2021a}, because of which the $g$-dominant counterparts are often not visible in the power spectral density profiles. It is important to note that currently, all studies have used only dipolar modes to constrain magnetic fields. According to selection rules mentioned in Section~\ref{Lorentz_stress_tensor}, only the $s=0,2$ Lorentz-stress components are constrained using dipolar modes. Using quadrupolar modes, as and when detectable magnetic asymmetries are available, would enable us to constrain $s=0,2,4$ components of Lorentz stress. Since a dipolar magnetic field $s_0=1$ can only induce $s=0,2$ Lorentz-stress components \citep[see Appendix D.2 of][]{Dasetal20}, having access to perturbations in quadrupolar modes could lend insight into whether pure dipolar models are sufficient to explain observed asymmetries.

\section{Supplementary Calculations}

\subsection{Showing that for $\ell=2$, the splitting between $m=0$ and $2$ components is always 4 times that of the splitting between $m=0$ and $1$ for poloidal + toroidal magnetic fields}
\label{l2_factor_of_4}
For slow rotators with rotation axis aligned along the magnetic field symmetry axis, the splittings, as discussed above, originate only from $s=2$ terms of the sensitivity kernels. Let us consider the $m$-independent part of the $\mu,\sigma$ part of the kernel for a multiplet $(n,2)$ is $\mathcal{C}_{2}^{\mu\sigma}$. Then,
\begin{align}
    r^2 \mathcal{B}_{20}^{\mu\sigma}(n,2,\pm m) &= (-1)^{m}
    \begin{pmatrix}
        2 & 2 & 2
        \\
        -m & 0 & m
    \end{pmatrix} \mathcal{C}_{2}^{\mu\sigma}.
\end{align}
Therefore:
\begin{align}
    r^2 \mathcal{B}_{20}^{\mu\sigma}(n,2,\pm 1) - r^{2} \mathcal{B}_{20}^{\mu\sigma}(n,2,0)
    &=
    \mathcal{C}_{2}^{\mu\sigma} \times \dfrac{1}{\sqrt{70}}, &
    r^2 \mathcal{B}_{20}^{\mu\sigma}(n,2,\pm 2) - r^{2} \mathcal{B}_{20}^{\mu\sigma}(n,2,0)
    &= \mathcal{C}_{2}^{\mu\sigma} \times \dfrac{4}{\sqrt{70}}.
\end{align}

The expression for the corresponding ``split" in the angular frequencies are ($s=0$ terms in both frequencies are the same and hence their difference goes to 0, so we only deal with $s=2$):
\begin{align}
    \delta\omega_{n,2,\pm2}-\delta\omega_{n,2,0}
    &= \dfrac{1}{I_{n\ell}} \: 
    \left[ \int_0^{R_{\rm{star}}} dr \;
           \sum_{\mu,\sigma}
           \left\lbrace
           r^2 \mathcal{B}_{20}^{\mu\sigma}(r;n,2,\pm 2) 
           -
           r^2 \mathcal{B}_{20}^{\mu\sigma}(r;n,2,0) 
           \right\rbrace \; h_{20}^{\mu\sigma}(r)
    \right] = \dfrac{1}{I_{n\ell}} \: \dfrac{4}{\sqrt{70}} \: \int_0^{R_{\rm{star}}} dr \, \sum_{\mu,\nu} \: \mathcal{C}_{2}^{\mu\sigma} h_{20}^{\mu\sigma} ,
    \\
    \delta\omega_{n,2,\pm1}-\delta\omega_{n,2,0}
    &= \dfrac{1}{I_{n\ell}} \: 
    \left[ \int_0^{R_{\rm{star}}} dr\;
           \sum_{\mu,\sigma}
           \left\lbrace
           r^2 \mathcal{B}_{20}^{\mu\sigma}(r;n,2,\pm 1) 
           -
           r^2 \mathcal{B}_{20}^{\mu\sigma}(r;n,2,0) 
           \right\rbrace \; h_{20}^{\mu\sigma}(r)
    \right] = \dfrac{1}{I_{n\ell}} \: \dfrac{1}{\sqrt{70}} \: \int_0^{R_{\rm{star}}} dr \, \sum_{\mu,\nu} \: \mathcal{C}_{2}^{\mu\sigma} h_{20}^{\mu\sigma} .
\end{align}

Thus, the ratio in the angular frequency splittings is,
\begin{align}
    \dfrac{ \delta\omega_{n,2,\pm2}-\delta\omega_{n,2,0} }{ \delta\omega_{n,2,\pm1}-\delta\omega_{n,2,0} }
    &= 4.
\end{align}

This has been shown to hold for $g$ and $p$-dominant modes separately in \cite{MathisBugnetetal2021}. Still, we explicitly show that this is true for any model with an axisymmetric magnetic field where terms only up to $s=2$ exist, e.g. a combined poloidal and toroidal field.

Similarly, we can also say that if one can calculate $\delta\omega_{n,\ell,0}$ and $\delta\omega_{n,\ell,\pm 1}$ for the type of magnetic field topology being discussed here, they can precisely estimate the value of $\delta\omega_{n,\ell, m}$ for all $m\in\lbrace -\ell, -\ell+1, ..., \ell-1, \ell \rbrace$ and $|m|>1$ by using the prescription mentioned above, which requires the calculation of certain Wigner 3-j symbols only. This can save a significant amount of computation time.

\subsection{Determining the critical value of Lorentz-stress GSH components from kernel weighted averages}
\label{app:Critical_ratio_approach}
For $\langle h_{s0,\square}^{\mu\sigma} \rangle$  to contribute to the measured mode frequencies at a comparable amount as $\langle h_{00,\rm{H-shell}}^{00} \rangle$ requires
\begin{align}
    \langle h_{s0,\square}^{\mu\sigma} \rangle \: 
    \left|\frac{1}{I_{n\ell}} \: \int_{\square} dr \; r^{2} \: \mathcal{B}_{s0}^{\mu\sigma}(r;n,\ell,m)\right|
    \geq
    \langle h_{00,\rm{H-shell}}^{00} \rangle \:
    &\left|
    \frac{1}{I_{n\ell}} \: \int_{\rm{H-shell}} dr \; r^{2} \: \mathcal{B}_{00}^{00}(r;n,\ell,m)
    \right|,
    \\\implies\label{ineq:h_crit_inequality}
    \dfrac{ \langle h_{s0,\square}^{\mu\sigma} \rangle }{ \langle h_{00,\rm{H-shell}}^{00} \rangle }
    \geq
    \left|
    \dfrac{ \int_{\rm{H-shell}} dr \; r^{2} \: \mathcal{B}_{00}^{00}(r;n,\ell,m) }{ \int_{\square} dr \; r^{2} \: \mathcal{B}_{s0}^{\mu\sigma}(r;n,\ell,m) } 
    \right|
    &=
    \left|
    \dfrac{ \int_{\rm{H-shell}} dr \; r^{2} \: \mathcal{B}_{00}^{00}(r;n,\ell,0) }{ \int_{\square} dr \; r^{2} \: \mathcal{B}_{s0}^{\mu\sigma}(r;n,\ell,0) } \right| \Psi_{s n \ell m}.
\end{align}
where 
$\Psi_{s n \ell m}=\left| \begin{pmatrix}
    \ell & s & \ell
    \\
    0 & 0 & 0
\end{pmatrix} \Bigr/ \begin{pmatrix}
    \ell & s & \ell
    \\
    -m & 0 & m
\end{pmatrix} \right|$, 
which is either 1 or 2 (of order unity) for the different possible $m$ belonging to $\ell\in\lbrace 1,2 \rbrace$. Since we are concerned more with the order of magnitudes of the splittings, we looked only at the $m=0$ component of each multiplet, for which $\psi_{n\ell m}=1$, and have hence defined $\langle h_{s0,\square}^{\mu\sigma} \rangle_{\rm{crit}}$ as the minimum value of $\langle h_{s0,\square}^{\mu\sigma} \rangle$ which solves the inequation~\ref{ineq:h_crit_inequality}.

Since the $B_r^2$ component is a more tangible form of magnetism as compared to $h_{s0,\rm{H-shell}}^{\mu\sigma}$ we have defined the critical ratio $\mathcal{R}_{s0}^{\mu \sigma}$ in Section~\ref{sec:Sensitivity_ratios} with respect to the kernel-weighted average of the horizontally averaged squared radial magnetic field in the H-burning shell region $\langle\bar{B}_{r}^{2}\rangle_{\rm{H-shell}}$. In order to relate $h_{00,\rm{H-shell}}^{00}$ to $\langle\bar{B}_{r}^{2}\rangle_{\rm{H-shell}}$, we can obtain the following mathematical steps (refer Eqn.~\eqref{Br2_bar}):
\begin{align}
    \frac{1}{4\pi}\iint B_{r}^2 \sin\theta \: d\theta \: d\varphi &= \dfrac{1}{2 \sqrt{\pi}} \: h_{00}^{00}(r),
    \\\implies
    \frac{1}{4\pi} \: \left|\dfrac{ \int_{\rm{H}}\: dr \: r^2 \:  \mathcal{B}_{00}^{00}(r;n,l,m) \: \iint B_{r}^2 \sin\theta \: d\theta \: d\varphi }{ \int_{\rm{H}}\: dr \: r^2 \:  \mathcal{B}_{00}^{00}(r;n,l,m) }\right| &= \dfrac{1}{2 \sqrt{\pi}} \: \left|\dfrac{ \int_{\rm{H}}\: dr \: r^2 \:  \mathcal{B}_{00}^{00}(r;n,l,m) \: h_{00}^{00}(r) }{ \int_{\rm{H}}\: dr \: r^2 \:  \mathcal{B}_{00}^{00}(r;n,l,m) }\right|,
    \\\implies
    \frac{1}{4\pi} \: \left|\dfrac{ \int_{\rm{H}}\: dr \: r^2 \:  \mathcal{B}_{00}^{00}(r;n,l,0) \: \iint B_{r}^2 \sin\theta \: d\theta \: d\varphi }{ \int_{\rm{H}}\: dr \: r^2 \:  \mathcal{B}_{00}^{00}(r;n,l,0) }\right| &= \dfrac{1}{2 \sqrt{\pi}} \: \langle h_{00,\rm{H-shell}}^{00} \rangle = \langle \bar{B}_{r}^{2}\rangle_{\rm{H-shell}} \, . \label{eq: h00_to_Br}
\end{align}

\subsection{Showing that the asymmetry parameter defined in Li et al. 2022 matches with Das et al. 2020 for dipolar + toroidal field in a non-rotating star}

From Eqn.~\eqref{eq15}, we have:
\begin{align}
    h_{00}^{00}(r) &= 3\gamma_0 (B_{10}^0)^2 \: (-1)^{0+0} 
    \begin{pmatrix}
        1 & 0 & 1
        \\
        0 & 0 & 0
    \end{pmatrix}
    \begin{pmatrix}
        1 & 0 & 1
        \\
        0 & 0 & 0
    \end{pmatrix} = 3\gamma_0 (B_{10}^0)^2 \; \dfrac{1}{3},
    \\
    h_{20}^{00}(r) &= 3\gamma_2 (B_{10}^0)^2 \: (-1)^{0+0} 
    \begin{pmatrix}
        1 & 2 & 1
        \\
        0 & 0 & 0
    \end{pmatrix}
    \begin{pmatrix}
        1 & 2 & 1
        \\
        0 & 0 & 0
    \end{pmatrix}  = 3\gamma_2 (B_{10}^0)^2 \; \dfrac{2}{15}.
\end{align}

This yields a relation between $h_{00}^{00}$ and $h_{20}^{00}$:
\begin{align}
    \label{h02_ratio}
    \dfrac{h_{00}^{00}(r)}{h_{20}^{00}(r)} &= \dfrac{\gamma_0}{\gamma_2}\: \dfrac{\frac{1}{3}}{\frac{2}{15}} = \sqrt{\frac{5}{4}}
\end{align}

The asymmetry parameter $a$ as defined in Eqn.~49 of the Supplementary Information of \cite{GangLietal2022} is:
\begin{align}
    a &= \dfrac{
    \int_{r_{\rm{in}}}^{r_{\rm{out}}} \: dr \; K(r) \iint B_{r}^{2} P_{2}(\cos\theta)\: \sin\theta \: d\theta \: d\varphi
    }{\int_{r_{\rm{in}}}^{r_{\rm{out}}} \: dr \; K(r) \iint B_{r}^{2} \: \sin\theta \: d\theta \: d\varphi},
\end{align}
where $K(r)$ is the approximate sensitivity kernel computed for splittings to $\frac{1}{4\pi}\:\iint B_{r}^{2} \: \sin\theta \: d\theta d\varphi$ and $P_{2}$ is the second order Legendre polynomial.
Now for the poloidal + toroidal mixed field, according to Eqn.~39 of \cite{Dasetal20}:
\begin{align}
    \label{Br2}
    B_{r}^2 &= h_{00}^{00}(r)Y_{00}^{0} + h_{20}^{00}(r) Y_{20}^{0}.
\end{align}

Therefore we obtain:
\begin{align}
    \iint B_{r}^{2} \sin\theta \: d\theta \: d\varphi 
    &= 
    h_{00}^{00}(r)\times 2\pi \times \int_{0}^{\pi} Y_{00}^{0} \sin \theta \: d\theta
    +
    h_{20}^{00}\times 2\pi \times \int_{0}^{\pi} Y_{20}^{0} \sin \theta \: d\theta,
    \\\implies\label{Br2_bar}
    \iint B_{r}^{2} \sin\theta \: d\theta \: d\varphi
    &= 
    h_{00}^{00}(r)\times 2\pi \times \dfrac{1}{\sqrt{\pi}}.
\end{align}
Similarly,
\begin{align}
    \iint B_{r}^{2} \: P_{2}(\cos\theta) \: \sin\theta \: d\theta \: d\varphi 
    &= 
    h_{00}^{00}(r)\times 2\pi \times \int_{0}^{\pi} Y_{00}^{0} \: P_{2}(\cos\theta) \: \sin \theta \: d\theta
    +
    h_{20}^{00}(r)\times 2\pi \times \int_{0}^{\pi} Y_{20}^{0} \: P_{2}(\cos\theta) \: \sin \theta \: d\theta,
    \\\implies\label{Br2_P2}
    \iint B_{r}^{2} \: P_{2}(\cos\theta) \: \sin\theta \: d\theta \: d\varphi 
    &=  
    h_{20}^{00}(r)\times 2\pi \times \dfrac{1}{\sqrt{5\pi}}.
\end{align}
Expressions~\eqref{Br2_bar} and~\eqref{Br2_P2} lead us to the expression for $a$, where writing $h_{20}^{00}$ in terms of $h_{00}^{00}$ using Eqn.~\eqref{h02_ratio} yields:
\begin{align}
    a &= \frac{1}{\sqrt{5}}\dfrac{ \int dr \: K(r) \: h_{20}^{00}(r) }{ \int dr \: K(r) \: h_{00}^{00}(r) } = \frac{1}{\sqrt{5}}\dfrac{ \int dr \: K(r) \: h_{00}^{00}(r)  \times \frac{2}{\sqrt{5}} }{ \int dr \: K(r) \: h_{00}^{00}(r)  } = \dfrac{2}{5}.
\end{align}
This matches the value of $a$ as obtained by \cite{GangLietal2022}

\section{Frequency splittings for different magnetic field configurations}
\label{appC}
    \subsection{Choice of a model of the magnetic field in the stellar interior}
    \label{B_topologies}
    \begin{figure}
        \centering
        \includegraphics[width=0.4\linewidth]{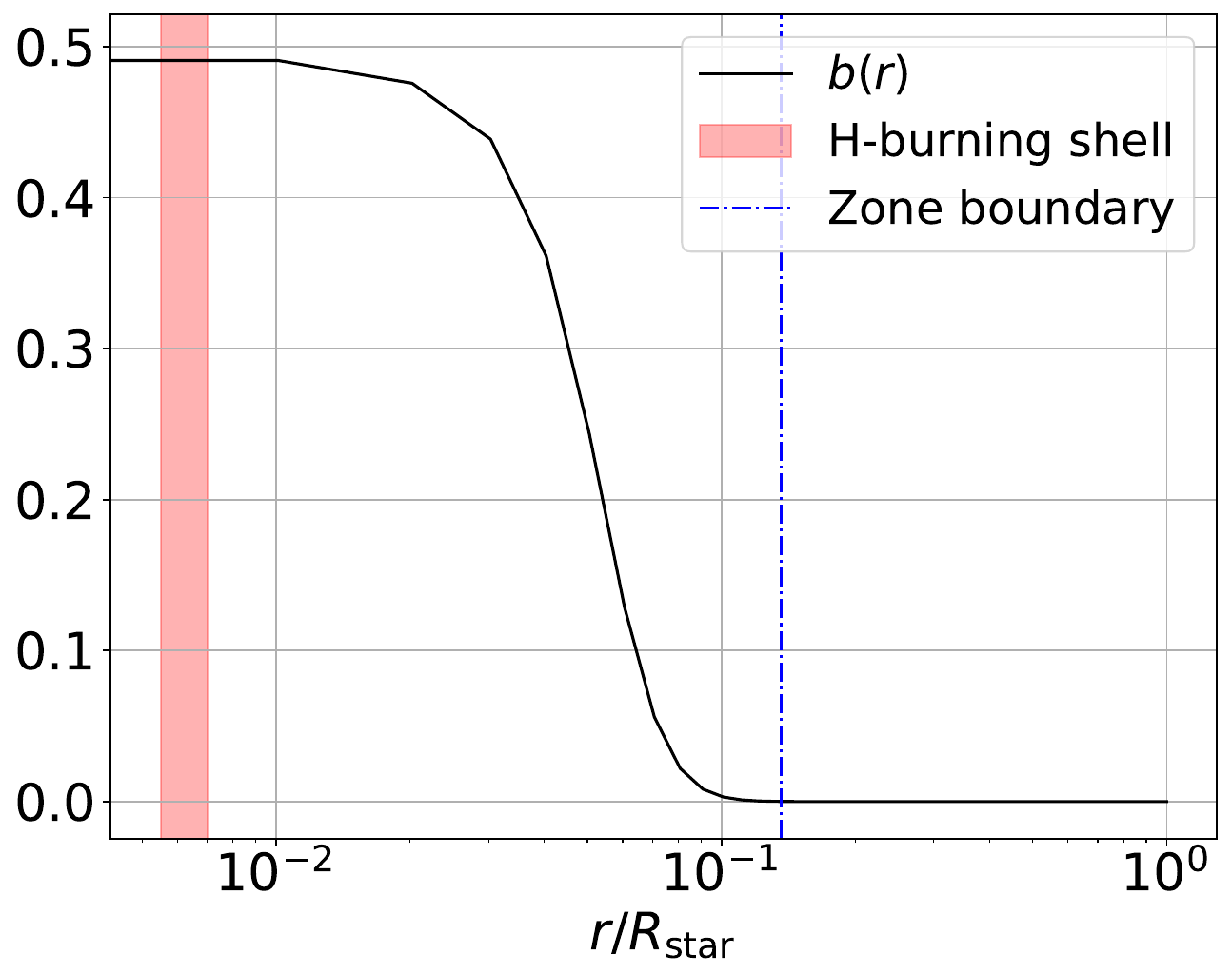}
        \caption{Plot of $b(r)$ (refer Eqn.~\ref{b_profile}), which we choose for this study. We note that the radial profiles of $\boldsymbol{B}_{\rm{tor}}$ and $\boldsymbol{B}_{\rm{pol}}$ constructed using $b(r)$ are almost constant at and beneath the H-burning shell.}
        \label{fig:Bugnet_field}
    \end{figure}

In this work, we have chosen a few particularly simple toy topologies and one mixed stable topology for the magnetic field \citep[following][]{Duez2010A,Bugnetetal2021} and studied their corresponding effect on the spectra of red-giant branch stars. 
\label{sec:def_B_tor}
    The magnetic field taken for the first case is purely toroidal:
    $\boldsymbol{B}_{\rm{tor}} = -B_{0} \:
    b(r)\sin\theta \; \hat{\varphi}$,
where:
\begin{align}
\label{b_profile}
b(r)=\frac{1}{4}\left[ 1-\tanh \left\lbrace 50 \: \left(\frac{r}{R_{\rm{star}}}-0.05\right) \right\rbrace \right]. 
\end{align}
As may be noted from the plot of $b(r)$ in Figure \ref{fig:Bugnet_field}, the toroidal magnetic field near the H-shell burning region is almost constant (with $0.494\:B_0$ field along the equator), gradually dropping to nearly zero as it reaches the boundary between the radiative and the convective zones. 

\label{sec:def_B_pol}
    The magnetic field taken for the second case is purely poloidal: 
    $\boldsymbol{B}_{\rm{pol}} = B_{0}
    \left[
        2b(r)\cos\theta \; \hat{r} - \frac{1}{r} \frac{\partial}{\partial r}\left\lbrace r^{2}b(r) \right\rbrace \sin\theta \; \hat{\theta}
    \right]$,
where again $b(r)$ is given by Eqn. \eqref{b_profile}, same as for the toroidal field discussed above. In this case, the maximum radial magnetic field obtained at points lying on the magnetic axis inside the core and in the vicinity of the H-burning is $0.988\:B_{0}$, which is almost $\approx B_0$. These two toy models help enable us to study how different magnitudes of magnetic fields in the core and the envelope influence the power spectrum one can observe for such a star. It was not practical to use a step function because the large derivative of $b(r)$ would result in the formation of very high magnetic fields along $\theta$ and the perturbative approach would not be a feasible one.

The third topology we use in our study is a semi-analytical approximation of the relaxed stable poloidal + toroidal magnetic fields as deduced by \cite{MathisZahn2005}. The same topology was also used by \cite{Bugnetetal2021} and \cite{MathisBugnetetal2021} for calculating splittings due to magnetic fields using approximations of $U_{n\ell}(r)$ and $V_{n\ell}(r)$ in the asymptotic $p$ and $g$ modes limits of mixed modes. Note that the formalism used here allows us to calculate the magnetic field splittings for our stellar model using the full kernels without resorting to asymptotics. The expression for this model of the magnetic field inside the radiative zone is:
\begin{align}
    \label{B_Bugnet_eqn}
    \boldsymbol{B}_{\rm{rad,Mixed}}
    &=
    B_{0}\bigr|_{\rm{Mixed}}
    \left[ 
         \frac{2A(r)}{r^2}\cos\theta \: \hat{r}
         -
         \frac{1}{r}\frac{\partial A(r)}{\partial r} \sin\theta \: \hat{\theta}
         +
         \frac{2.80 \: A(r)}{r \: R_{\rm{rad}}}\sin\theta \: \hat{\varphi}
    \right],
\end{align}
where $A(r)=-r \: j_{1}(2.80\:r/R_{\rm{rad}})\int_{r}^{R_{\rm{rad}}} y_{1}(2.80\:x/R_{\rm{rad}}) \rho x^3 \: dx -r \: y_{1}(2.80\:r/R_{\rm{rad}})\int_{0}^{r} j_{1}(2.80\:x/R_{\rm{rad}}) \rho x^3 \: dx  $, $j_1$ and $y_1$ are the first order spherical Bessel functions of the first and second kind respectively. The magnetic field in the convective zone is set to 0 in this topology. Our whole analysis is restricted to those values of $B_{0}$ and $B_{0}\bigr|_{\rm{Mixed}}$ where the perturbative approach up to the first order in Lorentz stress remains valid and the eigenfunctions in the presence of the magnetic field do not alter significantly.

\subsection{Nature of the frequency splittings in the RG star}
\label{subsec:Values_Splittings}
    
We now show our results for $\ell=1,2$ mixed-mode frequency splittings (calculated using the Eqn.~\ref{eq10}) for the same red-giant star's model (with $\nu_{\rm{max}}=160.928\:\mu{\rm{Hz}}$) using custom-designed pure toroidal and pure poloidal fields dominant in the core, followed by a stable mixed (poloidal+toroidal) magnetic field topology derived in the Appendix of \cite{Bugnetetal2021}, and two fields dominating in the convective envelope.

From Eqns.~\eqref{eq10}, \eqref{eq14}, \eqref{eq15}, we find that the scaling relation between the frequency splitting $\delta\nu_{n\ell m}$ and field amplitude scale $B_0$ is:
    $\delta\nu_{n\ell m} \propto B_{0}^2$.

    \subsubsection{Splittings due to toroidal field in the core}
\begin{figure}[]
    \centering
    \subfigure[]
    {\includegraphics[width=0.49\linewidth]{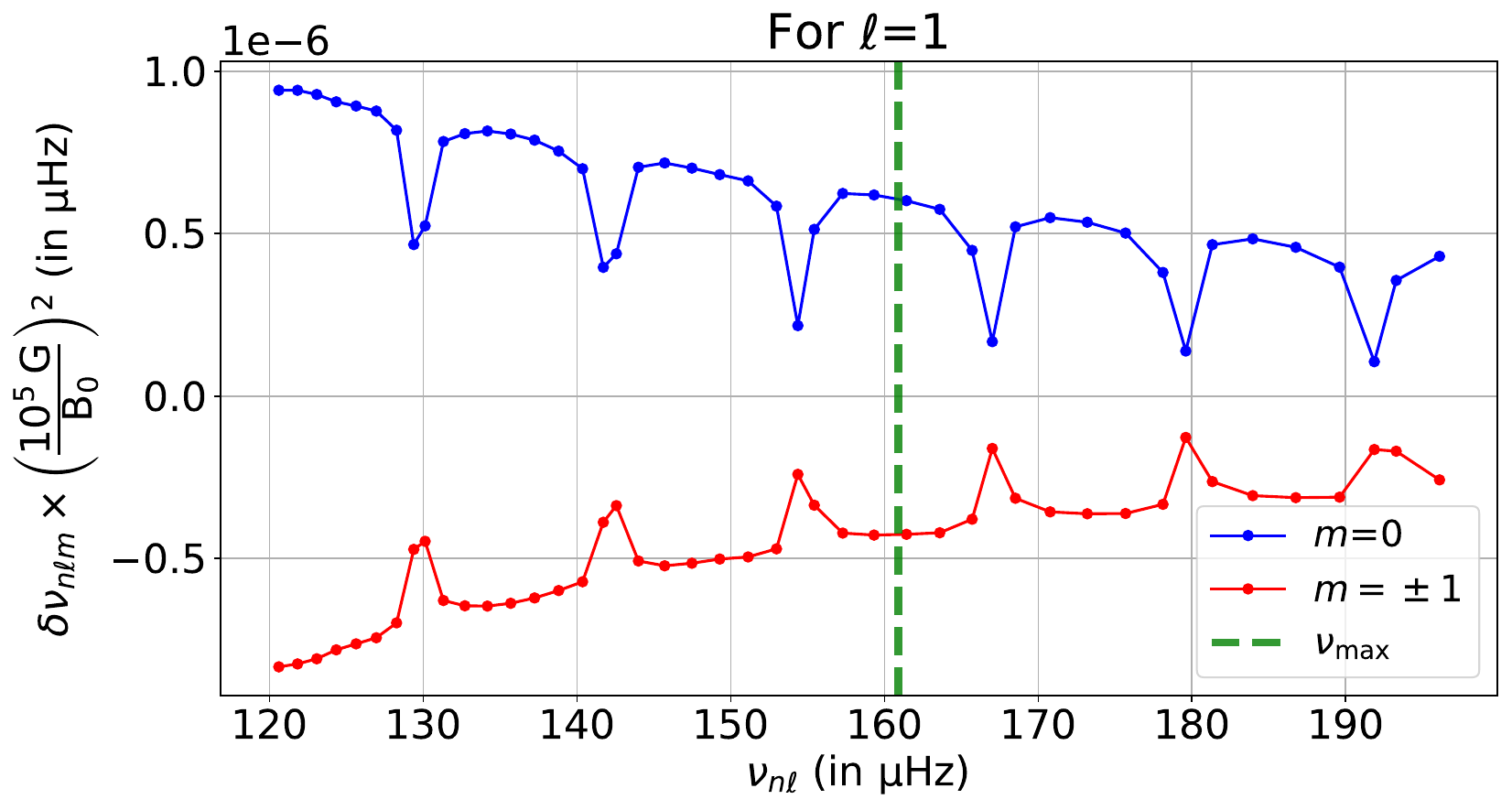}
    \label{fig:l1_tor}}
    \subfigure[]
    {\includegraphics[width=0.468\linewidth]{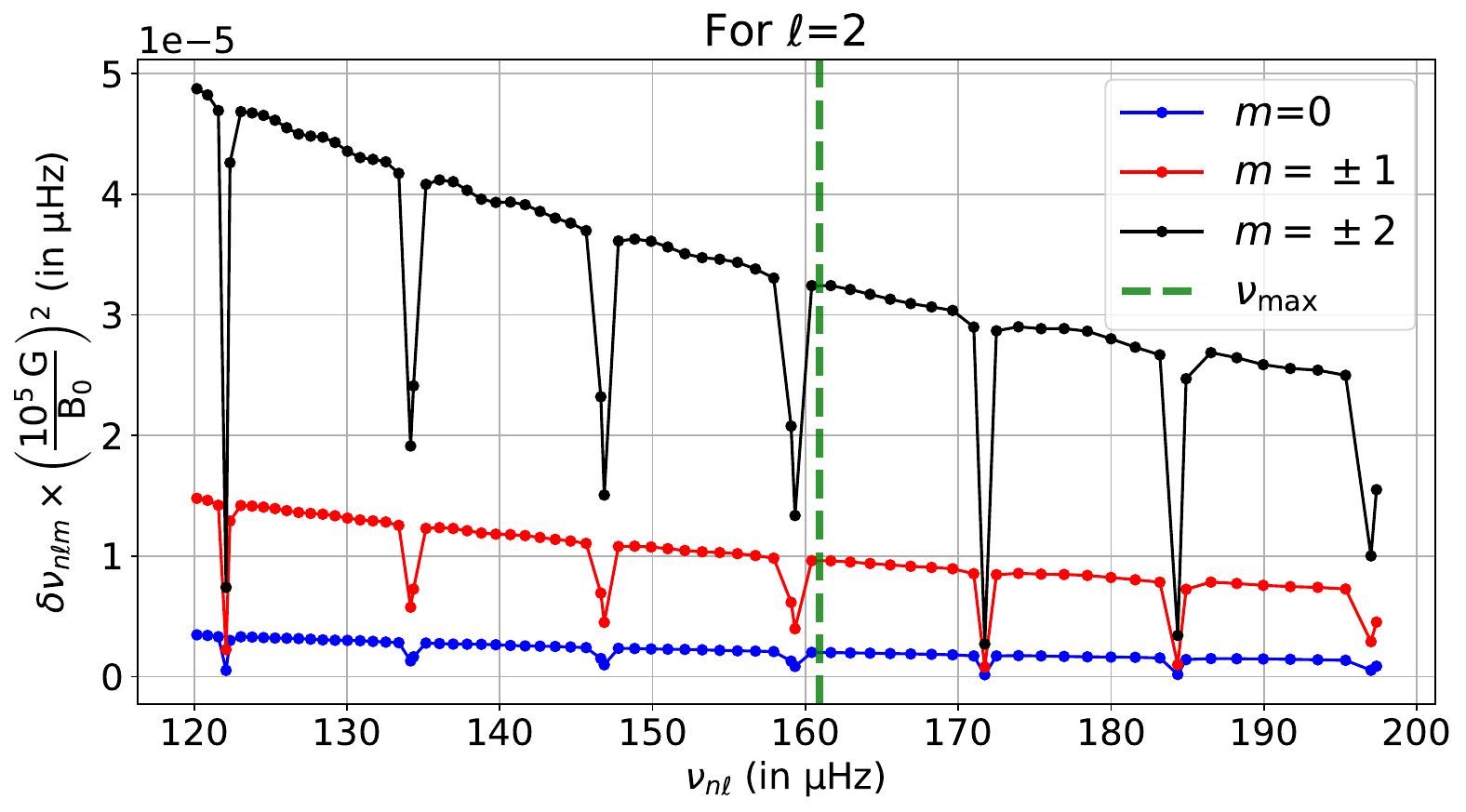}    
    \label{fig:l2_tor}}
    \caption{Frequency splittings obtained for the (a) $\ell=1$ and (b) $\ell=2$ mixed modes of the used star due to a pure toroidal magnetic field with field amplitude scaling factor $B_0$. When power spectra of solar-like stars are plotted, the region with the most prominent peaks is usually fitted with a Gaussian envelope whose maximum occurs at the frequency $\nu_{\rm{max}}$, indicating that normal modes in the vicinity of this frequency have large mode amplitudes, and also SNRs high enough to perform precise asteroseismic analysis.}
    \label{fig:dnu_tor}
\end{figure}

We first study how a pure toroidal magnetic field dominantly trapped in the radiative interior as constructed in Section:\ref{sec:def_B_tor} affects the degenerate mode splittings for the mixed modes of this star. We find that the splittings (as shown in Figure \ref{fig:dnu_tor}) for mixed modes with high \sbh{$\zeta$} tend to follow an approximately $1/\nu_{n\ell}^{1.5}$ trend. The values of frequency splittings as obtained for $B_0=10^5$ G are on the order of $10^{-7}\sim 10^{-5} \rm{\mu Hz}$. We infer from these values that the toroidal field in the core has to be very high to be detectable by \textit{Kepler} (\sbh{dipolar $g$-dominated mode closest to $\nu_{\rm{max}}$ requires $B_0\gtrsim 10^7$ G}) as also expected from \cite{MathisBugnetetal2021}.

\subsubsection{Splittings due to poloidal field in the core}

\begin{figure}[]
\centering
\begin{subfigure}[]
  {\includegraphics[width=0.48\linewidth]{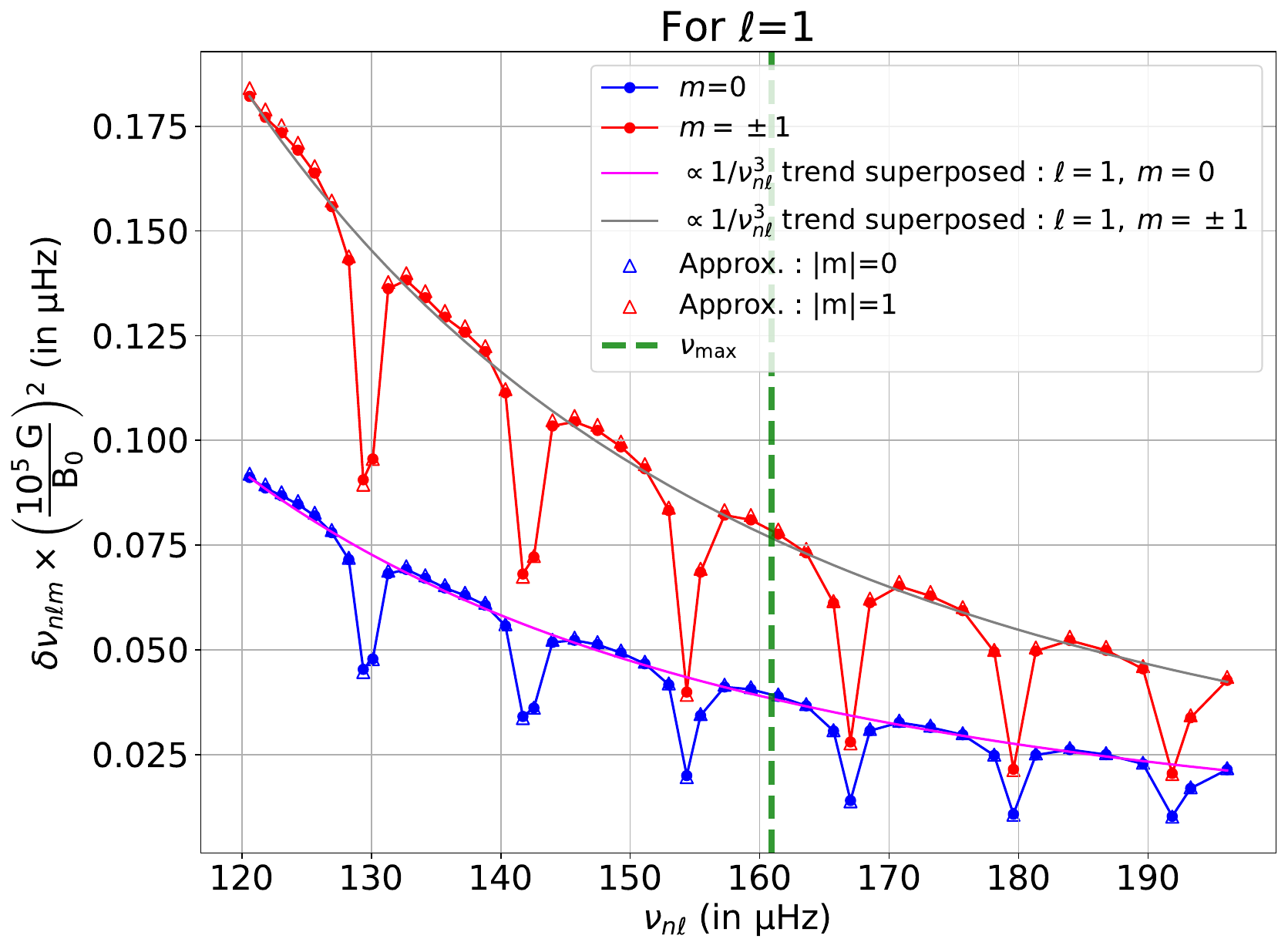}
  \label{fig:l1_pol}}
\end{subfigure}%
\begin{subfigure}[]
  {\includegraphics[width=0.47\linewidth]{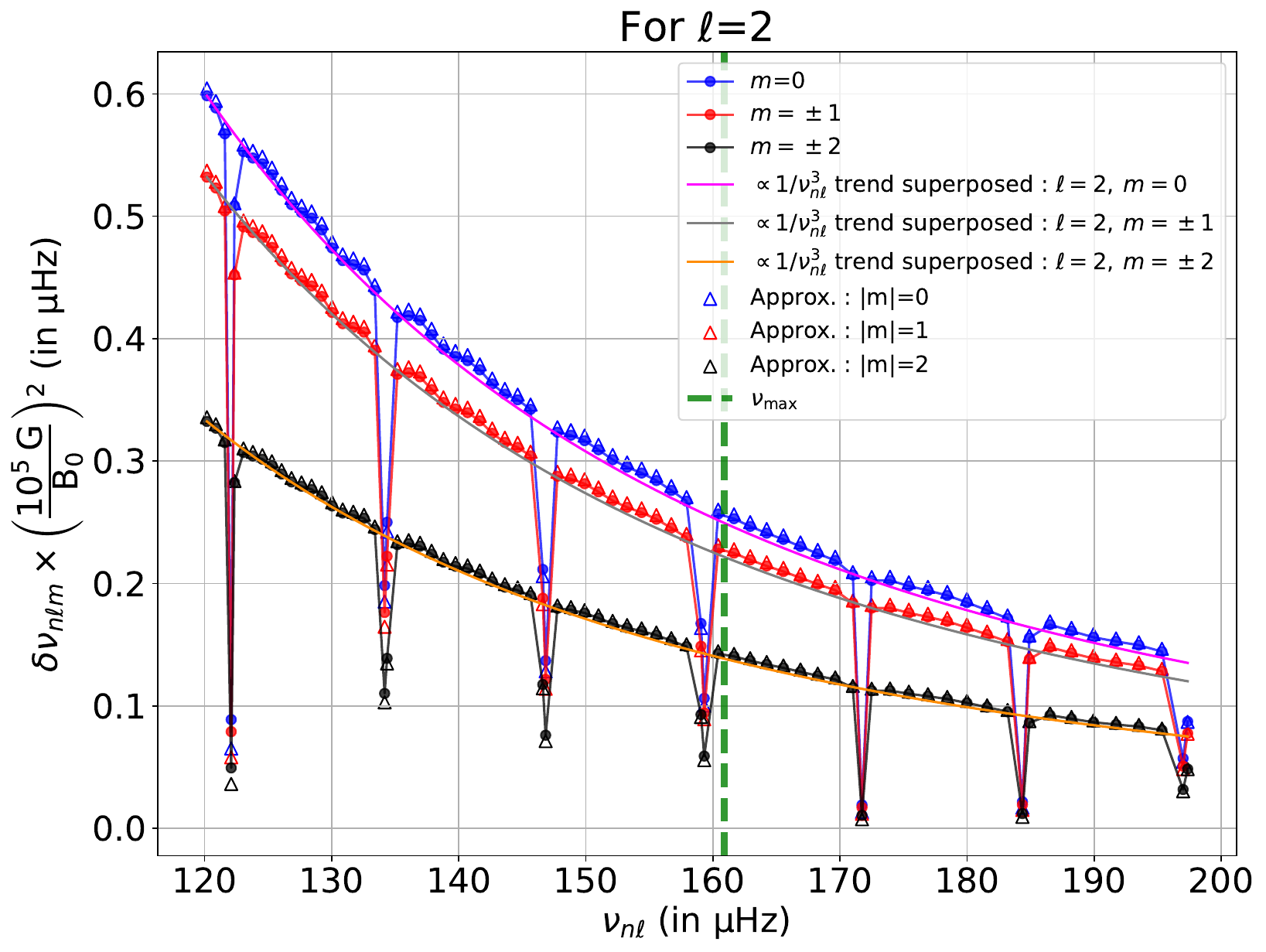}
  \label{fig:l2_pol}}
\end{subfigure}
\caption{Frequency splittings obtained for (a) $\ell=1 \;\rm{and}$ (b) $\ell=2$ mixed modes due to pure poloidal magnetic field with field amplitude scaling factor $B_0$.
The splittings calculated using a modification of the formula constructed by \cite{Hasanetal05}, which is given by Eqn.~\eqref{Approx}, are indicated by  \textbf{Approx} mentioned in the legends.}
\label{fig:dnu_pol}
\end{figure}
\begin{figure}[]
\centering
\begin{subfigure}[]
  {\includegraphics[width=0.48\linewidth]{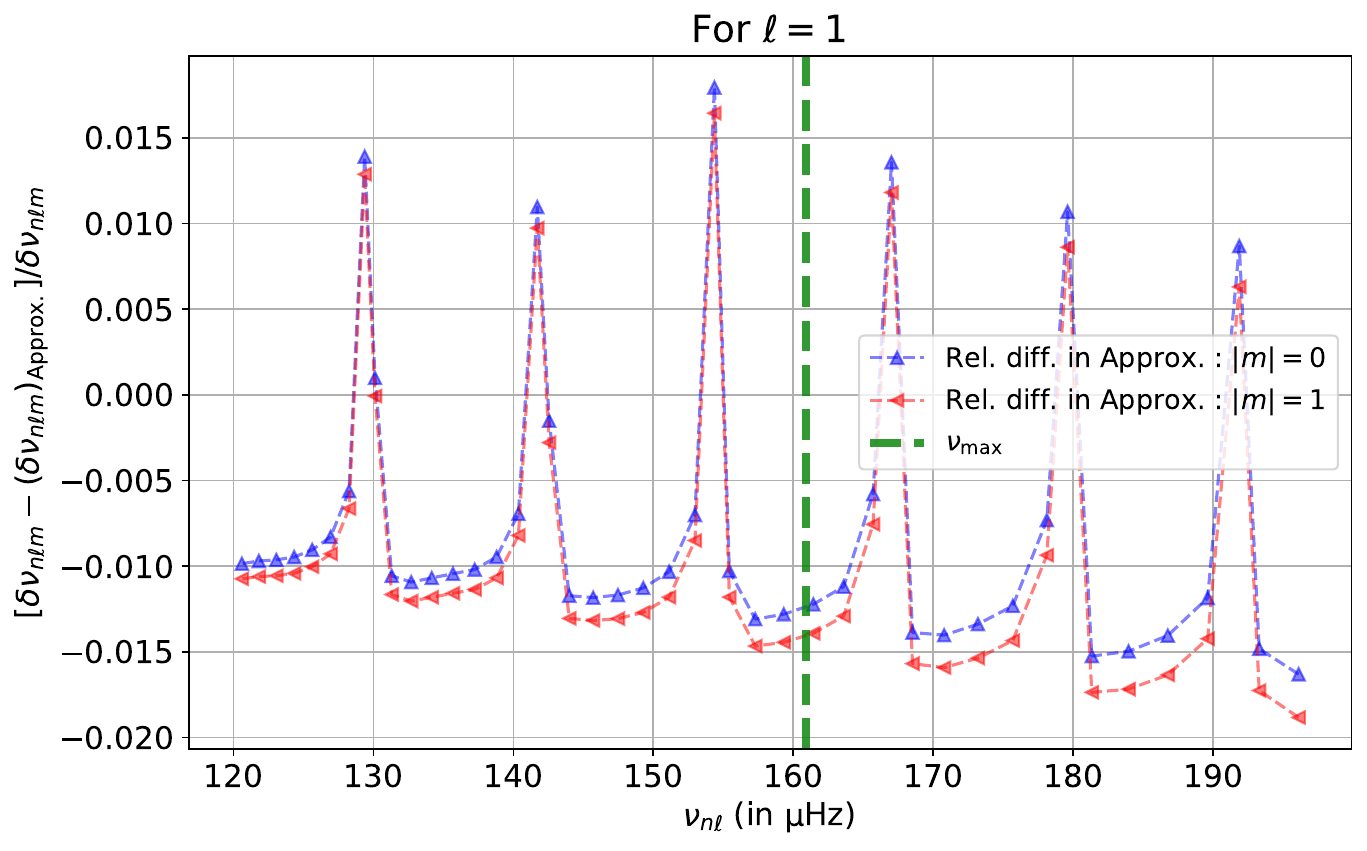}
  \label{fig:rel_diff_l1}}
\end{subfigure}
\begin{subfigure}[]  {\includegraphics[width=0.468\linewidth]{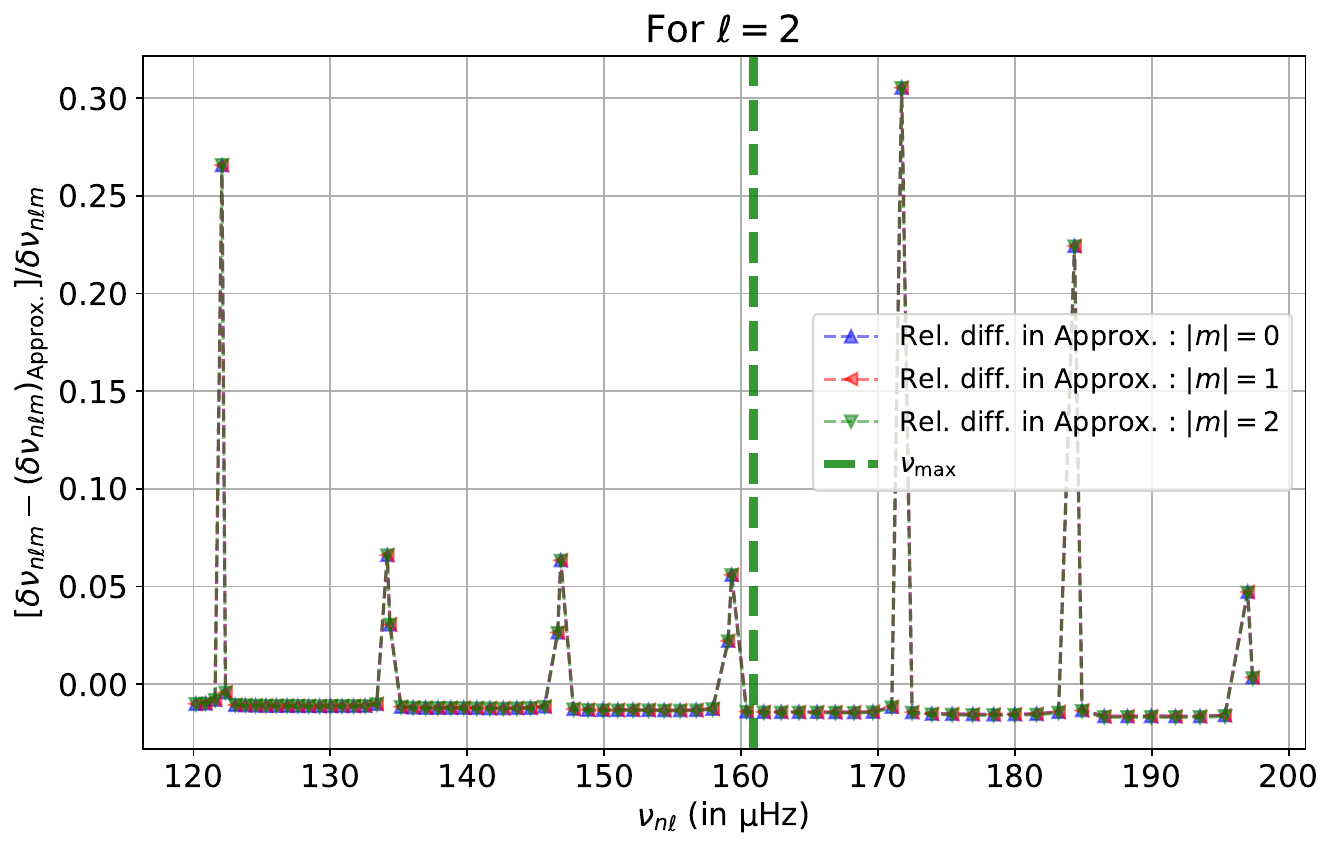}
  \label{fig:rel_diff_l2}}
\end{subfigure}
\caption{Relative difference between the approximate asymptotic magnetic splittings $(\delta\nu_{n\ell m})_{\rm{Approx.}}$ in Eqn.~\eqref{Approx} and their values $\delta\nu_{n\ell m}$ from the expression as in Eqn.~\eqref{eq10} for different components of (a) dipolar and (b) quadrupolar modes respectively in the presence of the poloidal magnetic field.}
\label{fig:rel_diff}
\end{figure}

Next, we study how a pure divergenceless poloidal magnetic field dominating the core affects the degenerate mode splittings. We see from Figure~\ref{fig:dnu_pol} that the mixed modes with high $\zeta_{n\ell}$ tend to follow the $1/\nu_{n\ell}^{3}$ trend, as demonstrated in \cite{Bugnetetal2021, MathisBugnetetal2021}. Also, the values of frequency splittings as obtained for $B_0=10^5$ G are of the order of $10^{-2}\sim 10^{-1} \rm{\mu Hz}$, which are detectable for mixed modes close to $\nu_{\rm{max}}$ from the power spectral density profile for red-giants with 4-year observational windows of \textit{Kepler} (\sbh{$g$-dominated dipole mode closest to $\nu_{\rm{max}}$ requires $B_0\gtrsim 32$ kG to be detectable, as obtained in Section~\ref{section:Detectability}}). This is in agreement with past theoretical studies on the detectability of internal magnetic fields \citep[e.g.][]{Bugnetetal2021, GangLietal2022}.

Comparing splittings due to pure toroidal and poloidal fields in the core, we find that the toroidal counterpart of a similar magnitude shows splittings which are orders of magnitudes smaller than the other's.

\cite{Hasanetal05} have shown that for higher order $g$-modes, the approximate magnetic frequency splittings are:
\begin{align}
    \left( \delta\omega_{n\ell m} \right)_{g} &= \dfrac{1}{8\pi\omega_{n\ell}}
    \:
    \dfrac{B_0^2}{R_{\rm{star}}^2} \: C_{\ell,m} \: \dfrac{\int_0^1 \left| 2\:\frac{d}{dx}\left[ x\:b(R_{\rm{star}}x) V_{n\ell} \right] \right|^{2}\:dx}{\int_0^1 V_{n\ell}^2 \: \rho \: x^2 \: dx},
\end{align}
where $\rho$ is the equilibrium density profile of the star, $x=r/R_{\rm{star}}$, and
\begin{align}
    C_{\ell,m} &= \dfrac{ \int_0^{\pi} d\theta \: \sin\theta \; \left[ \left| \cos\theta \: \partial_{\theta}Y_{\ell m}(\theta,\varphi) \right|^2
    +
    m^2 \left| \cot\theta \: Y_{\ell m}(\theta,\varphi) \right|^2
    \right] }{\ell(\ell+1)\int_{0}^{\pi} d\theta \: \sin\theta \: \left| Y_{\ell m}(\theta,\varphi) \right|^2}.
\end{align}

For mixed modes, we approximate that the frequency splittings (which are computed and marked \textbf{Approx.} in Figures \ref{fig:l1_pol} and \ref{fig:l2_pol}) are given as
\begin{align}
    \label{Approx}
    \left( \delta\omega_{n\ell m} \right)_{\rm{Approx.}} &\approx \zeta_{n\ell}\left( \delta\omega_{n\ell m} \right)_{g}.
\end{align}

We find that for cases where $\zeta_{n\ell}\rightarrow 1$, splitting values, upon calculating with Eqns.~\eqref{eq10} and \eqref{Approx} and comparing them, are extremely close for both $\ell=1$ (up to $\sim$2\% relative deviation) and $2$ (up to $\sim$2\% relative deviation). 
For unperturbed frequencies with small $\zeta_{n\ell}$, we find that there are noticeable deviations (relative deviation of up to $\sim$2\% for $\ell=1$ and up to $\sim 30\%$ for $\ell=2$) of the approximated splittings from the splittings calculated using the formulae provided by \cite{Dasetal20}.

\subsubsection{Splittings due to mixed stable magnetic field in the core}
\label{App:Stable_field_splitting}
\begin{figure}[]
\centering
\begin{subfigure}[]
  {\includegraphics[width=0.47\linewidth]{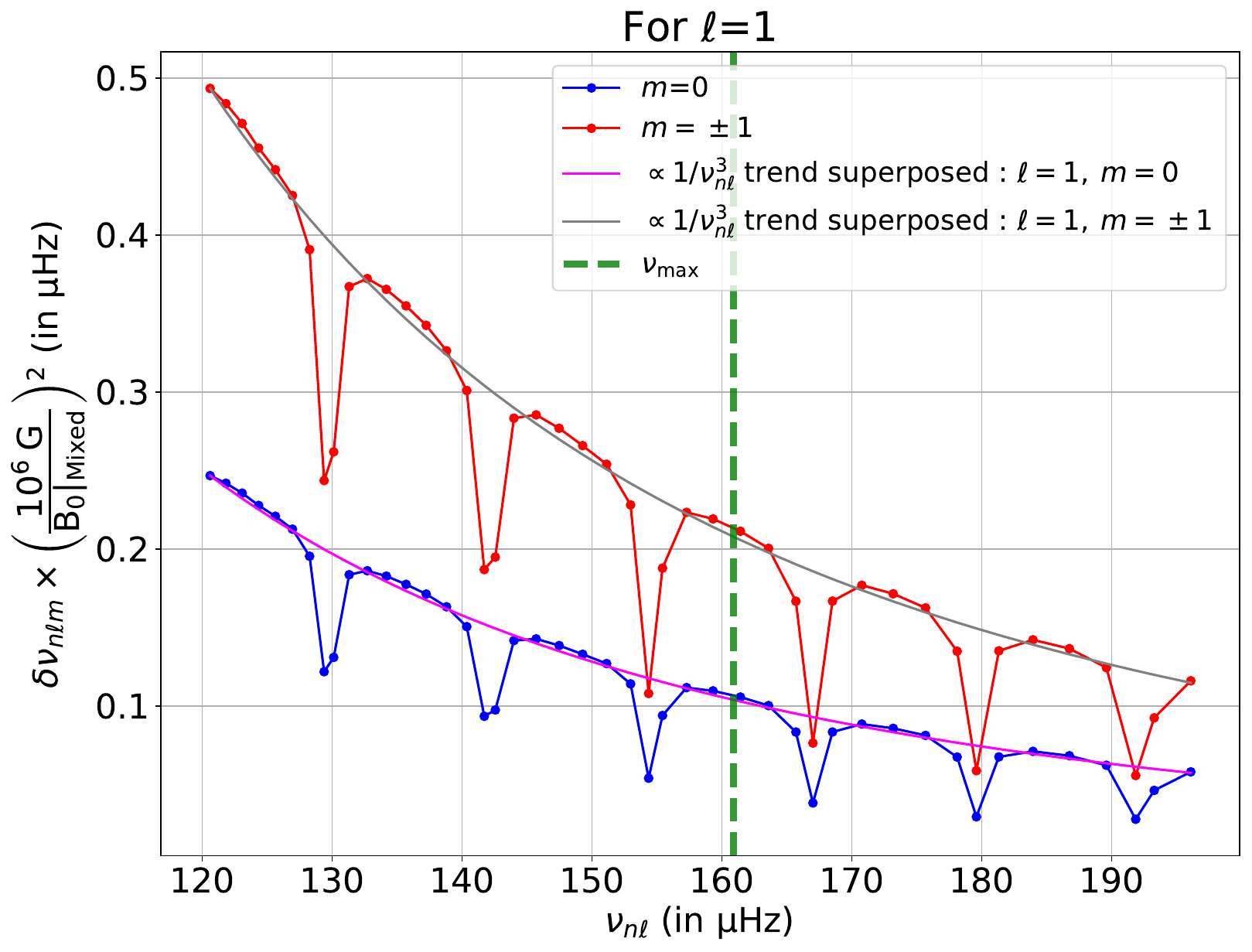}
  \label{fig:l1_Bugnet}}
\end{subfigure}%
\begin{subfigure}[]
  {\includegraphics[width=0.475\linewidth]{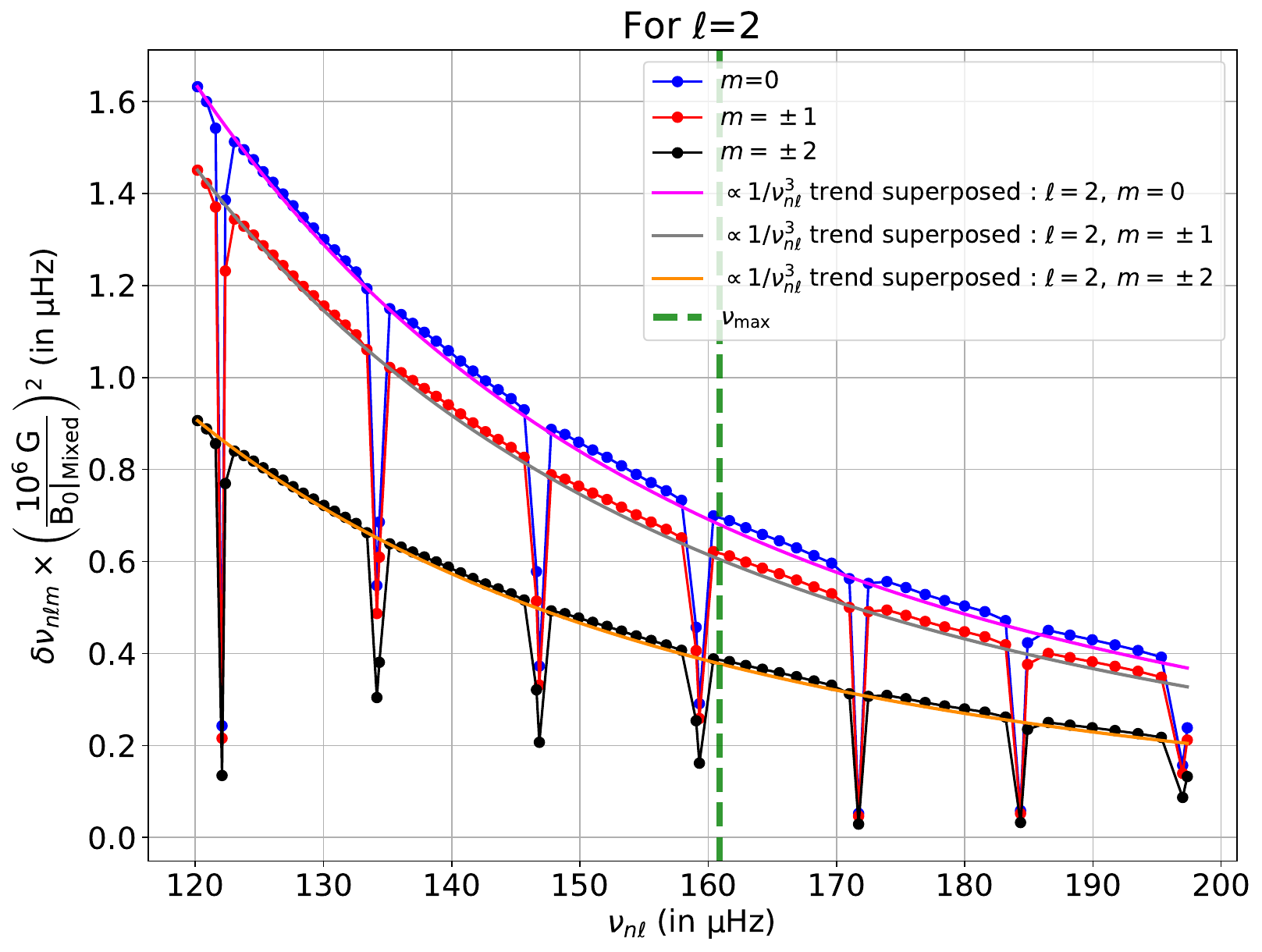}
  \label{fig:l2_Bugnet}}
\end{subfigure}
\caption{Frequency splittings obtained for (a) $\ell=1 \;\rm{and}$ (b) $\ell=2$ mixed modes due to the magnetic field configuration as prescribed by \cite{Bugnetetal2021} with field amplitude scaling factor $B_0\bigr|_{\rm{Mixed}}$.}
\label{fig:dnu_Bugnet}
\end{figure}

Next, we present in Figure~\ref{fig:dnu_Bugnet} the splittings expected due to the stable magnetic field configuration inside the radiative core of an RGB star as described in Section~\ref{B_topologies} above. 

We have also computed the splittings considering only the poloidal part of this field and found that there is no noticeable difference between the two. This again verifies the fact that the toroidal part has no significant contribution to the observable. The $g$-dominant modes still have a very prominent $\nu_{n\ell}^{-3}$ trend. \sbh{If the RG model was observed by \textit{Kepler} for 4 years ($\nu_{\rm{res}}\approx7.9\:\rm{nHz}$) and this mixed field topology was used, then the splittings for the $g$-dominated dipole mode closest to $\nu_{\rm{max}}$ would be detectable for $B_0\gtrsim 0.2$ MG}. 

\cite{MathisBugnetetal2021} have demonstrated in their work the dependence of the frequency splittings for $g$-dominated and $p$-dominated mixed modes on the different magnetic field components. Upon inclusion of the approximation that the horizontal motions of the $g$-dominant modes and radial motions of the $p$-dominant modes are much larger in comparison to their complements, they also found that $g$-dominated modes are impacted significantly by the radial magnetic field within the core whereas the $p$-dominated modes are affected by the horizontal (tangential) components of the magnetic field in the envelope.

\begin{figure}
\centering
\begin{subfigure}[For $\ell=1$]
  {\includegraphics[width=0.455\linewidth]{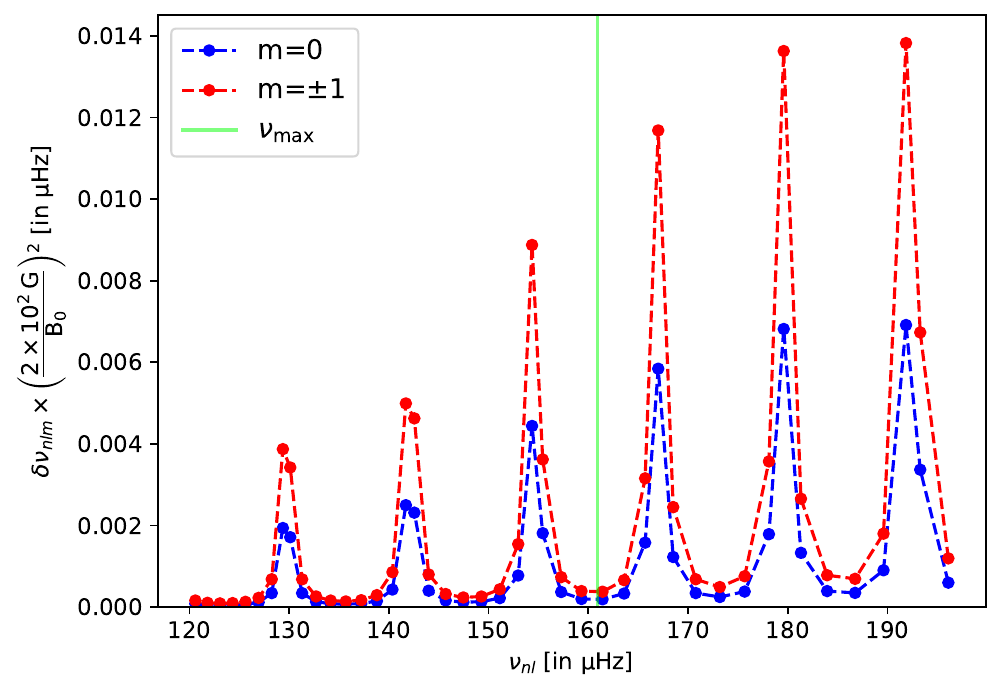}}
\end{subfigure}%
\begin{subfigure}[For $\ell=1$, and scaled by $\nu_{n\ell}^{-2.5}(1-\zeta_{n\ell})^{-1}$]
  {\includegraphics[width=0.435\linewidth]{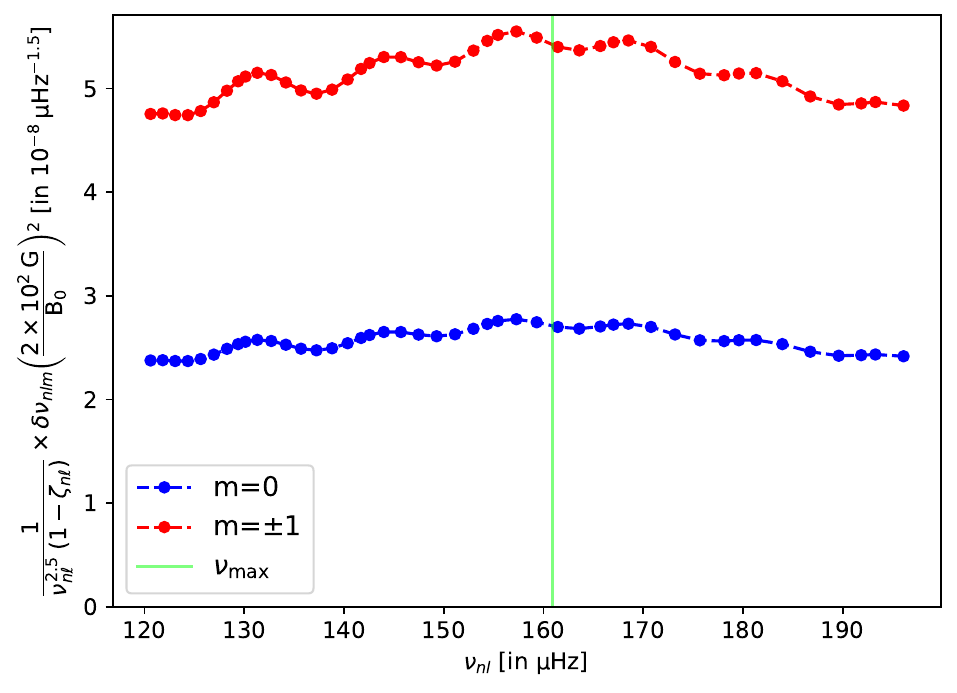}}
\end{subfigure}
\begin{subfigure}[For $\ell=2$]
  {\includegraphics[width=0.46\linewidth]{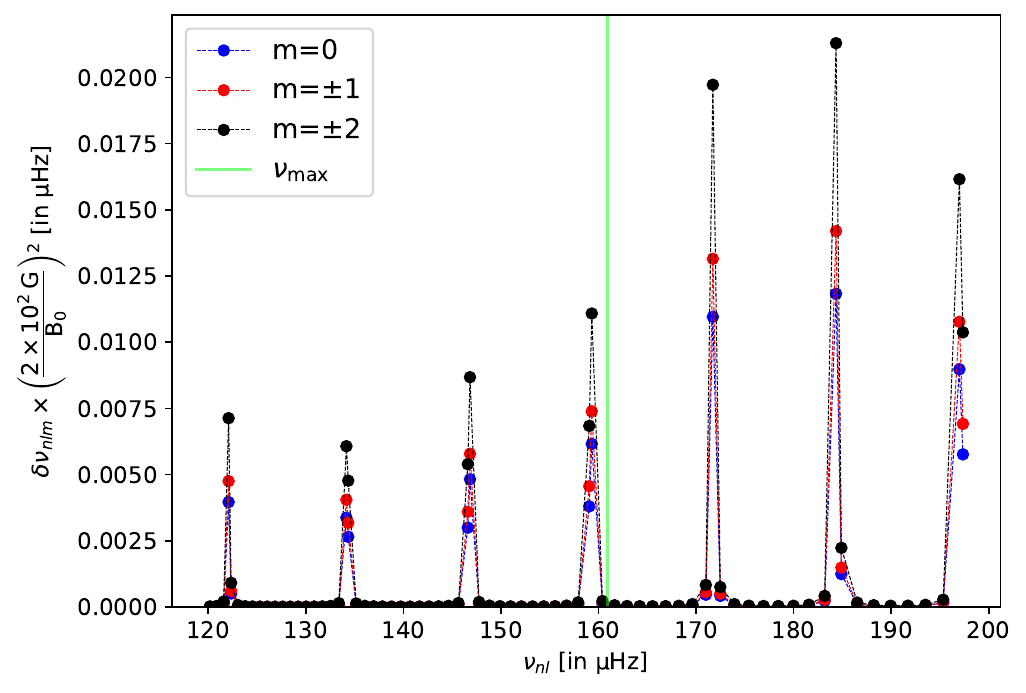}}
\end{subfigure}%
\begin{subfigure}[For $\ell=2$, and scaled by $\nu_{n\ell}^{-2.5}(1-\zeta_{n\ell})^{-1}$]
  {\includegraphics[width=0.435\linewidth]{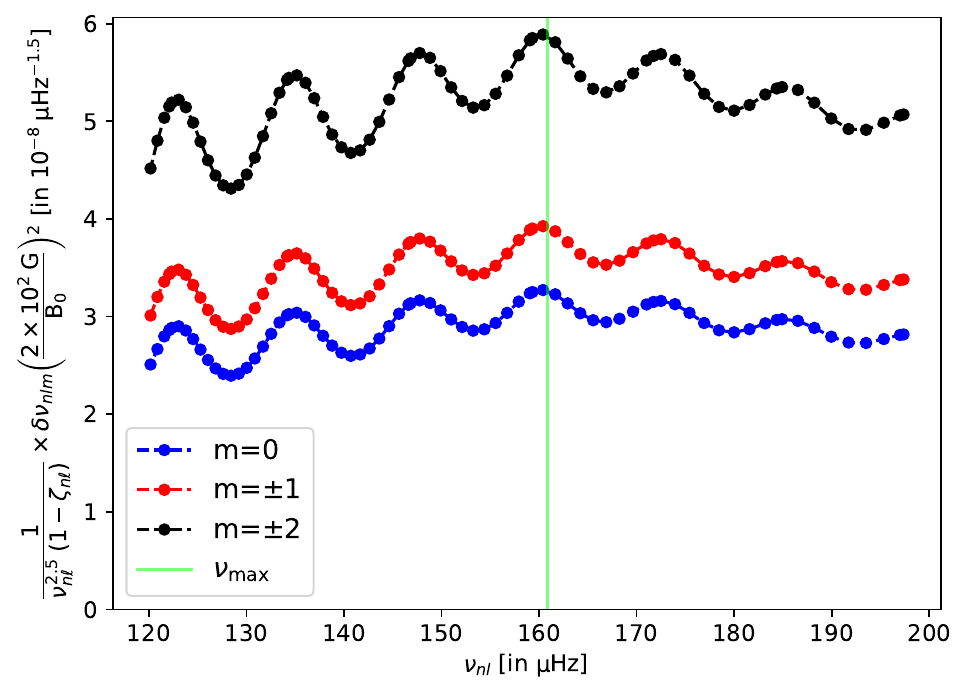}}
\end{subfigure}
\caption{Splittings for Case (C.2.4.1).}
\label{fig:env_tor}
\end{figure}

\subsubsection{More magnetic field topologies} 
\label{appendix:b}

\begin{deluxetable}{||c|c||}[ht]
\label{Table2}
\tablewidth{290pt}
\tabletypesize{\small}

\tablecaption{The magnetic field used in 2 cases}

\tablenum{1}\label{table:appendix_1}

\tablehead{\colhead{\textbf{Case}} & \colhead{\textbf{B-field dominating in envelope} ($0\leq r \leq 1$)}  } 

\startdata
(C.2.4.1) & Toroidal $\alpha(r)$: $B_0\times \frac{1}{4}\left[ 1+\tanh\lbrace 50(r/R_{\rm{star}}-0.15) \rbrace \right]$ \\
(C.2.4.2) & Poloidal $\beta(r)$: $B_0\times \frac{1}{4}\left[ 1+\tanh\lbrace 50(r/R_{\rm{star}}-0.15) \rbrace \right]$   \\
\enddata
\end{deluxetable}

Finally, we calculate the splittings in the normal modes as would be observed for two magnetic field configurations where the fields dominate in the envelope of the star. The two topologies taken into consideration are shown in Table~\ref{table:appendix_1}.

\begin{itemize}

\item \textbf{Case (C.2.4.1)}: The splittings are almost 0 for $g$-dominant modes but bigger for the $p$-dominant modes due to the presence of $B_\varphi$ near the surface, as shown in Figure~\ref{fig:env_tor}. The splittings are found to be approximately proportional to $(1-\zeta_{n\ell})\nu^{2.5}$. \sbh{The detectability of this topology has been discussed in Section~\ref{section:Detectability}}.

\item \textbf{Case (C.2.4.2)}: Here, we again find that the splittings are almost 0 for $g$-dominant modes, as there is no radial magnetic field pressure within the core. The $B_{\theta}$ component of the poloidal B-field in the envelope impacts the $p$-dominant modes and the splittings are nearly 4 times that of the values in case C.2.4.1 for the same value of $B_0$, and follow the same approximate relation as well. Both of these show that the high-frequency $p$-dominated modes can be indeed used to probe the near-surface tangential magnetic fields, where effects due to the radial magnetic field in the core are also significantly smaller.

\end{itemize}

\section{Kernels for the Red-giant Phase}

\label{appendix:l0_mode}

\begin{figure}[H]
    \centering
    \subfigure{
    \includegraphics[width=1.0\textwidth]{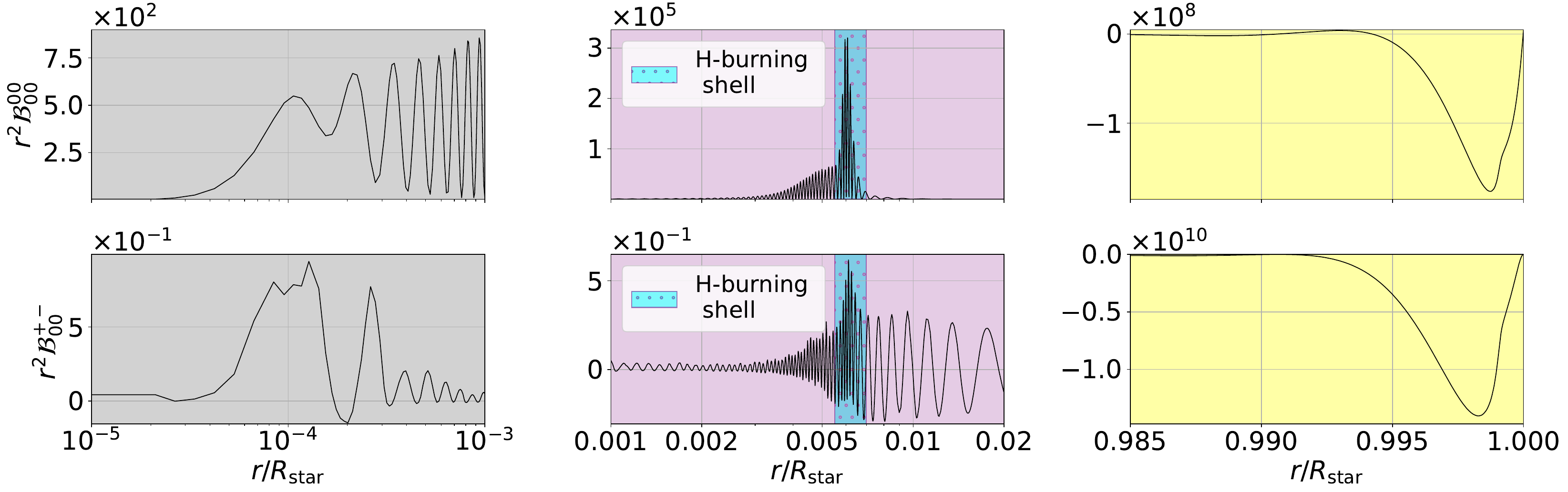}
    \label{fig:f14a}
    }
    \subfigure{
    \includegraphics[width=1.0\textwidth]{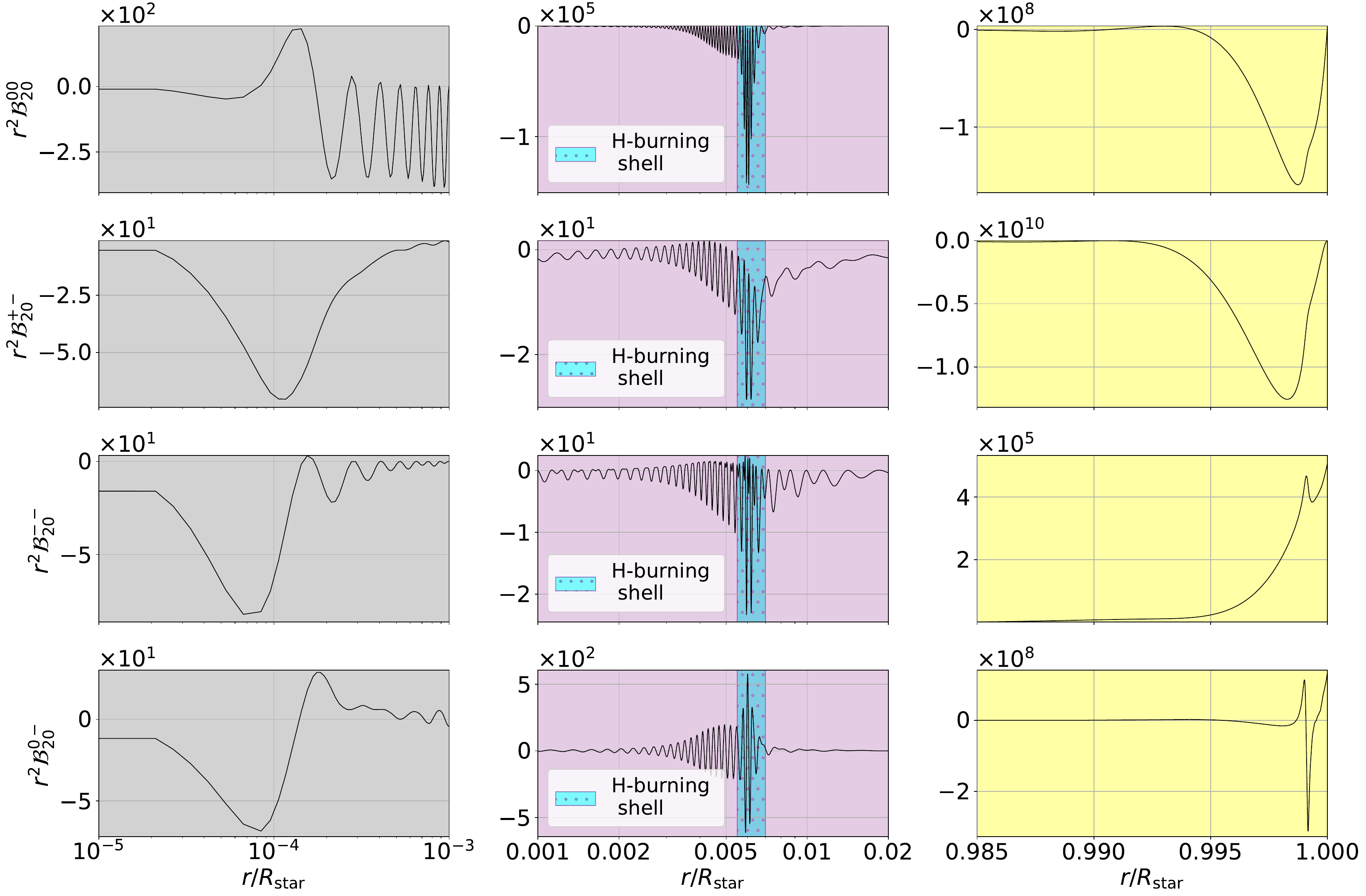}
    \label{fig:f14b}
    }
    \caption{Same as Figure~\ref{fig:f17}, but for \textit{p}-dominated $\ell=1$ mode with an unperturbed frequency of $179.598\:\mu\rm{Hz}$ (point $\color{red}p_1$ in Figure~\ref{Zeta_plot}).}
    \label{fig:f14}
\end{figure}

\begin{figure}[H]
    \centering
    \subfigure{
    \includegraphics[width=1.0\textwidth]{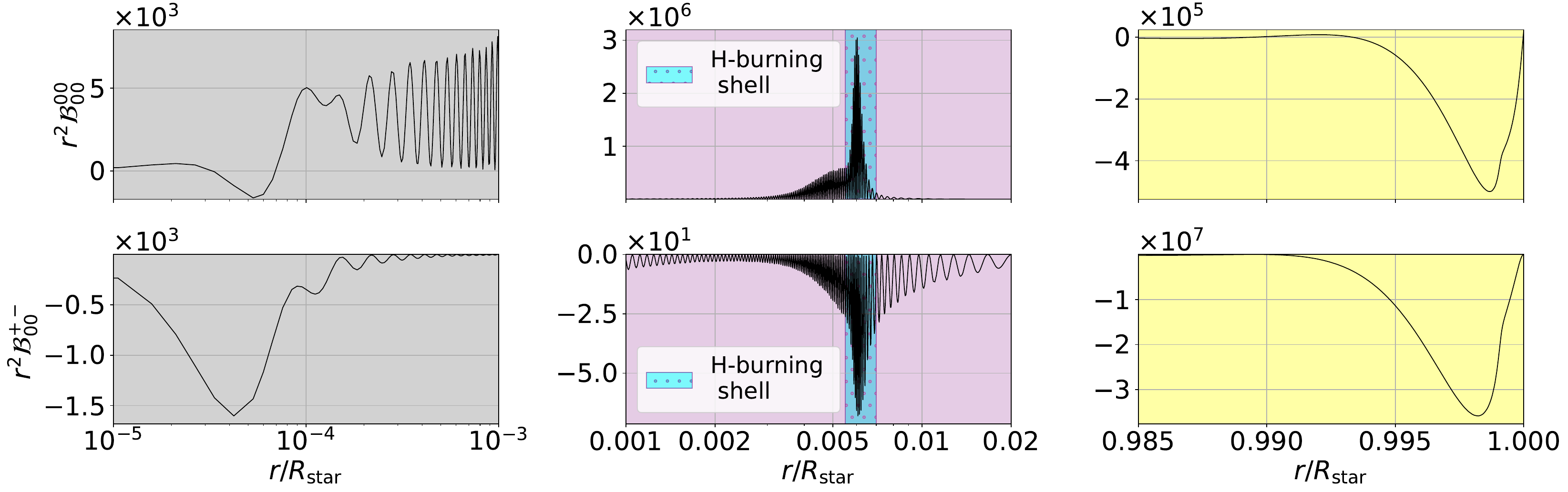}
    \label{fig:f61a}
    }
    \subfigure{
    \includegraphics[width=1.0\textwidth]{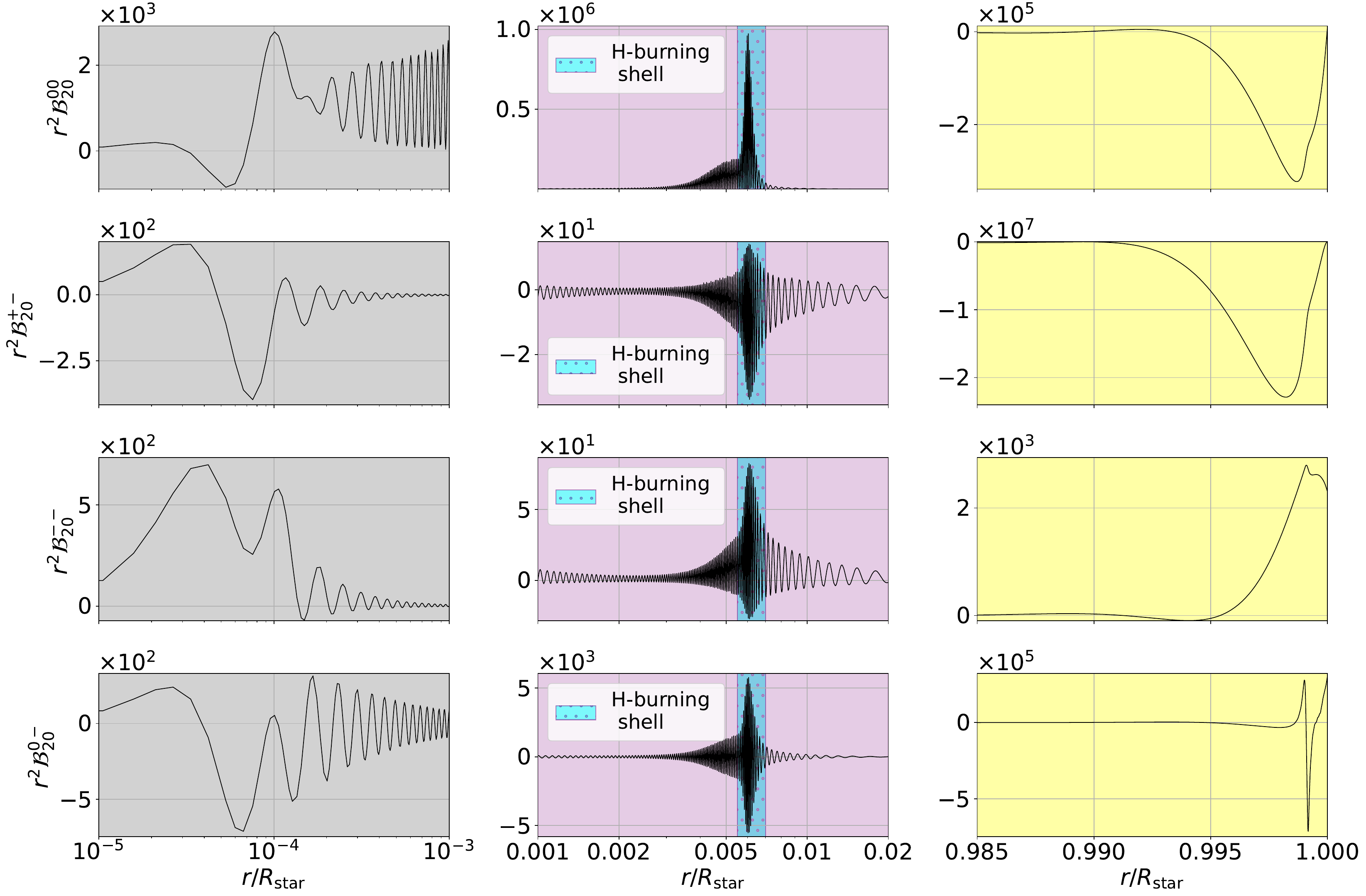}
    \label{fig:f61b}
    }
    \caption{Same as Figure~\ref{fig:f17}, but for \textit{g}-dominated $\ell=2$ mode with an unperturbed frequency of $168.247\:\mu\rm{Hz}$ (point $\color{blue}g_2$ in Figure~\ref{Zeta_plot}).}
    \label{fig:f61}
\end{figure}

\begin{figure}[H]
    \centering
    \subfigure{
    \includegraphics[width=1.0\textwidth]{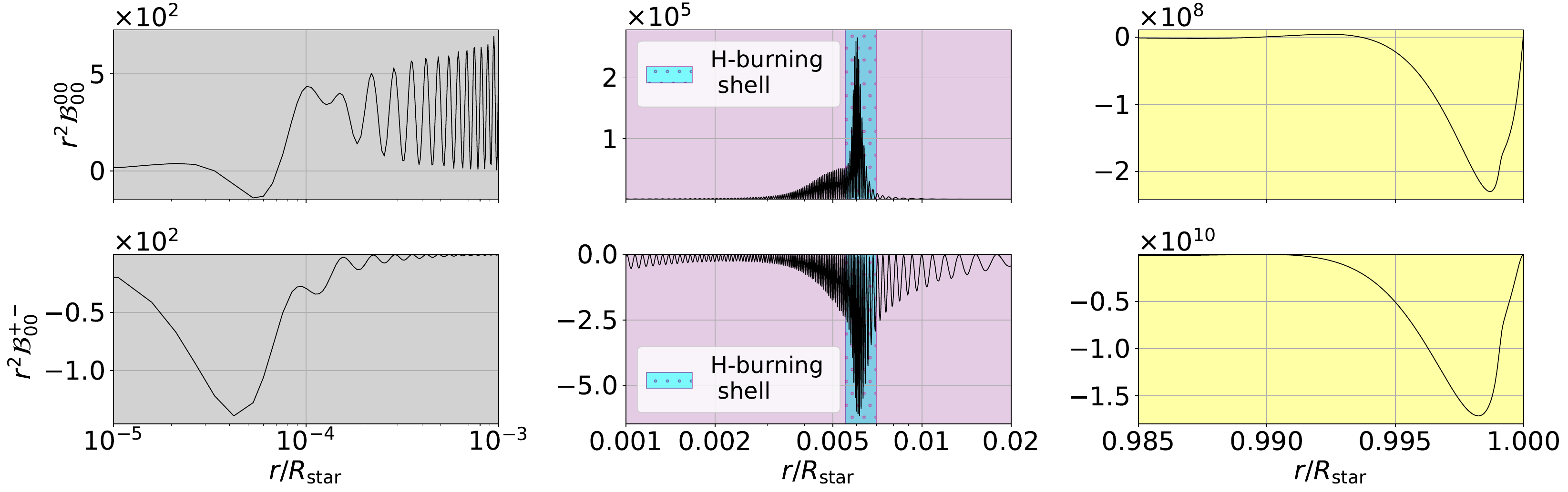}
    \label{fig:f58a}
    }
    \subfigure{
    \includegraphics[width=1.0\textwidth]{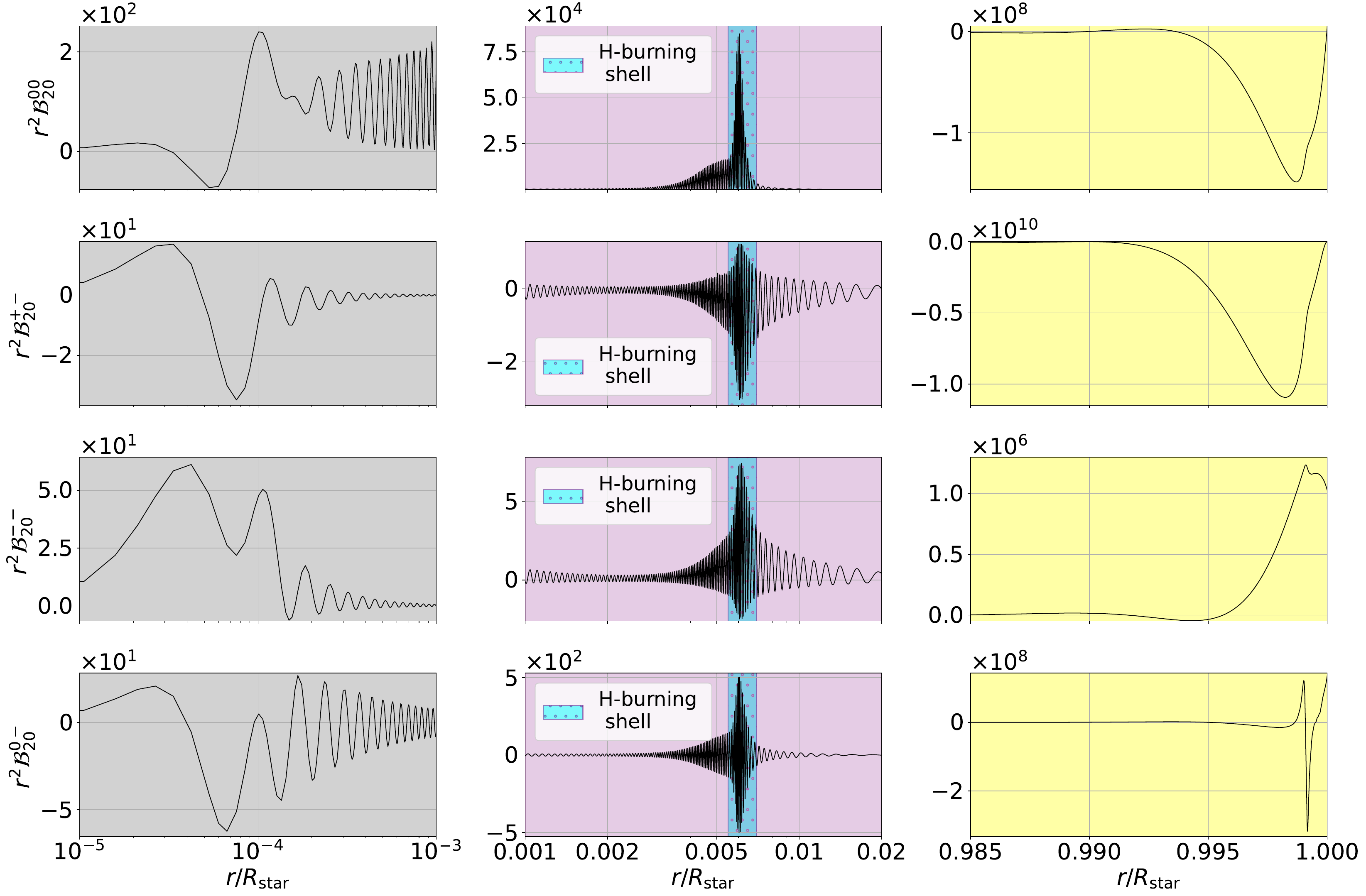}
    \label{fig:f58b}
    }
    \caption{Same as Figure~\ref{fig:f17}, but for \textit{p}-dominated $\ell=2$ mode with an unperturbed frequency of $171.729\:\mu\rm{Hz}$ (point $\color{blue}p_2$ in Figure~\ref{Zeta_plot}).}
    \label{fig:f58}
\end{figure}

\begin{figure}[H]
\centering
\subfigure{
\includegraphics[width=1.0\textwidth]{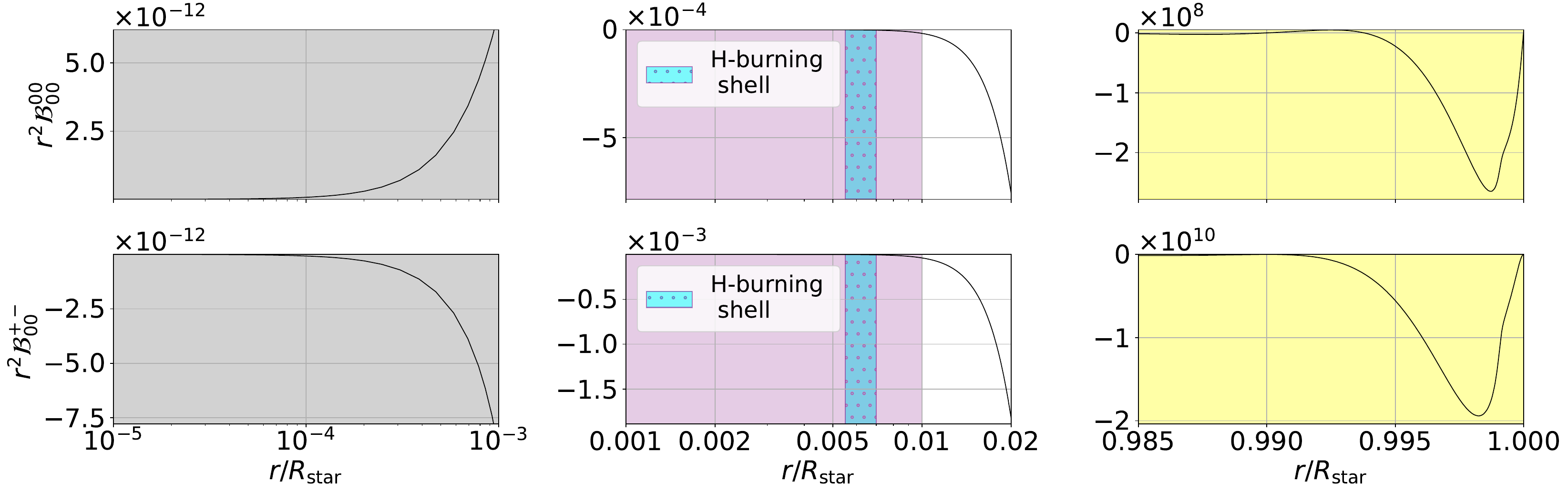}
\label{fig:f3a}
}
\caption{Same as Figure~\ref{fig:f17}, but an $\ell=0$ mode with an unperturbed frequency of $173.203\:\mu\rm{Hz}$ (point $\color{gray}p_0$ in Figure~\ref{Zeta_plot}). We do not plot the kernels for $s=2$ as we have already proven that they are always 0 for $\ell=0$. The splittings are much more sensitive to the sub-surface magnetic fields than those inside the core.}
\label{fig:f3}
\end{figure}

\section{Kernels for the Sub-giant Phases}

\begin{figure}[H]
    \centering
        \begin{subfigure}{ \includegraphics[width=\linewidth]{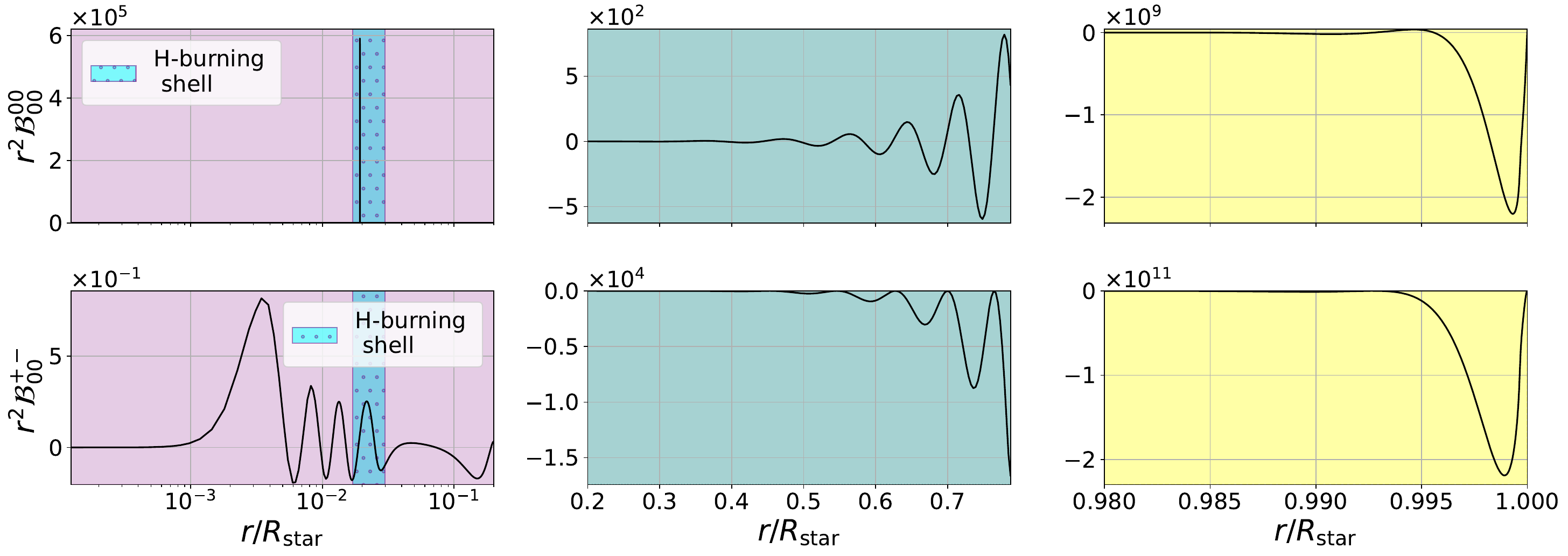} \label{fig:kern_mid_f8_0}
    }
    \end{subfigure}
    \begin{subfigure}{ \includegraphics[width=\linewidth]{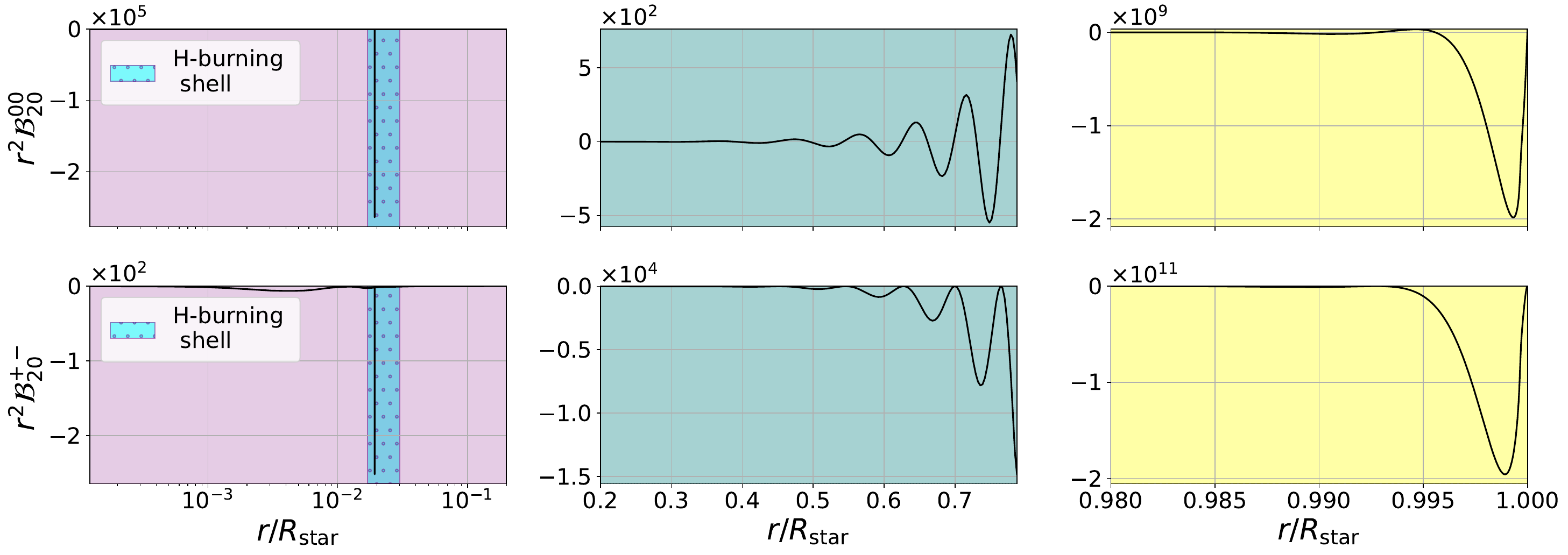}\label{fig:kern_mid_f8_2}} 
    \end{subfigure}
    \caption{Prominent trends in the magnetic field sensitivity kernels $r^{2}\mathcal{B}_{s0}^{00}$ and $r^{2}\mathcal{B}_{s0}^{+-}$ in $\rm{Hz}^{2}cm^{-1}G^{-2}$ for the MSG phase. In this figure, we choose $p$-dominated $\ell=1$ mode with an unperturbed frequency $\nu_{n\ell}$ of $588.108\:\mu\rm{Hz}$. The top two rows and bottom two rows show the $s=0$ and $s=2$ components of the kernels respectively in the deep regions of the radiative core which contains the inner core with the H-burning shell (\textit{left-most column}), outer radiative core (\textit{middle column}) and near the surface (\textit{right-most column}).}
    \label{fig:kern_mid}
\end{figure}
\begin{figure}[H]
    \begin{subfigure}{ \includegraphics[width=\linewidth]{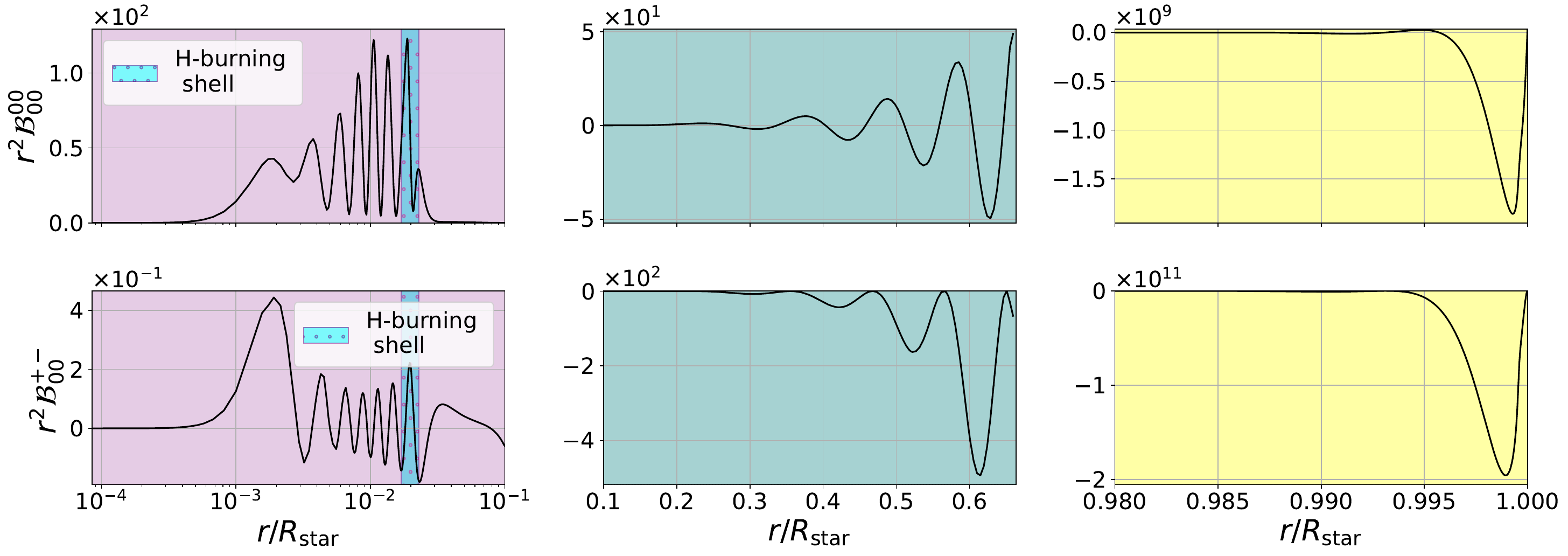} \label{fig:kern_late_f15_0}
    }  
    \end{subfigure}
    \begin{subfigure}{ \includegraphics[width=\linewidth]{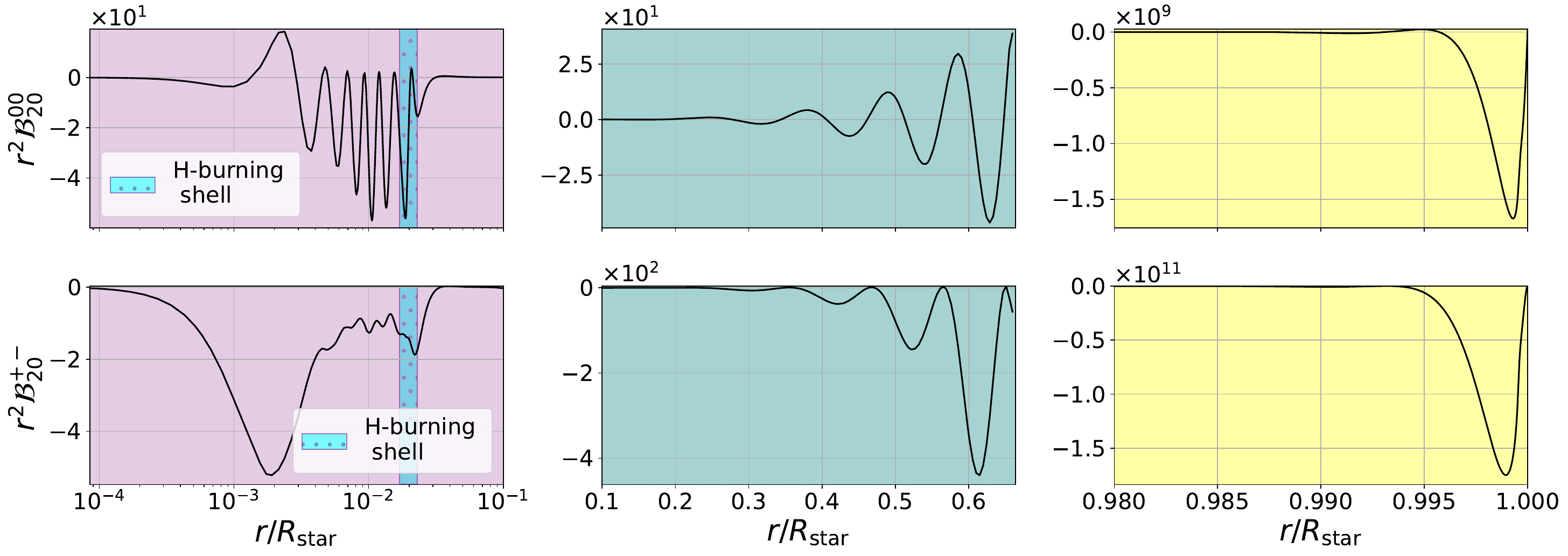}\label{fig:kern_mid_f15_2}}
    \end{subfigure}
    \caption{Similar to Figure~\ref{fig:kern_mid}, but for a $p$-dominated $\ell=1$ mode in the LSG phase with an unperturbed frequency $\nu_{n\ell}$ of $554.267\:\mu\rm{Hz}$}
    \label{fig:kern_late}
\end{figure}

\end{document}